\documentclass[%
reprint,
amsmath,amssymb,
aps,
pra,
]{revtex4-2}

\usepackage{graphicx}% Include figure files
\usepackage{dcolumn}% Align table columns on decimal point
\usepackage{bm}% bold math
\begin{document}

%\preprint{APS/123-QED}

\title{Spin and Orbital Angular Momentum of Coherent Photons in a Waveguide}% Force line breaks with \\

\author{Shinichi Saito}
 \email{shinichi.saito.qt@hitachi.com}
\affiliation{Center for Exploratory Research Laboratory, Research \& Development Group, Hitachi, Ltd. Tokyo 185-8601, Japan.}%Lines break automatically or can be forced with \\
%\author{Isao Tomita}%
%\affiliation{% Department of Electrical and Computer Engineering, National Institute of Technology, Gifu College, 2236-2 Kamimakuwa, Motosu, Gifu 501-0495, Japan.}%

\date{\today}% It is always \today, today, %  but any date may be explicitly specified

\begin{abstract}
Spin angular momentum of a photon corresponds to a polarisation degree of freedom of lights, and such that various polarisation properties are coming from macroscopic manifestation of quantum-mechanical properties of lights.
An orbital degree of freedom of lights is also manipulated to form a vortex of lights with orbital angular momentum, which is also quantised.
However, it is considered that spin and orbital angular momentum of a photon cannot be split from the total orbital angular momentum in a gauge-invariant way.
Here, we revisit this issue for a coherent monochromatic ray from a laser source, propagating in a waveguide. 
We obtained the helical components of spin and orbital angular momentum by the correspondence with the classical Ponyting vector.
By applying a standard quantum field theory using a coherent state, we obtained the gauge-independent expressions of spin and orbital angular momentum operators.
During the derivations, it was essential to take a finite cross-sectional area into account, which leads the finite longitudinal component along the direction of the propagation, which allows the  splitting.
Therefore, the finite mode profile was responsible to justify the splitting, which was not possible as far as we are using plane-wave expansions in a standard theory of quantum-electrodynamics (QED).
Our results suggest spin and orbital angular momentum are well-defined quantum-mechanical freedoms at least for coherent photons propagating in a waveguide and in a vacuum with a finite mode profile.
\end{abstract}
%Max 600 characters in PRL, 500 words for PRA & PRB

\maketitle
%\tableofcontents

%\begin{figure}[h]
%\begin{center}
%\includegraphics[width=8.6cm]{Fig1.pdf}
%\caption{
 %New $3D$ carbon allotrope, Hopfene.
%(a) Crystal structure. 
%}
%\end{center}
%\end{figure}

\section{Introduction}
Newton recognised the polarisation degree of freedom in lights and called it as {\it "sides"}\cite{Newton1730}, whose properties were successfully elucidated by Stokes \cite{Stokes51} and Poincar\'e \cite{Poincare92} within the framework of classical mechanics \cite{Max99,Jackson99,Yariv97}.
Later, the discoveries of Plank and Einstein led to the establishment of quantum mechanics, and the wave-particle duality is unified in the form of a light quanta, a photon \cite{Dirac30,Baym69,Sakurai14,Sakurai67}.
From a quantum mechanical point of view, the polarisation is understood as spin of a photon  \cite{Baym69,Sakurai14,Sakurai67}.
There are a lot of experimental evidences to believe that spin of a photon is 1 in the unit of Dirac constant, $\hbar$, which is the Plank constant, $h$, divided by $2\pi$ \cite{Dirac30,Baym69,Sakurai14,Sakurai67}.
The most standard justification of spin 1 nature of a photon is the selection rule of absorption and emission of a photon by electrons in an atom \cite{Dirac30,Baym69,Sakurai14,Sakurai67}. 
Spin $1/2$ nature of an electron and the integer quantisation of orbital angular momentum of electrons in a spherical potential are well-established, and the absorption and emission of a photon involves the change of $\hbar$ in the orbital angular momentum of electronic states \cite{Dirac30,Baym69,Sakurai14,Sakurai67}.
Spin 1 of a photonic state implies that there are potentially 3 orthogonal states, quantised along the direction of the propagation. 
However, a photon is propagating at the speed of light, $c$, in the vacuum, and it is described by a transverse wave.
Consequently, electromagnetic fields of photons are oscillating perpendicular to the direction of the propagation, such that we can observe only 2 orthogonal polarisation modes and the spin 0 component is not observed \cite{Sakurai67}.
As a result, the polarisation state of a photon \cite{Goldstein11,Gil16,Pedrotti07,Hecht17} is described as a quantum-mechanical 2-level system using the SU$(2)$ Lie algebra \cite{Jones41,Payne52,Max99,Yariv97,Baym69,Sakurai14,Jackson99,Yariv97,
Collett70,Luis02,Luis07,Bjork10,Castillo11,Sotto18,Sotto18b,Sotto19}.
Therefore, it is natural to believe that a photon has inherent spin 1 as a quantum-mechanical degree of freedom.

It was rather recently that orbital angular momentum \cite{Allen92,Enk94,Leader14,Barnett16,Yariv97,Jackson99,Grynberg10,Bliokh15} 
of a light is considered in addition to spin.
Allen and his co-workers demonstrated that the orbital angular momentum of the Laguerre-Gauss mode of a light is quantised in the unit of $\hbar$ \cite{Allen92}.
In their derivation, the classical electromagnetic wave in the Laguerre-Gauss mode under Lorentz gauge is used and the orbital angular momentum was calculated by using the classical Poynting vector, and the quantisation of electromagnetic fields as photons were taken into account at the end of the calculation to estimate the orbital angular momentum per photon \cite{Allen92}.
In this pioneering work, they obtained that the orbital angular momentum of a photon is quantised in the unit of $\hbar$ \cite{Allen92}.
This suggests that the orbital angular momentum is also well-defined quantum-mechanical degree of freedom in addition to spin. 

However, this native expectation is subsequently denied, because the gauge-independent expressions of spin and orbital angular momentum for photons were not obtained  \cite{Enk94,Leader14,Barnett16,Chen08,Ji10}.
It is now generally believed that spin and orbital angular momentum of photons are not separately well-defined in a proper unique gauge invariant way \cite{Enk94,Leader14,Barnett16,Chen08,Ji10}.
We will revisit this grand challenge for a monochromatic coherent ray of photons travelling in a waveguide, because lasers \cite{Yariv97} are ubiquitously available these days.
We are interested in laser optic experiments, so that we have not considered the Lorentz invariance, which is important for high-energy physics such as Quantum Chromo-Dynamics (QCD) \cite{Leader14}.
Nevertheless, we have employed the field theory of Quantum-Electro-Dynamics (QED), tailored to consider the Laguerre-Gauss mode in a GRaded-INdex (GRIN) fibre \cite{Kawakami68,Yariv97}.
We show that it is essential to consider the finite size of the mode profile to derive appropriate expressions for spin and optical angular momentum operators.

\section{Classical Electro-Magnetic Waves with Optical Angular Momentum}

Before showing our final results, it would be instructive to start from reviewing classical results for electromagnetic waves and adding some complexities gradually to address what was the potential issue \cite{Allen92,Enk94,Leader14,Barnett16,Yariv97,Jackson99,Grynberg10,Bliokh15}. 
First, we review orbital angular momentum described by a Laguerre-Gauss mode in a free space under the Lorentz gauge \cite{Allen92}.
Then, we confirm that the same result can be obtained by using the Coulomb gauge and compare the difference of gauges.
We also review the impacts of polarisation on optical angular momentum by using a horizontally polarised mode and a circularly polarised mode.
Finally, we extend the analysis for the GRIN waveguide for both polarisations.

\subsection{Lorentz gauge in homogeneous media}
\subsubsection{Lorentz gauge}
Here, we consider a uniform transparent material with the dielectric constant of $\epsilon$ and the permeability of $\mu_0$.
The velocity of the light in the material is given by $v_0=1/\sqrt{\epsilon \mu_0}=c/n_0$, where $n_0=\sqrt{\epsilon/\epsilon_0}$ is the refractive index of the material and $c=1/\sqrt{\epsilon_0 \mu_0}$ is the velocity of the light in a vacuum with the dielectric constant of $\epsilon_0$.
The permeability of the material barely changes in a non-magnetic material, and in the limit of $\epsilon \rightarrow \epsilon_0$ the material is equivalent to the vacuum.
The vector potential ${\bf A}$ and the scalar potential ${\Phi}$ under Lorentz gauge satisfy the following equations \cite{Yariv97,Jackson99}

\begin{eqnarray}
\left (
{\bf \nabla}^2
-
\frac{1}{v_0^2}
\partial_t^2
\right )
{\bf A} 
&=&0 \\
\left (
\nabla^2 
-
\frac{1}{v_0^2}
\partial_t^2
\right )
\Phi
&=&0 \\
{\bf \nabla} \cdot {\bf A} 
+
\frac{1}{v_0^2}
\partial_t
\Phi
&=&0.
\end{eqnarray}
The electric field, ${\bf E}$, and magnetic induction, ${\bf B}$, are obtained by
\begin{eqnarray}
{\bf E}
&=&
-
\nabla \Phi 
-
\partial_t
{\bf A} \label{Eq:E} \\
{\bf B}
&=&
\nabla 
\times
{\bf A},  \label{Eq:B}
\end{eqnarray}
respectively, which immediately gives the electric displacement field ${\bf D}=\epsilon {\bf E}$ and the magnetic field  ${\bf H}={\bf B}/\mu_0$.
We can confirm that Maxwell equations \cite{Jackson99}, 
\begin{eqnarray}
\nabla 
\cdot 
{\bf B}
&=&0 \\
\nabla 
\cdot 
{\bf D}
&=&
0\\
\nabla 
\times
{\bf E}
&=&
-\dot{\bf B}\\
\nabla 
\times
{\bf H}
&=&
\frac{\partial {\bf D}}{\partial t},
\end{eqnarray}
in the absence of the charge $\rho=0$ and the current ${\bf J}=0$ are satisfied under the Lorentz gauge by directly inserting Eqs. (\ref{Eq:E}) and (\ref{Eq:B}).

\subsubsection{Paraxial approximation}
First, let's briefly see an expected electromagnetic wave in a cylindrical coordinate.
From Maxwell equations, we obtain the Helmholtz 
\begin{eqnarray}
\nabla^2 {\bf E}
=
\mu_0 \epsilon 
\frac{\partial^2}{\partial t^2}{\bf E},
\end{eqnarray}
whose solution, polarised along the horizontal direction as an example, is expected to be
\begin{eqnarray}
{\bf E}({\bf r})
&=&
{\bf E}(r,z)\\
&=&
E_0
\psi (r,\phi,z)
{\rm e}^{i(kz-\omega t)}
\hat{\bf x},
\end{eqnarray}
where ${\bf r}=(x,y,z)$ is the Cartesian coordinate, $z$ is the axis along the direction of the propagation, $k=k_{n_0}$ is the wavenumber, $\omega$ is the angular frequency, $t$ is time, $r=\sqrt{x^2+y^2}$ is the radius in the cylindrical coordinate $(r,\phi,z)$, and $\hat{\bf x}$ is the unit vector along the $x$ axis.
$\psi (r,\phi,z)$ describes the mode profile of the field.
If the ray is predominantly propagating along $z$ as an almost collimated beam, we can use a paraxial approximation \cite{Allen92,Yariv97}
\begin{eqnarray}
\frac{\partial^2 \psi}{\partial z^2}
\ll
k \frac{\partial \psi}{\partial z}
, \ 
k^2 \psi ,
\end{eqnarray}
and the Helmholtz equation becomes
\begin{eqnarray}
i\frac{\partial }{\partial z}
\psi
=
-
\frac{1}{2k}
\left(
\frac{\partial^2 }{\partial x^2} 
+
\frac{\partial^2 }{\partial y^2} 
\right)
\psi,
\end{eqnarray}
which is the same form with the non-relativistic Sch\"odinger equation \cite{Simon93,Barnett16b,Baym69,Sakurai14,Chuang09}.
The solution in a cylindrical coordinate \cite{Allen92,Yariv97} is obtained as 
\begin{eqnarray}
\psi(r,\phi,z)
&=&
\sqrt{
\frac{2}{\pi}
\frac{n!}{(n+|m|)!}
}
\frac{1}{w}
\left(
\frac{\sqrt{2}r}{w}
\right)^{|m|}
L_n^{|m|} 
\left(
2
\left(
  \frac{r}{w}
\right)^2 
\right) \nonumber \\
&&
{\rm  e}^{-\frac{r^2}{w^2}}
{\rm  e}^{ik\frac{r^2}{2 R}}
{\rm  e}^{im\phi}
{\rm  e}^{-i(2n+|m|+1)\tan^{-1}(z/z_0)},
\end{eqnarray}
where $L_n^{|m|}$ is the associate Laguerre function, $n$ is the radial number of nodes, $m$ is the quantum number for orbital angular momentum, $\phi$ is the angle in the cylindrical coordinate, the dispersion is give by $\omega=v_0 k=v_0 k_{n_0}=v_0 n_0 k_0=c k_0$ with the wavenumber in the vacuum, $k_0=2\pi/\lambda$ for the wavelength of $\lambda$, the beam waist is given by $w=w(z)=w_0\sqrt{1+ (z/z_0)^2}$, where $w_0$ is the waist at the origin $z=0$, the Rayleigh length (the confocal parameter) is $z_0=k_{n_0} w_0^2/2$, and the radius of the spherical phase is $R(z)=z+z_0^2/z$.

\subsubsection{Topological charge}
In the mode profile of $\psi(r,z)$, the phase factor of ${\rm  e}^{im\phi}$ is very important to describe the optical orbital angular momentum of $\hbar m$ \cite{Allen92}.
Another important feature of the Laguerre-Gauss mode is the Gouy phase \cite{Pancharatnam56,Berry84,Tomita86,Allen92,Simon93,Hamazaki06,Bliokh09}
\begin{eqnarray}
\phi_{\rm G}=(2n+|m|+1)\tan^{-1}(z/z_0).
\end{eqnarray}
The phase of the Laguerre-Gauss mode as a scalar field of $\psi(r,z)$ is given by
\begin{eqnarray}
\phi_{\rm LG} (r, \phi , z)
=
k_{n_0} \frac{r^2}{2 R(z)}
+
m \phi
-
\phi_{\rm G}(z).
\end{eqnarray}
We consider the gradient of the phase in the cylindrical coordinate $(r,\phi,z)$
\begin{eqnarray}
\nabla \phi_{\rm LG}
=
\left (
\frac{k_{n_0}  r}{R}, 
\frac{m}{r}, 
\frac{\partial \phi_{\rm LG} }{\partial z}
\right ),
\end{eqnarray}
where the unit vectors along $r$ and $\phi$ are obtained by a rotation of the unit vectors in $(x,y)$ coordinate (Fig. 1) as 
\begin{eqnarray}
\left (
  \begin{array}{c}
   \hat{\bf r}
\\
   \hat{\bf \Phi}
  \end{array}
\right)
= 
\left (
  \begin{array}{cc}
   \cos \phi & \sin \phi
\\
   -\sin \phi & \cos \phi
  \end{array}
\right)
\left (
  \begin{array}{c}
   \hat{\bf x}
\\
   \hat{\bf y}
  \end{array}
\right).
\end{eqnarray}
In particular, it is important to be aware that the unit vectors $\hat{\bf r}=\hat{\bf r}(\phi)$ and $ \hat{\bf \Phi}= \hat{\bf \Phi}(\phi)$ depend on $\phi$.

\begin{figure}[h]
\begin{center}
\includegraphics[width=4cm]{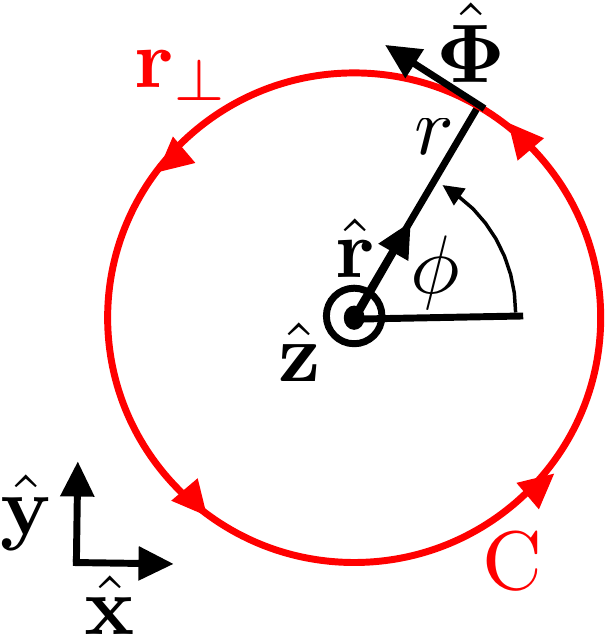}
\caption{
Topological charge by optical angular momentum.
The contour integral along the closed circle ${\rm C}$ in a cylindrical coordinate $(r,\phi,z)$ is considered, which gives the winding number, called the topological charge.
Note that the existence of the node at the origin is required to support the topological charge.
We assume that the light is propagating along $z$ and the direction of rotation is defined to be positive, if the rotation is anti-clock-wise seen from the top of the $z$ axis in the detector side.
In this definition, the left-circular {\it vortexed} state, shown above, corresponds to the positive winding number, corresponding to the quantised optical angular momentum pointing towards the positive $z$ direction.  
}
\end{center}
\end{figure}

We consider the contour integral for the closed path ${\rm C}$ (Fig. 1) for the gradient of the phase as 
\begin{eqnarray}
I
&=&
\frac{1}{2 \pi}
\oint_{\rm C}
\nabla \phi_{\rm LG}
\cdot
d {\bf r}_{\perp} \\
&=&
\frac{1}{2 \pi}
\int_{0}^{2\pi} 
\frac{m}{r} 
r d\phi
 \\
&=&m,
\end{eqnarray}
which is the {\it winding number}, called the {\it topological charge}.
Please note that $I$ is the dimensionless number, such that it is confusing to call it as charge.
The winding number would be a more precise word, instead.
Nevertheless, the existence of the finite $I$ is responsible for twisting lights to form a vortex with optical angular momentum, such that it works like a source of generating a vortex of the electric field, similar to charge, which is the source of divergence of the electric field.
In order to sustain the vortex, it is essential to have a node within the inside of the contour, ${\rm C}$.
Otherwise, the integration of the gradient simply becomes zero as 
\begin{eqnarray}
\frac{1}{2 \pi}
\int_{\bf r_0}^{\bf r_1}
\nabla \phi_{\rm LG}
\cdot
d {\bf r}
&=&
\frac{1}{2 \pi}
\left (
\phi_{\rm LG} ({\bf r_1})
-
\phi_{\rm LG} ({\bf r_0})
\right ) \\
&\rightarrow&
0
\end{eqnarray}
in the limit of closed integration circle, ${\bf r_1}\rightarrow{\bf r_0}$.
This means that there is a node required at the centre of the beam in order to sustain non-zero topological charge, which is guaranteed in the Laguerre-Gauss mode with a power of $r^{|m|}$ for $m \neq 0$.
Please also note that {\it there is no singularity in the electric field but there is a node (zero point)}.
In other words, the amplitude becomes zero, such that it is impossible to define a phase at the node.
Therefore, we can also claim that there is a singularity in the phase, if we try to define the phase at the node.
This is consistent with the view that we should not expect singularities in observables like electric and magnetic fields. 
The topological charge simply corresponds to a node.

Another important source of an unnecessary confusion is the definition of the direction of the rotation of the vortex.
Depending on whether we are observing the vortex from the detector side or from the source side, the rotation will become opposite.
In our paper, we define the positive rotation for the left-circular vortex, seen from the detector side, which corresponds to the positive topological charge, $m>0$ (Fig. 1).
We usually use the right-handed coordinate for Cartesian coordinate of $(x,y,z)$, and we are assuming that the light is propagating towards the positive $z$ direction.
In the descriptions of the rotation of the vortex and the polarisation ellipse, we think it is natural to describe in the $(x,y)$ plane, seen from the top of the $z$ axis, corresponding to seeing from the detector side for a ray pointing towards $z$ (Fig. 1).
In the cylindrical coordinate, a standard definition of the angle $\phi$ is measured from the $x$ axis in the anti-clock-wise direction, such that $x=\cos \phi$ and $y=\sin \phi$.
In this coordinate, the left-circulation (anti-clock-wise) of the contour corresponds to the positive topological charge, and we will confirm that this corresponds to the quantised orbital angular momentum of $\hbar m {\mathcal N}$, pointing towards the direction of the propagation $z>0$, where ${\mathcal N}$ is the number of photons in the ray.
Consequently, if the rotation of the vortex rotates in the opposite direction, which is the right-circular (clock-wise) rotation, seen from the detector side, the orbital angular momentum of the vortex becomes negative, as $\hbar m {\mathcal N}<0$.

Similar to the polarised lights, we would like to propose to call as {\it vortexed} lights for the ray with a vortex of non-zero topological charge.

\subsubsection{Convention of the time average}
The time dependence of the ray, we are considering in this paper, is simply described by ${\rm e}^{-i \omega t}$.
Strictly, both {\bf E} and {\bf B} must be real, since these are observables but it is easier to use complex valuables, instead, and to make a convention to take the real part at the end of the calculations \cite{Yariv97}.
In this convention, it is important to take a factor of $2$ for the products, because the time average of $\cos^2 (\omega t)$ or $\sin^2 (\omega t)$ must be $1/2$.
This is important when we consider the momentum of the electromagnetic wave
\begin{eqnarray}
{\bf P}_{\rm Field}
&=&
\epsilon
\left(
{\bf E} \times {\bf B}
\right) \\
&=&
\frac{1}{v_0^2}
  {\bf S}_{\rm Poynting}
\end{eqnarray}
and the Poynting vector 
\begin{eqnarray}
{\bf S}_{\rm Poynting}
=
{\bf E} \times {\bf H},
\end{eqnarray}
whose time averages are obtained as
\begin{eqnarray}
\bar{\bf P}_{\rm Field}
&\equiv&
\langle 
  {\bf P}_{\rm Field}
\rangle_t \\
&=&
\frac{1}{v_0^2}
\frac{1}{2}
\Re
\left [
{\bf E}^{*} \times {\bf H}
\right ] \\
&=&
\frac{1}{v_0^2}
\bar{\bf S}_{\rm Poynting},
\end{eqnarray}
and 
\begin{eqnarray}
\bar{\bf S}_{\rm Poynting}
&\equiv&
\langle 
  {\bf S}_{\rm Poynting}
\rangle_t
 \\
&=&
\frac{1}{2}
\Re
\left [
{\bf E} \times {\bf H}^{*}
\right ]
 \\
&=&
\frac{1}{2}
\Re
\left [
{\bf E}^{*} \times {\bf H}
\right ],
\end{eqnarray}
respectively.
The Poynting vector describes the flux flow of the energy by photons, such that
\begin{eqnarray}
\bar{\bf S}_{\rm Poynting}
=
v_0
\bar{U}
\hat{\bf z},
\end{eqnarray}
where $\bar{U}_{\rm Field}=\hbar \omega {\mathcal N}/V$ is the energy density of photons, where $V$ is the volume of the system.

\subsection{Horizontally polarised Laguerre-Gauss mode in Lorentz gauge}
Next, we review the horizontally polarised Laguerre-Gauss mode in Lorentz gauge \cite{Allen92} using the vector potential, 
\begin{eqnarray}
{\bf A}(r,\phi,z)
&=&
(A(r,\phi,z),0,0)\\
&=&
A(r,\phi,z)
\hat{\bf x}\\
&=&
A_0
\psi (r,\phi,z)
{\rm e}^{i(kz-\omega t)}
\hat{\bf x}\\
&=&
A_0
u (r,z)
{\rm e}^{i m \phi}
{\rm e}^{i(kz-\omega t)}
\hat{\bf x} \\
&=&
A_0
\Psi(r,\phi,z)
\hat{\bf x},
\end{eqnarray}
where the total mode profile and the propagation is described by the wavefunction
\begin{eqnarray}
\Psi(r,\phi,z)
&=
\psi (r,\phi,z)
{\rm e}^{i(kz-\omega t)} \\
&=
u (r,z)
{\rm e}^{i m \phi}
{\rm e}^{i(kz-\omega t)}.
\end{eqnarray}

In this case, we obtain ${\rm B}$ and ${\rm E}$ as a function of ${\bf A}=(A,0,0)$.
It is straightforward to obtain
\begin{eqnarray}
{\bf B}
&=
\left(
0,
\partial_z
A,
-\partial_y
A
\right),
\end{eqnarray}
where we have abbreviated as $\partial_x=\partial/\partial x$, $\partial_y=\partial/\partial y$, $\partial_z=\partial/\partial z$, and $\partial_t=\partial/\partial t$.
The Lorentz condition becomes
\begin{eqnarray}
\partial_x
A
+
\frac{1}{v_0^2}
\partial_t
\phi
&=&
\partial_x
A
-
\frac{i \omega}{v_0^2}
\phi \\
&=&0,
\end{eqnarray}
from which we obtain
\begin{eqnarray}
\Phi
=
\frac{v_0^2}{i \omega}
\frac{\partial A}{\partial x}.
\end{eqnarray}
Then, we obtain
\begin{eqnarray}
{\bf E}
=
i \omega
\left(
\frac{v_0^2}{\omega^2}
\partial_x^2 A 
+A, 
\frac{v_0^2}{\omega^2}
\partial_x \partial_y A , 
\frac{v_0^2}{\omega^2}
\partial_x \partial_z A 
\right).
\end{eqnarray}
In the paraxial approximation, we can neglect as
\begin{eqnarray}
\left|
\partial_x^2 A
\right|
{\rm , \ }
\left|
\partial_y^2 A
\right|
\ll
\left|
\partial_x \partial_z A
\right|,
\end{eqnarray}
and we use $\partial_z \rightarrow i k_{n_0}$ and $\omega=v_0  k_{n_0}$.
Then, we obtain
\begin{eqnarray}
{\bf E}
&\approx&
i \omega
\left(
A, 
0, 
\frac{v_0^2 i k_{n_0}}{\omega^2}
\partial_x A 
\right) \\
&\approx&
i \omega
\left(
A, 
0, 
\frac{i v_0}{\omega}
\partial_x A 
\right).
\end{eqnarray}
For the calculations of Poynting vector and the momentum, we calculate 
\begin{eqnarray}
{\bf E}^{*}
\times 
{\bf B}
\approx
i \omega
\left(
A \partial_x A^{*}, 
-A^{*} \partial_y A ,
-A^{*} \partial_z A 
\right), 
\end{eqnarray}
which yields
\begin{eqnarray}
&&\Re
\left [
{\bf E}^{*}
\times 
{\bf B}
\right ] 
=
\frac{1}{2}
\left( 
{\bf E}
\times 
{\bf B}^{*}
+
{\bf E}^{*}
\times 
{\bf B}
\right )\\
&&=
\frac{\omega}{2i}
\left(
A^{*} \partial_x A - A \partial_x A^{*} ,
A^{*} \partial_y A - A \partial_y A^{*} ,
A^{*} \partial_z A - A \partial_z A^{*} 
\right) \nonumber \\
&&=
\frac{\omega}{2i}
\left(
A^{*} \overrightarrow{\nabla} A - A \overrightarrow{\nabla} A^{*} 
\right) , 
\end{eqnarray}
where $\overrightarrow{\nabla}=
\left(
\partial_x ,
\partial_y ,
\partial_z  
\right) $.
This is very similar to the expression of the quantum mechanical expectation value \cite{Allen92}.
By defining a standard quantum-mechanical momentum operator
\begin{eqnarray}
\widehat{{\bf p}}
&=&
\frac{\hbar}{i}
\overrightarrow{\nabla}\\
&=&
\frac{\hbar}{i}
\left(
\partial_x ,
\partial_y ,
\partial_z  
\right) 
\end{eqnarray}
The naive expectation value of the momentum would be $A^{*} \hat{{\bf p}} A$. However, this becomes a complex value.
The real expectation value of the momentum would be
\begin{eqnarray}
\frac{1}{2}
A^{*}
\left(
 \hat{{\bf p}} 
+
\hat{{\bf p}}^{\dagger} 
\right) A
&=&
\frac{1}{2}
\left(
A^{*} \hat{{\bf p}} A
+
A^{*} (\hat{{\bf p}})^{\dagger} A
\right) \\
&=&
\frac{1}{2}
\left(
A^{*} \hat{{\bf p}} A
-
A^{*} (\frac{\hbar}{i} \overleftarrow{\nabla}) A
\right)
\\
&=&
\frac{1}{2}
\left(
A^{*} \hat{{\bf p}} A
-
A (\frac{\hbar}{i} \overrightarrow{\nabla}) A^{*}
\right)
\\
&=&
\frac{1}{2}
\left(
A^{*} \hat{{\bf p}} A
-
A \hat{{\bf p}} A^{*}
\right).
\end{eqnarray}
Thus, we obtain
\begin{eqnarray}
&&\Re
\left [
{\bf E}^{*}
\times 
{\bf B}
\right ] 
=
\frac{\omega}{2\hbar}
A^{*}
\left(
 \hat{{\bf p}} 
+
\hat{{\bf p}}^{\dagger} 
\right) A.
\end{eqnarray}
Therefore, we realise that the vector potential is essentially an wavefunction.
In fact, it can also be re-written as
\begin{eqnarray}
\Re
\left [
{\bf E}^{*}
\times 
{\bf B}
\right ] 
=
\frac{\omega}{2i}
|A_0|^2
\left(
\Psi^{*} \overrightarrow{\nabla} \Psi - \Psi \overrightarrow{\nabla} \Psi^{*} 
\right). 
\end{eqnarray}
For taking the time average, it becomes
\begin{eqnarray}
\bar{\bf P}_{\rm Field}
&=&
\frac{\epsilon}{2}
\left \langle
\Re
\left [
{\bf E}
\times 
{\bf B}^{*}
\right ] 
\right \rangle_t \\
&=&
\frac{\epsilon \omega}{2}
|A_0|^2
\left \langle
\frac{1}{2i}
\left(
\Psi^{*} \overrightarrow{\nabla} \Psi - \Psi \overrightarrow{\nabla} \Psi^{*} 
\right) 
\right \rangle_t,
\end{eqnarray}
for which we expect the relationships, $|E_0| \approx \omega |A_0|$ and $\bar{U}_{\rm Field}
=\epsilon |E_0|^2 /2 = \epsilon \omega^2 |A_0|^2 /2$, and we obtain
\begin{eqnarray}
\bar{\bf P}_{\rm Field}
&=&
\frac{\bar{U}_{\rm Field}}{\omega}
\left \langle
\frac{1}{2i}
\left(
\Psi^{*} \overrightarrow{\nabla} \Psi - \Psi \overrightarrow{\nabla} \Psi^{*} 
\right) 
\right \rangle_t.
\end{eqnarray}
The dominant contribution of this value becomes
\begin{eqnarray}
\bar{\bf P}_{\rm Field}
\rightarrow
\frac{\bar{U}_{\rm Field} }{\omega}
k
\hat{\bf z}.
\end{eqnarray}
If we accept the coherent monochromatic light is quantised as photons, the average energy density simply becomes
\begin{eqnarray}
\bar{U}_{\rm Field}=\hbar \omega \frac{\mathcal N}{V}, 
\end{eqnarray}
which immediately yields
\begin{eqnarray}
\bar{\bf P}_{\rm Field}
\rightarrow
\hbar k \frac{\mathcal N}{V}
\hat{\bf z}.
\end{eqnarray}
This means that the total momentum density of the electromagnetic wave is the sum of the contributions from photons per unit volume, and each photon has the momentum of $p=\hbar k$.
This is also consistent with the Bose-Einstein condensation nature of the coherent ray of photons from a laser source, because the coherent photons occupy the same energy and momentum state.

In the above estimation, we have not considered the mode profile, coming from the Laguerre-Gauss mode, such that we calculate 
\begin{eqnarray}
\left \langle
\Re 
[
{\bf E}^{*}
\times 
{\bf B}
]
\right \rangle_t
=
|A_0|^2
\left \langle
\Re 
[
i \omega
\left(
\Psi \partial_x \Psi^{*}, 
-\Psi^{*} \partial_y \Psi ,
-\Psi^{*} \partial_z \Psi 
\right)
]
\right \rangle_t \nonumber \\
\end{eqnarray}
in more detail.
To do so, it is better to move to use the cylindrical coordinate.
The derivatives are converted to be 
\begin{eqnarray}
\partial_x
&=&
\cos \phi
\partial_r
-\frac{1}{r}
\sin \phi
\partial_\phi \\
\partial_y
&=&
\sin \phi
\partial_r
+\frac{1}{r}
\cos \phi
\partial_\phi,
\end{eqnarray}
for which we use 
\begin{eqnarray}
\partial_r 
{\rm e}^{i k_{n_0} \frac{r^2}{2R}}
=
i k_{n_0}
\frac{r}{R}
{\rm e}^{i k_{n_0} \frac{r^2}{2R}}
\end{eqnarray}
and 
\begin{eqnarray}
\partial_\phi 
{\rm e}^{i m \phi}
=
i l
{\rm e}^{i m \phi}.
\end{eqnarray}
Then, finally we obtain
\begin{eqnarray}
\bar{\bf P}_{\rm Field}
=
\frac{\bar{U}_{\rm Field} }{v_0}
\left(
\frac{r}{R} \cos \phi 
-\frac{m}{kr} \sin \phi,
\frac{r}{R} \sin \phi 
+\frac{m}{kr} \cos \phi,
1
\right), \nonumber \\
\end{eqnarray}
where we can also use the Plank's law for the quantisation of photons, $\bar{U}_{\rm field}V=\hbar \omega {\mathcal N}=\hbar v_0 k {\mathcal N}$, and thus $\bar{U}_{\rm Field}/v_0=\hbar \omega/v_0=\hbar k{\mathcal N}/V$.
In the cylindrical coordinate, the momentum density becomes \cite{Allen92}
\begin{eqnarray}
\bar{\bf P}_{\rm Field}
=
\hbar k
\frac{\mathcal N }{V}
\left(
\frac{r}{R} 
\hat{\bf r}
+
\frac{m}{k r}
\hat{\bf \Phi}
+
\hat{\bf z}
\right) .
\end{eqnarray}

After obtaining the momentum density, we can proceed to estimate orbital angular momentum, which is naturally expected as  \cite{Allen92}
\begin{eqnarray}
&&
\bar{\bf M}_{\rm Field} (r,\phi,z)
=
{\bf r} \times \bar{\bf P}_{\rm Field}
\\
&&=
\frac{\bar{U}_{\rm Field} }{v_0}
\left(
r \sin \phi
-\frac{r}{R} z \sin \phi
-\frac{m}{kr} z \cos \phi,
\right. \nonumber \\
&& \quad \quad % \quad \quad \quad \quad
- r \cos \phi
+
\frac{r}{R} z \cos \phi 
-\frac{m}{kr} z \sin \phi,
\ 
\left.
\frac{m}{kr}
\right) \\
&&=
\frac{\bar{U}_{\rm Field} }{v_0}
\left(
-
\frac{m}{kr}
z
\hat{\bf r}
-
r \frac{z_0^2}{z^2+z_0^2}
\hat{\bf \Phi}
+
\frac{m}{k}
\hat{\bf z}
\right) \\
&&=
\frac{\bar{U}_{\rm Field} }{\omega}m
\left(
-
\frac{z}{r}
\hat{\bf r}
-
\frac{kr}{m} \frac{z_0^2}{z^2+z_0^2}
\hat{\bf \Phi}
+
\hat{\bf z}
\right),
\end{eqnarray}
for which we can also use the quantisation condition to obtain
\begin{eqnarray}
\bar{\bf M}_{\rm Field} (r,\phi,z)
&=&
\hbar m \frac{\mathcal N}{V}
\left(
-
\frac{z}{r}
\hat{\bf r}
-
\frac{k r }{m}
\left( 1-\frac{z}{R} \right)
\hat{\bf \Phi}
+
\hat{\bf z}
\right).
\nonumber \\
\end{eqnarray}
This means that the major component of the optical orbital angular momentum is along $z$ direction, which is given by $\hbar m {\mathcal N}$.
This suggests that a photon with topological charge of $m$ carries the orbital angular momentum of $\hbar m$ along the direction of the propagation.

We can also calculate the magnitude of the optical orbital angular momentum density as \cite{Allen92}
\begin{eqnarray}
|
\bar{\bf M}_{\rm Field} 
|
&=&
\hbar m  \frac{\mathcal N}{V}
\sqrt{
1+
\left(
\frac{z}{r}
\right)^2
+
\left(
  \frac{k r }{m}
\right)^2
\left(
\frac{1}{1+(z/z_0)^2}
\right)^2
} . \nonumber \\
\end{eqnarray}

\subsection{Horizontally polarised Laguerre-Gauss mode in Coulomb gauge}
In the previous subsection, we have reviewed the original approach using the Lorentz gauge \cite{Allen92} for the preparations.
The results should not be dependent on the arbitrary choice of the gauge.
Here, we use the Coulomb gauge to confirm it.

In the Coulomb gauge \cite{Jackson99}, the vector potential satisfies the transversality condition
\begin{eqnarray}
{\bf \nabla}
\cdot
{\bf A}
=
0,
\end{eqnarray}
which yields
\begin{eqnarray}
{\bf B}
&=&
{\bf \nabla}
\times
{\bf A} \\
{\bf E}
&=&
-
\partial_t
{\bf A}.
\end{eqnarray}

One might naively think that the horizontally polarised Laguerre-Gauss mode is described by
\begin{eqnarray}
{\bf E}
&=&
E_0
u(r,z)
{\rm e}^{i m \phi}
{\rm e}^{i(k z-\omega t)}
\hat{\bf x} \\
&=&
E_x
\hat{\bf x},
\end{eqnarray}
however, this is wrong because this does not satisfy the transversality condition due to the $r$ and $\phi$ dependences of the vortexed mode ($\partial_x A \neq 0$ and $\partial_y A \neq 0$).

The correct form for the Coulomb gauge would be 
\begin{eqnarray}
{\bf E}
\approx
i \omega
\left(
A, 
0, 
\frac{i v_0}{\omega}
\partial_x A 
\right),
\end{eqnarray}
which is the same form for that in the Lorentz gauge.
Therefore, the small finite {\it longitudinal} component is responsible for guaranteeing the gauge-invariant solution.
Consequently, the vector potential in the Coulomb gauge is described as 
\begin{eqnarray}
{\bf A}
&=&
\frac{1}{i \omega}
{\bf E} \\
&=&
\left(
A, 
0, 
- 
\frac{v_0}{i \omega}
\partial_x A 
\right),
\end{eqnarray}
which is obviously different from that in the Lorentz gauge due to the existence of the longitudinal component of $A_z$.
We can double check that this satisfy the transversality condition, directly by calculating  
\begin{eqnarray}
{\bf \nabla}
\cdot
{\bf A}
&=&
\partial_x A
-\frac{v_0}{i \omega}
\partial_z \partial_x A \\
&=&
\left(
1-\frac{v_0 k}{\omega}
\right)
\partial_x A \\
&=& 0.
\end{eqnarray}

By using the vector potential and vanishing scalar potential in the Coulomb gauge, we obtain the same formulas for ${\bf E}$ and ${\bf B}$, compared with those obtained in the Lorentz gauge.
Therefore, ${\bf A}$ and $\Phi$ could depend on the choice of the gauges, while the observables such as ${\bf E}$ and ${\bf B}$ cannot be dependent \cite{Jackson99}.
The differences of the gauges are summarised in Table \ref{Table-I}.
In particular, the inclusions of the small {\it longitudinal} fields are indispensable for the considerations of the orbital angular momentum due to the spatial dependence of the mode profile.
This is a remarkable difference compared with the simple plane-wave expansion without considering the mode profile in the most of the theory of QED \cite{Dirac30,Sakurai67,Enk94,Leader14,Barnett16,Chen08,Ji10}.
This is one of the key considerations to enable the splitting of spin and orbital angular momentum, as we shall see in due course.
\begin{table}[h]
\caption{\label{Table-I}
Summary of fields in different gauges. The horizontal polarisation is assumed.
}
\begin{ruledtabular}
\begin{tabular}{ccc}
&
Lorentz gauge & Coulomb gauge \\
\colrule
Gauge &
${\bf \nabla} \cdot {\bf A} 
+
\frac{1}{v_0^2}
\partial_t
\phi
=0$ 
& 
${\bf \nabla}
\cdot
{\bf A}
=
0$ \\
Helmholtz eq. &
$\nabla^2 {\bf A}
=
\mu_0 \epsilon 
\frac{\partial^2}{\partial t^2}{\bf A}
$ & 
$\nabla^2 {\bf E}
=
\mu_0 \epsilon 
\frac{\partial^2}{\partial t^2}{\bf E}$ \\
Vector potential &
${\bf A}
=
(A,0,0)
$
&
${\bf A}
=
\left(
A, 
0, 
- 
\frac{v_0}{i \omega}
\partial_x A 
\right)
$ \\
Scalar potential &
$\Phi
=
\frac{v_0^2}{i \omega}
\frac{\partial A}{\partial x}$
&
$\Phi
=
0$ \\
Electric field &
$
i \omega
\left(
A, 
0, 
\frac{i v_0}{\omega}
\partial_x A 
\right)$
&
$
i \omega
\left(
A, 
0, 
\frac{i v_0}{\omega}
\partial_x A 
\right)$ \\
Magnetic induction &
$\left(
0, 
\partial_z A, 
-\partial_y A 
\right)$
&
$\left(
0, 
\partial_z A, 
-\partial_y A 
\right)$
\end{tabular}
\end{ruledtabular}
\end{table}

\subsection{Circularly polarised Laguerre-Gauss mode in Lorentz gauge}
Before we continue to consider the full quantum field theoretic treatment, it is further worth for learning from the historical work \cite{Allen92} for circularly polarised mode, because this shows how spin could appear in optical angular momentum.
Here, we will go back to the Lorentz gauge \cite{Allen92}, because now we understand that the choice of the gauge should not affect the final result at all.

For circularly polarised Laguerre-Gauss mode, we assume 
\begin{eqnarray}
{\bf A}
=
\frac{1}{\sqrt{2}}
(1,i \sigma, 0)
A,
\end{eqnarray}
where $\sigma=\sigma_z$ corresponds to the quantum number for spin pointing to the direction of the propagation ($z$).
In our preferred notation, shown in Fig. 1, the left-circularly polarised state corresponds to the anti-clock-wise rotation of the polarization circle, seen from the detector side, which corresponds to $\sigma=+1$ and spin angular momentum along $z$ for the photon is $+\hbar$.
The right-circulary polarised state rotates clock-wise, which corresponds to $\sigma=-1$ and spin angular momentum per photon is $-\hbar$.
$A=A(r,\phi,z)=A_0\Psi(r,\phi,z)=A_0u (r,z){\rm e}^{i m \phi}{\rm e}^{i(kz-\omega t)}$ is described by the Laguerre-Gauss mode, such that we have spatial profile with the non-zero derivatives.

It is straightforward to obtain the magnetic induction as
\begin{eqnarray}
{\bf B}
&=&
\left(
-\frac{i \sigma}{\sqrt{2}} \partial_z A,
 \frac{1}{\sqrt{2}} \partial_z A,
 \frac{i \sigma}{\sqrt{2}} \partial_x A
-\frac{1}{\sqrt{2}} \partial_y A
\right).
\end{eqnarray}

From the Lorentz condition, we obtain
\begin{eqnarray}
\partial_x
A
+
\frac{1}{v_0^2}
\partial_t
\Phi
&=&
\partial_x
A
-
\frac{i \omega}{v_0^2}
\Phi \\
&=&0,
\end{eqnarray}
which gives
\begin{eqnarray}
\Phi
=
\frac{v_0^2}{i \omega}
\left( 
\frac{1}{\sqrt{2}} \partial_x A
+\frac{i \sigma}{\sqrt{2}} \partial_y A
\right ).
\end{eqnarray}
In the paraxial approximation, we calculate 
\begin{eqnarray}
\nabla \Phi
\approx
v_0
\left(
0,
0,
\frac{1}{\sqrt{2}}
\partial_x A 
+
\frac{i \sigma}{\sqrt{2}}
\partial_y A 
\right),
\end{eqnarray}
and together with $\partial_t {\bf A}=-i \omega {\bf A}$, we obtain 
\begin{eqnarray}
{\bf E}
\approx
\left(
i \omega
\frac{1}{\sqrt{2}}
A, 
-\omega
\frac{\sigma}{\sqrt{2}}
A, 
-v_0
\left(
\frac{1}{\sqrt{2}}
\partial_x A
+
\frac{i \sigma}{\sqrt{2}}
\partial_y A
\right)
\right). \nonumber \\
\end{eqnarray}

Then, we can proceed for calculating the momentum and the optical angular momentum.
First, we calculate 
\begin{eqnarray}
\left(
{\bf E}^{*}
\times 
{\bf B}
\right)_x
&\approx&
\frac{\omega}{2i}
\left(
 A^{*} \partial_x A 
-A \partial_x A^{*} 
+i \sigma
(
 A^{*} \partial_y A 
+ A \partial_y A^{*} 
)
\right) \nonumber \\
\left(
{\bf E}^{*}
\times 
{\bf B}
\right)_y
&\approx&
\frac{\omega}{2i}
\left(
 A^{*} \partial_y A 
-A \partial_y A^{*} 
-i \sigma
(
 A^{*} \partial_x A 
+ A \partial_x A^{*} 
)
\right) \nonumber \\
\left(
{\bf E}^{*}
\times 
{\bf B}
\right)_z
&\approx&
- i \omega
 A^{*} \partial_z A, 
\end{eqnarray}
where the spin independent term is coming from the orbital component, which is the same as that in the horizontally polarised mode and is proportional to 
$\Im
(
\Psi^{*} \overrightarrow{\nabla} \Psi 
) $, 
while the spin dependent term is described by the components of 
$
\Re
(
\Psi^{*} \overrightarrow{\nabla} \Psi 
) 
$.
We have already calculated orbital angular momentum, such that we will focus on the contributions for spin angular momentum.
The extra factors for spin are 
\begin{eqnarray}
\delta \left(
{\bf E}^{*}
\times 
{\bf B}
\right)_x
&\approx&
\frac{\omega}{2}
\sigma 
\left(
 A^{*} \partial_y A 
+ A \partial_y A^{*} 
\right) \\
\delta \left(
{\bf E}^{*}
\times 
{\bf B}
\right)_y
&\approx&
\frac{\omega}{2}
\sigma
\left(
 A^{*} \partial_x A 
+ A \partial_x A^{*} 
\right).
\end{eqnarray}
For them, we evaluate the derivatives, 
\begin{eqnarray}
\partial_\phi 
{\rm e}^{im \phi}
&=&
i m
{\rm e}^{i m \phi} \\
\partial_\phi 
{\rm e}^{-i m \phi}
&=&
- i m
{\rm e}^{-i m \phi},
\end{eqnarray}
which will cancel each other for 
$
\Re
(
\Psi^{*} \overrightarrow{\nabla} \Psi 
) 
=
(
\Psi^{*} \overrightarrow{\nabla} \Psi 
+
\Psi \overrightarrow{\nabla} \Psi^{*} 
)/2 
$.
Therefore, we can drop $\partial_\phi $ as 
\begin{eqnarray}
\partial_x
&=&
\cos \phi
\partial_r
-\frac{1}{r}
\sin \phi
\partial_\phi\\
&\rightarrow&
\cos \phi
\partial_r\\
\partial_y
&=&
\sin \phi
\partial_r
+\frac{1}{r}
\cos \phi
\partial_\phi \\
&\rightarrow&
\sin \phi
\partial_r,
\end{eqnarray}
and we also use the identity
\begin{eqnarray}
\frac{1}{2}
\left(
u^{*} 
\partial_r
u
+
u 
\partial_r
u^{*}
\right)
=
\frac{1}{2}
\partial_r
|u|^{2}.
\end{eqnarray}
Finally, we obtain
\begin{eqnarray}
\delta
\bar{\bf P}_{\rm Field}
&=&
\frac{\epsilon}{2}
\delta 
\left \langle
\Re
\left [
{\bf E}^{*}
\times 
{\bf B}
\right ] 
\right \rangle
\\
&=&
\frac{\epsilon \omega}{2}
|A_0|^2
\sigma
\left(
\sin \phi
\frac{1}{2}
\partial_r |u|^2,
-
\cos \phi
\frac{1}{2}
\partial_r |u|^2,
0
\right) \nonumber \\
&=&
-
\frac{\epsilon \omega}{2}
|A_0|^2
\sigma
\frac{1}{2}
\partial_r |u|^2
\hat{\bf \Phi}.
\end{eqnarray}

This gives the angular momentum contribution from spin as
\begin{eqnarray}
&& \delta
\bar
{\bf M}_{\rm Field} (r,\phi,z)
 \nonumber \\
&&=
\frac{\bar{U}_{\rm Field} }{\omega}
\sigma
\left(
-z \cos \phi 
\frac{1}{2}
\partial_r |u|^2 , 
z \sin \phi 
\frac{1}{2}
\partial_r |u|^2 , 
-r \frac{1}{2}
\partial_r |u|^2 
\right). \nonumber \\
\end{eqnarray}
By averaging over the cross section, $x$ and $y$ components vanish, and we calculate
\begin{eqnarray}
\delta
\bar
\bar{\bf M}_{\rm Field} (z)
=
\frac{
\int_0^{\infty}
rdr
\int_0^{2\pi}
d \phi
\delta
\langle
{\bf M}_{\rm Field} (r,\phi,z)
\rangle
  }
  {
\int_0^{\infty}
rdr
\int_0^{2\pi}
d \phi
|u(r,\phi)|^2
  }
\hat{\bf z},
\end{eqnarray}
where we use the normalisation condition
\begin{eqnarray}
-\int_0^{\infty}
rdr
\frac{r}{2}
\partial_r |u|^2
&=&
\left [
-\frac{r^2}{2}
|u|^2
\right ]_{0}^{\infty}
+
\int_0^{\infty}
dr
r |u|^2 \nonumber \\
&=&1
\end{eqnarray},
and we obtain \cite{Allen92}
\begin{eqnarray}
\delta
\bar
{\bf M}_{\rm Field}
&=&
\frac{\bar{U}_{\rm Field} }{\omega}
\sigma_z
\hat{\bf z}\\
&=&
\hbar 
\frac{{\cal N}}{V}
\sigma_z
\hat{\bf z}.
\end{eqnarray}
Therefore, the circular polarised ray carries the spin angular momentum, and the single photon contributes with the amount of $\hbar \sigma_z$ along the direction of the polarisation.
In our convention (Fig. 1), the left-circularly polarised photon ($\sigma_z=+1$) brings $\hbar$, while the right-circularly polarised photon ($\sigma_z=-1$) brings $-\hbar$ \cite{Allen92}, as we expected.

\subsection{GRIN fibre for a Laguerre-Gauss mode}
Next, we consider a GRIN fibre \cite{Kawakami68,Yariv97}, which has a quadratic dependence of the dielectric constant profile on $r$, described as
$\epsilon(r) \mu
=
n(r)^2/c^2
=
(1-g^2 r^2)/v_0^2
$, which is equivalent to the refractive index dependence of $n(r)^2=n_0^2(1-g^2 r^2)$.
We consider that the distribution of the dielectric constant is sufficiently uniform, such that we can neglect the derivative, $\nabla \epsilon \approx 0$.
The advantages to consider a GRIN fibre doe not reside purely in practical availabilities, but we can solve the Helmholtz equations exactly without employing the paraxial approximation. 
Therefore, it is a quite useful model to consider a theoretically sensitive issue like the splitting of spin and orbital angular momentum from the total angular momentum.
Here, we consider a Laguerre-Gauss mode in a GRIN fibre within the classical electromagnetic treatment \cite{Yariv97} for the application to the angular momentum.

We continue to use the Lorentz gauge in this subsection, and the Helmholtz equation in a GRIN fibre becomes
\begin{eqnarray}
\left[
\nabla^2
-
\frac{1}{v_0^2} 
\left(
1-g^2 r^2
\right)
\partial_t^2
\right]
{\bf A}
=0.
\end{eqnarray}
For the horizontally polarised mode, the solution would be in the form of ${\bf A}
=
{\bf A}(r,\phi,z)
=
A(r,\phi,z)
\hat{\bf x}
=
A_0
\Psi (r,\phi,z)
\hat{\bf x}
=
A_0
\psi (r,\phi,z)
{\rm e}^{i(kz-\omega t)}
\hat{\bf x}
=
A_0
u (r,z)
{\rm e}^{i m \phi}
{\rm e}^{i(kz-\omega t)}
\hat{\bf x}$.
The solution becomes \cite{Yariv97} 
\begin{eqnarray}
\psi
&=&
\sqrt{
\frac{2}{\pi}
\frac{n!}{(n+|m|)!}
}
\frac{1}{w_0}
\left(
\frac{\sqrt{2}r}{w_0}
\right)^{|m|} \nonumber \\
&&
L_n^{|m|} 
\left(
2
\left(
  \frac{r}{w_0}
\right)^2 
\right) 
{\rm  e}^{-\frac{r^2}{w_0^2}}
{\rm  e}^{i m \phi}, 
\end{eqnarray}
where the beam waist becomes constant, 
$w_0=\sqrt{2/(gk_{n_0})}$, with $k_{n_0}=k_0 n_0=2\pi n_0/\lambda \neq k$, and the dispersion relationship, $\omega=\omega(k)$, is given by
\begin{eqnarray}
\omega (k)
&=
\sqrt{v_0^2 k^2+ \delta \omega_0^2 (2n+m+1)^2}
+
\delta \omega_0 (2n+m+1) , \nonumber \\
\end{eqnarray}
where $\delta \omega_0 =v_0 g$.
The radius of the spherical phase diverges, $R\rightarrow \infty$, so that the beam is perfectly collimated to propagate in a GRIN fibre for a long distance without focussing or de-focussing within the fibre.
The important point, here, is that the profile of the Laguerre-Gauss mode works as an envelop function, $\psi (r,\phi,z)$, against the total wavefunction, $\Psi (r,\phi,z)$.
In the simple plane-wave expansion, the approximation of $\psi (r,\phi,z) \rightarrow 1$ is employed, but this is not acceptable when we consider the orbital angular momentum, due to the vortexed beam shape with a node, characterised by topological charge.

The Lorentz condition becomes
\begin{eqnarray}
{\bf \nabla} \cdot {\bf A} 
+
\frac{1}{v_0^2}
(
1-g^2 r^2
)
\partial_t
\Phi
=0.
\end{eqnarray}
By inserting the horizontally polarised form, ${\bf A}=(A,0,0)$, we obtain
\begin{eqnarray}
\partial_x
A
-
\frac{i \omega}{v_0^2}
(1-g^2 r^2)
\Phi
=0,
\end{eqnarray}
which gives
\begin{eqnarray}
\Phi
=
\frac{v_0}{k_{n_0}}
\frac{1}{1-g^2 r^2}
\frac{1}{i}
\partial_x A.
\end{eqnarray}
Therefore, we can approximate $\nabla \Phi
\approx
v_0
\left(
0,0,
\partial_x A 
\right)$.
Together with this and 
$\partial_t {\bf A}
=
-i \omega 
(A,0,0)$, 
we obtain 
\begin{eqnarray}
{\bf E}
\approx
\left(
i \omega
A, 
0, 
-v_0\partial_x A 
\right).
\end{eqnarray}
We also obtain
\begin{eqnarray}
{\bf B}
&=
\left(
0,
\partial_z
A,
-\partial_y
A
\right).
\end{eqnarray}

Then, we can proceed for calculating the momentum and angular momentum.
For that, we need to estimate
\begin{eqnarray}
\Re 
[
{\bf E}^{*}
\times 
{\bf B}
]
=
|A_0|^2
\left \langle
\Re 
[
i \omega
\left(
\Psi \partial_x \Psi^{*}, 
-\Psi^{*} \partial_y \Psi ,
-\Psi^{*} \partial_z \Psi 
\right)
]
\right \rangle_t. \nonumber \\
\end{eqnarray}
By evaluating derivatives, 
\begin{eqnarray}
\Im 
[
\Psi^{*} \partial_r \Psi
]
&\rightarrow& 0 \\
\Im 
[
\Psi^{*} \partial_\phi \Psi
]
&\rightarrow& m \\
\Im 
[
\Psi^{*} \partial_z \Psi
]
&\rightarrow &
k_{n_0},
\end{eqnarray}
we obtain
\begin{eqnarray}
\bar
{\bf P}_{\rm Field}
&=&
\frac{\epsilon}{2}
\omega
|A_0|^2
\left(
-\frac{m}{r} \sin \phi,
 \frac{m}{r} \cos \phi,
k_{n_0}
\right)
\\
&=&
\frac{\bar{U}_{\rm Field}}{v_0}
\left(
-\frac{m}{k_{n_0} r} \sin \phi,
 \frac{m}{k_{n_0} r} \cos \phi,
1
\right).
\end{eqnarray}
Using the quantisation of the energy for photons, 
we obtain
\begin{eqnarray}
\bar
{\bf P}_{\rm Field}
&=&
\hbar k_{n_0} \frac{\mathcal N}{V} 
\left(
\frac{m}{k_{n_0} r}
\hat{\bf \Phi}
+
\hat{\bf z}
\right).
\end{eqnarray}

Finally, we obtain the angular momentum
\begin{eqnarray}
\bar
{\bf M}_{\rm Field} (r,\phi,z)
&=&
\frac{\bar{U}_{\rm Field} }{v_0}
\left(
r \sin \phi
-\frac{m}{k_{n_0}r} z \cos \phi,
\right. \nonumber \\
&& \quad 
- r \cos \phi
-\frac{m}{k_{n_0}r} z \sin \phi,
\ 
\left.
\frac{m}{k_{n_0}r}
\right) \nonumber \\
&=&
\frac{\bar{U}_{\rm Field} }{v_0}
\left(
-
\frac{m}{k_{n_0}r}
z
\hat{\bf r}
-
r 
\hat{\bf \Phi}
+
\frac{m}{k_{n_0}}
\hat{\bf z}
\right)
\nonumber \\
&=&
\hbar m {\cal N}
\left(
-
\frac{z}{r}
\hat{\bf r}
-
\frac{k_{n_0} r }{m}
\hat{\bf \Phi}
+
\hat{\bf z}
\right)
\\
\end{eqnarray}

For the circular polarised state, we can follow exactly the same procedure to obtain the spin contribution to the angular momentum as
\begin{eqnarray}
\delta
\bar
{\bf M}_{\rm Field} (r,\phi,z)
&=&
\frac{\bar{U}_{\rm Field} }{\omega}
\sigma
\hat{\bf z}\\
&=&
\hbar \frac{\mathcal N}{V}
\sigma_z
\hat{\bf z}.
\end{eqnarray}

These results are the same as those obtained by taking the limit of $R \rightarrow \infty$ in the formulas obtained for the free space.

\section{Quantum field theory for photons with spin \& orbital angular momentum}
In the previous section, we have reviewed the important discovery of Allen and collaborators for optical angular momentum \cite{Allen92}.
While it was intriguing to obtain the quantised angular momentum, solely by accepting the fact that the energy of the optical ray is quantised by photon at the end of the calculation, it is not conclusive whether spin and orbital angular momentum are really fundamental quantum degrees of freedom of photons or not \cite{Allen92,Enk94,Leader14,Barnett16,Yariv97,Jackson99,Grynberg10,Bliokh15,Chen08,Ji10}.
In particular, it is highly questionable whether we can derive a full quantum-mechanical expression solely by using Poynting vector and the classical expectation for the angular momentum, ${\bf r} \times \bar{\bf P}_{\rm Field}$, because $\hbar$ is not included in classical mechanics as a fundamental constant.
In particular, spin is inherent quantum degree of freedom without a classical counterpart.
Therefore, we need to employ full quantum field theory to understand the quantum nature of spin and orbital angular momentum of photons.

\subsection{Problems of plane-wave expansions in QED}

\subsubsection{Motivation to consider a plane-wave}
First, we clarify the problems of using plane-waves for the description of the coherent monochromatic ray of photons emitted from a laser source.
Historically, the quantum mechanics was developed to explain black-body radiation, such that it would be natural for physicists at that time to consider photons of all possible modes under thermal equilibrium with the Plank distribution function at finite temperature \cite{Dirac30,Baym69,Sakurai14,Sakurai67}. 
Therefore, a standard theory of QED is based on the plane-wave expansions of the field, imposing the commutation relationship to field operators as Bosons for photons  \cite{Dirac30,Baym69,Sakurai14,Sakurai67}.
However, photons are barely interacting each other due to the absence of charge, and a coherent ray of photons from a laser source is described by a single mode \cite{Yariv97,Grynberg10,Fox06} essentially similar to the Bose-Einstein condensation, in a sense that the macroscopic number of photons are occupying the same state. 
Due to the absence of the Coulomb interaction between photons, photons can be treated purely quantum mechanically without considering the ensemble average \cite{Sakurai67,Grynberg10,Fox06}, such that the temperature for photons are equivalent to zero temperature, even if the measurements are conducted at room temperature.  

In that sense, it is not suitable for light by using a plane-wave for discussing the nature of orbital angular momentum.
Even lights from sun are not spreading to the entire universe like plane-waves, and lights are predominantly propagating along uni-direction with finite spreading as wave-packets.
Moreover, the plane-wave cannot sustain the vortexed lights, as we have shown in the previous section due to the lack of the node at the centre of the vortex.
Even without the orbital angular momentum ($m=0$), the plane-wave description is not suitable for the light propagating with the finite mode profile for discussing the nature of spin of photons, as we shall see below.
Nevertheless, in this subsection, we intentionally use the plane-wave to understand what was the problem to elucidate the nature of the angular momentum of photons. 

\subsubsection{Many-body theory for photons}
In this subsection, we explain our notation on the use of the quantum field theory for photons.
The use of the plane wave corresponds to the flat nodeless mode profile, which spreads the entire volume of the system, which is described by an envelop function
\begin{eqnarray}
\psi(r,\phi,z)
=
1,
\end{eqnarray}
and the full single wavefunction for a photon is 
\begin{eqnarray}
\Psi(r,\phi,z)
&=&
{\rm e}^{i(kz-\omega t + \beta_0)} \\
&=&
{\rm e}^{i \beta},
\end{eqnarray}
where $\beta=kz-\omega t + \beta_0$ describes the standard phase evolution for a photon, propagating along $z$ and $\beta_0$ is the arbitrary $U(1)$ global phase.
Here, we consider a propagation in a uniform material, such that the dispersion is $\omega=v_0 k$.
The normalisation over the volume, $V$, is included in 
$A_0 \propto 
{\mathcal N}/V
$, 
or the electric field strength, 
$E_0
=
\sqrt{
2 \hbar \omega {\mathcal N}/(\epsilon V)
}$
The factor of the average number of photons, ${\mathcal N}$, is coming after taking the quantum-mechanical average over the coherent state, such that the electric field strength per photon, 
$e_0
=
\sqrt{
2 \hbar \omega/(\epsilon V)
}$, is used to define the complex electric field operator,
\begin{eqnarray}
\bm{\hat{\mathcal{E}}}(z,t)=
e_0
{\rm e}^{i \beta}
\left(
  \hat{a}_{\rm H}
  \hat{\bf x}
  +\hat{a}_{\rm V}
  \hat{\bf y}
\right),
\end{eqnarray}
whose complex conjugate (adjoint) is 
\begin{eqnarray}
\bm{\hat{\mathcal{E}}}^{\dagger}(z,t)=
e_0
{\rm e}^{-i \beta}
\left(
  \hat{a}_{\rm H}^{\dagger}
  \hat{\bf x}
  +\hat{a}_{\rm V}^{\dagger}
  \hat{\bf y}
\right),
\end{eqnarray}
where $\hat{a}_{\rm H}^{\dagger}$ ($\hat{a}_{\rm H}$) and $\hat{a}_{\rm V}^{\dagger}$ ($\hat{a}_{\rm V}$) are creation (annihilation) operators for photons in horizontally (H) and vertically (V) polarised modes \cite{Sakurai67,Grynberg10,Fox06,Parker05}.
Creation and annihilation operators must satisfy the commutation relationships for Bosons \cite{Sakurai67,Grynberg10,Fox06,Parker05},
\begin{eqnarray}
&&[\hat{a}_{\sigma},\hat{a}_{\sigma^{'}}] = 0 \\
&&[\hat{a}_{\sigma},\hat{a}_{\sigma^{'}}^{\dagger}] = \delta_{{\sigma},{\sigma}^{'}},
\end{eqnarray}
where $\sigma$ and ${\sigma}^{'}$ describe the polarisation, and $\delta_{{\sigma},{\sigma}^{'}}$ is the Kronecker delta, which gives 1 for the same mode and 0 for the orthogonal mode.

The observable electric field operator is given by
\begin{eqnarray}
\hat{
{\bf E}
}
&=&
\frac{1}{2}
\left (
\bm{\hat{\mathcal{E}}}
+
\bm{\hat{\mathcal{E}}}^{\dagger}
\right) \\
&=&
\frac{e_0}{2}
\left(
\left(
  \hat{a}_{\rm H}
  {\rm e}^{i \beta}
  +
  \hat{a}_{\rm H}^{\dagger}
  {\rm e}^{-i \beta}
\right)
  \hat{\bf x}
+
\left(
  \hat{a}_{\rm V}
  {\rm e}^{i \beta}
  +
  \hat{a}_{\rm V}^{\dagger}
  {\rm e}^{-i \beta}
\right)
  \hat{\bf y}
\right), \nonumber \\
\end{eqnarray}
which always satisfy the transversality condition
\begin{eqnarray}
{\bf \nabla}
\cdot
\hat{
{\bf E}
}
&=&
\partial_z
\hat{E_z} \nonumber\\
&=&
0.
\end{eqnarray}
One would recognise that this is already a big problem when we consider orbital angular momentum, because of the lack of the small longitudinal component along $z$ (Table I), which was responsible to guarantee the gauge condition.
Nevertheless, let's continue to see what happens to spin angular momentum under the plane wave expansion.

Please also note that we have not summed up over all possible electromagnetic modes in a waveguide, because we are considering a single mode of a monochromatic coherent ray from a laser source. 

The transversality condition for the Coulomb gauge, ${\bf \nabla} \cdot \hat{\bf A}=0$, also yields the vector potential
\begin{eqnarray}
\hat{\bf A}
&= &
\frac{1}{i \omega}
\frac{e_0}{2}
\left(
  (\hat{a}_{\rm H} {\rm e}^{i \beta}
  -\hat{a}_{\rm H}^{\dagger} {\rm e}^{-i \beta}
  ) \hat{\bf x}
  +
  (\hat{a}_{\rm V} {\rm e}^{i \beta}
  -\hat{a}_{\rm V}^{\dagger} {\rm e}^{-i \beta}
  ) \hat{\bf y}
\right) \nonumber \\
&= &
\frac{a_0}{2}
\left(
  (\hat{a}_{\rm H} {\rm e}^{i \beta}
  -\hat{a}_{\rm H}^{\dagger} {\rm e}^{-i \beta}
  ) \hat{\bf x}
  +
  (\hat{a}_{\rm V} {\rm e}^{i \beta}
  -\hat{a}_{\rm V}^{\dagger} {\rm e}^{-i \beta}
  ) \hat{\bf y}
\right), \nonumber \\
\end{eqnarray}
which gives the amplitude of the vector potential per photon, 
$
a_0
=
e_0/(i \omega)
=
\sqrt{
2 \hbar \omega/(\epsilon V)
  }/(i \omega)
$, 
corresponding to the average amplitude for the vector potential of
$
A_0
=
a_0 \sqrt{\mathcal N}
=
e_0/(i \omega)
=
\sqrt{
2 \hbar \omega {\mathcal N}/(\epsilon V)
  }/(i \omega)
$.

The magnetic induction operator is calculated as 
\begin{eqnarray}
\hat{\bf B}
&=&
\nabla \times \hat{\bf A} \\
&=&
(0,0,\partial_z)
\times
(\hat{A_x},\hat{A_y},0) \\
&=&
(-\partial_z \hat{A_y},
\partial_z \hat{A_x},0) \\
&=&
\frac{1}{v_0}
\frac{e_0}{2}
\left(
  -
  (\hat{a}_{\rm V} {\rm e}^{i \beta}
  +\hat{a}_{\rm V}^{\dagger} {\rm e}^{-i \beta}
  ) \hat{\bf x}
+
  (\hat{a}_{\rm H} {\rm e}^{i \beta}
  +\hat{a}_{\rm H}^{\dagger} {\rm e}^{-i \beta}
  ) \hat{\bf y}
\right) \nonumber \\
\end{eqnarray}
This corresponds to the average amplitude of the magnetic induction of 
$|B_0|
=
\mu_0
|H_0|
=
|E_0|/v_0
=
\sqrt{\epsilon \mu_0}
|E_0|$, 
which gives the ratio between the magnetic field and the electric field, $\eta=|H_0|/|E_0|=\sqrt{
\mu_0/\epsilon}$.
In the vacuum, the last value becomes $\eta_0=\sqrt{\mu_0/\epsilon_0} \approx
377 \ {\rm \Omega}$.

We think it is worth for clarifying our definition of the polarisation for electromagnetic waves (Fig. 2). As we explained in Fig. 1, we define our rotation seen from the detector side, and the positive rotation is for the anti-clock-wise direction.
The electric field and magnetic induction operators are summarised as
\begin{eqnarray}
\hat{
{\bf E}
}
&=&
\hat{E_x}
  \hat{\bf x}
+
\hat{E_y}
  \hat{\bf y} \\
\hat{
{\bf B}
}
&=&
\frac{1}{v_0}
\left(
-\hat{E_y}
  \hat{\bf x}
+
\hat{E_x}
  \hat{\bf y}
\right),
\end{eqnarray}
where the components of the electric field operator are defined as
\begin{eqnarray}
\hat{E_x}
&=&
\frac{e_0}{2}
\left(
  \hat{a}_{\rm H}
  {\rm e}^{i \beta}
  +
  \hat{a}_{\rm H}^{\dagger}
  {\rm e}^{-i \beta}
\right ) \\
\hat{E_y}
&=&
\frac{e_0}{2}
\left(
  \hat{a}_{\rm V}
  {\rm e}^{i \beta}
  +
  \hat{a}_{\rm V}^{\dagger}
  {\rm e}^{-i \beta}
\right ).
\end{eqnarray}
The relative vectorial relationships are schematically depicted in Fig. 2. 
In our definition, the vectorial direction of the magnetic induction is obtained by rotating the electric filed with the amount of $90^{\circ}$ along $z$.
This corresponds to the application of the optical rotator, which rotates the polarisation state described by Jones vector in the Poincar\'e sphere with the amount of $180^{\circ}$ along $S_3$, which converts the horizontal linear polarisation to the vertical one or the diagonal linear polarisation to the anti-diagonal one, while keeping the circular polarised states for both left and right circulations.
\begin{figure}[h]
\begin{center}
\includegraphics[width=5cm]{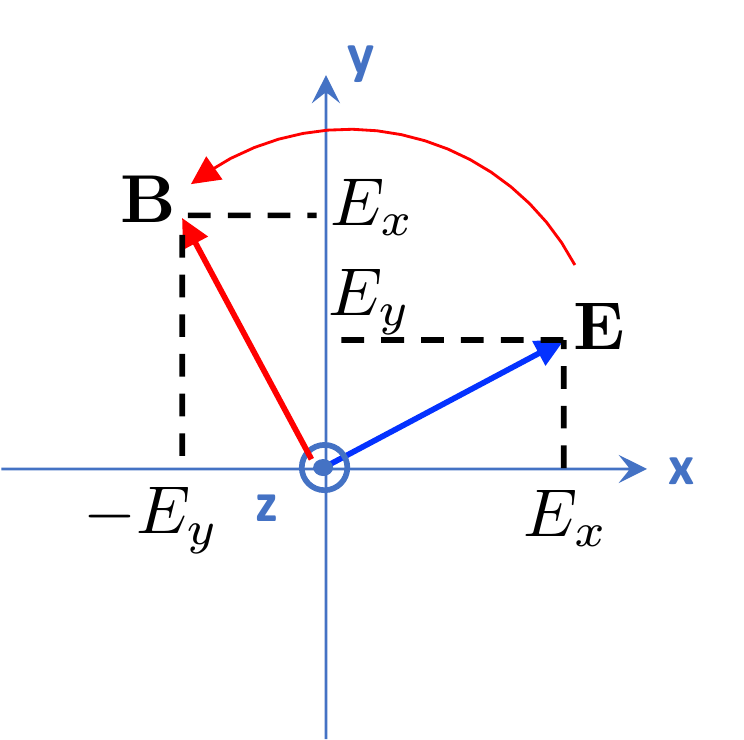}
\caption{
Orthogonality between electric and magnetic fields.
The vectorial direction of ${\bf B}$ is obtained by rotating ${\bf E}$ with the amount of $90^{\circ}$ along $z$.
}
\end{center}
\end{figure}

The Hamiltonian operator is expected to be
\begin{eqnarray}
\hat{H}
=
\int 
d^3{\bf r}
\left [
\frac{1}{2}
\hat{
{\bf E}
}
\cdot
\hat{
{\bf D}
}
+
\frac{1}{2}
\hat{
{\bf B}
}
\cdot
\hat{
{\bf H}
}
\right].
\end{eqnarray}
Upon inserting field operators, $\hat{\bf E}$ and $\hat{\bf B}$, we consider a boundary condition
\begin{eqnarray}
\int_0^{L}
{\rm e}^{\pm i k z}
dz
=
\pm
\frac{1}{ik}
\left[
{\rm e}^{\pm i k z}
\right]_0^{L}
=
\pm
\frac{1}{ik}
\left(
{\rm e}^{\pm i k L}
-1
\right)
=0, \nonumber \\
\end{eqnarray}
which is equivalent to the longitudinal phase-matching condition 
$k=2 \pi n/L$ with an integer $n \in \mathbb{Z}$ for a laser in a cavity with the length of $L_z=L$.
The actual boundary condition depends on the experimental preparation, but as far as a coherent ray is emitted from a laser source, we can assume that the phase is coherent and a similar boundary condition is satisfied.
This also gives 
\begin{eqnarray}
\int_0^{L}
{\rm e}^{\pm i 2k z}
dz
=
\pm
\frac{1}{i2k}
\left[
{\rm e}^{\pm i 2k z}
\right]_0^{L}
=
\pm
\frac{1}{i2k}
\left(
{\rm e}^{\pm i 2k L}
-1
\right)
=0, \nonumber \\
\end{eqnarray}
and the special integration gives the volume $V=L_x L_y L_z$, which will be cancelled with the contribution from $e_0^2$. 
Finally, we obtain 
\begin{eqnarray}
\hat{H}
&=&
\hbar \omega\left [
\left(
\hat{a}_{\rm H}^{\dagger}\hat{a}_{\rm H}
+
\frac{1}{2}
\right)
+
\left(
\hat{a}_{\rm V}^{\dagger}\hat{a}_{\rm V}
+
\frac{1}{2}
\right)
\right ],
\end{eqnarray}
where zero-point fluctuations of $\hbar \omega/2$ per polarisation degree of freedom are successfully included.

The momentum density operator for photons is given by
\begin{eqnarray}
\hat{\bf p}_{\rm Field}
&=&\epsilon (\hat{\bf E} \times \hat{\bf B}) \\
&=&\frac{1}{v_0^2}  \hat{\bf S}, 
\end{eqnarray}
where the Ponynting vector operator is
\begin{eqnarray}
\hat{\bf S}=\hat{\bf E} \times \hat{\bf H}.
\end{eqnarray}
The integrated total momentum operator becomes

\begin{eqnarray}
\hat{\bf P}_{\rm Field}
&=&
\int d^3 {\bf r} \ \hat{\bf p}_{\rm Field} \\
&=&
\hbar k
\hat{\bf z}
\left [
\left(
\hat{a}_{\rm H}^{\dagger}\hat{a}_{\rm H}
+
\frac{1}{2}
\right)
+
\left(
\hat{a}_{\rm V}^{\dagger}\hat{a}_{\rm V}
+
\frac{1}{2}
\right)
\right ], 
\end{eqnarray}
where the zero-point oscillations are included.
If we consider a ray, propagating in an opposite direction, the zero-pint oscillations cancel each other among photons with $+k$ and $-k$.

\subsubsection{Problems to derive angular momentum operators}
Then, we proceed to calculate the angular momentum operator
\begin{eqnarray}
\hat{\bf M}
&=&
\int d^3 {\bf r} \ {\bf r} \times \hat{\bf p}_{\rm field}\\
&=&
\epsilon
\int d^3 {\bf r} \ {\bf r} \times (\hat{\bf E} \times \hat{\bf B})
\end{eqnarray}
using plane-wave basis.
We use identities \cite{Jackson99},
\begin{eqnarray}
&&{\bf r} \times (\hat{\bf E} \times (\nabla \times \hat{\bf A}))
=
\hat{\bf E} ({\bf r}  \cdot (\nabla \times \hat{\bf A}))
-
({\bf r} \cdot \hat{\bf E}) (\nabla \times \hat{\bf A}) \nonumber \\
&&{\bf r}  \cdot (\nabla \times \hat{\bf A})
=
r_{i} \epsilon_{ijk} \partial_{j} \hat{A}_k
=
\epsilon_{ijk} r_{i} \partial_{j} \hat{A}_k
=({\bf r} \times \nabla)\cdot \hat{\bf A}, \nonumber
\end{eqnarray}
and {\it split} the total angular momentum operator \cite{Enk94,Leader14,Barnett16,Chen08,Ji10} into the orbital angular momentum operator $\hat{\bf L}$ and the spin angular momentum operator  $\hat{\bf S}$
\begin{eqnarray}
\hat{\bf M}
=
\hat{\bf L}
+
\hat{\bf S},
\end{eqnarray}
where 
\begin{eqnarray}
\hat{\bf L}
=
\epsilon
\int d^3 {\bf r} \ 
\hat{\bf E} 
({\bf r} \cdot \hat{\bf B})
\end{eqnarray}
and
\begin{eqnarray}
\hat{\bf S}
&=&
-
\epsilon
\int d^3 {\bf r} \ 
({\bf r} \cdot \hat{\bf E}) \hat{\bf B} \\
&=&
-
\epsilon
\int d^3 {\bf r} \ 
({\bf r} \cdot \hat{\bf E}) (\nabla \times \hat{\bf A}).
\end{eqnarray}
But, this is not extremely successful, because $\hat{\bf L}$ vanishes
\begin{eqnarray}
\hat{\bf L}
&=&
\epsilon
\int d^3 {\bf r} \ 
\hat{\bf E} 
({\bf r} \cdot \hat{\bf B}) \\
&=&
\frac{\epsilon}{v}
\int d^3 {\bf r} \ 
(\hat{E}_x ,\hat{E}_y ,0)
\left(
-x \hat{E}_y 
+
y \hat{E}_x 
\right)\\
&=&0,
\end{eqnarray}
due to the odd parity symmetry of $x$ and $y$ against the origin, while $\hat{E}_y$ and $\hat{E}_x$ are parity even for plane-waves.
Consequently, this proves that plane-waves cannot sustain the orbital angular momentum.
This result for $\hat{\bf L}$ may not be a big concern, because we have confirmed in the previous section, a node in the core of the wave, as topological charge, is required for a vortexed ray with orbital angular momentum.
However, we have the same problem for $\hat{\bf S}$, as  
\begin{eqnarray}
\hat{\bf S}
&=&
-
\epsilon
\int d^3 {\bf r} \ 
({\bf r} \cdot \hat{\bf E}) \hat{\bf B} \\
&=&
-
\epsilon
\int d^3 {\bf r} \ 
\left(
x \hat{E}_x 
+
y \hat{E}_y 
\right)
(- \hat{E}_y ,\hat{E}_x ,0) \\
&=&0,
\end{eqnarray}
due to the same argument on the parity symmetries of the integrand.
Moreover, if we continue to use $\hat{\bf S}$, anyway, we can attempt to integrate like   
\begin{eqnarray}
\hat{\bf S}_{i}
&=&
-
\epsilon
\int d^3 {\bf r} \ 
\epsilon_{ilm}
(r_j \hat{E}_j)
(\partial_l \hat{A}_m)\\
&=&
-\epsilon
\left [
(r_j \hat{E}_j)
\hat{A}_m
\right ]_{-\infty}^{\infty}
+
\epsilon
\int d^3 {\bf r} \ 
\epsilon_{ilm}
\hat{A}_m
\partial_l(r_j \hat{E}_j), \nonumber \\
\end{eqnarray}
where the first term might {\it vanish} \cite{Leader14,Chen08,Ji10}, if we consider the mode vanishes at the boundary of the waveguide.
The tactic of the introduction of the vanishing boundary condition \cite{Leader14,Chen08,Ji10} can be justified, if we consider a mode profile, which is not properly taken into account for plane-waves.
Then, we obtain the only finite component along the direction of the propagation ($i=z$),
\begin{eqnarray}
\hat{\bf S}_{z}
&=
\epsilon
\int d^3 {\bf r} \ 
(\hat{\bf E}\times \hat{\bf A}),
\end{eqnarray}
which apparently depends on the choice of the gauge \cite{Enk94,Leader14,Barnett16,Chen08,Ji10}.
We obtained this expression by using the Coulomb gauge, and one might be able to justify to take only the transversal component of the vector potential to justify this formula \cite{Enk94,Barnett16}.
However, it is still questionable to retain the finite operator contribution, which has vanished in the symmetry argument.
Nevertheless, if we continue to proceed to express $\hat{\bf S}_z$ in creation and annihilation operators, we obtain
\begin{eqnarray}
\hat{\bf S}_{z}
&=&
(-i)\hbar 
\hat{\bf z}
\left(
  \hat{a}_{\rm H}^{\dagger} \hat{a}_{\rm V}
  -
  \hat{a}_{\rm V}^{\dagger} \hat{a}_{\rm H}
\right) ,
\end{eqnarray}
which makes reasonable sense \cite{Enk94}.
Although the derivation, we have reviewed, in this subsection is not acceptable, the final result is intriguing.

Next, we show that the problems were coming from the choice of the expansions of the field by plane-waves.
Our goal is to justify the splitting between spin and orbital angular momentum and to get more insights for obtaining full quantum operators for spin and orbital angular momentum.
We will achieve this goal by using a Laguerre-Gauss mode and a standard quantum-field theory for a vortexed coherent monochromatic ray.

\section{Spin and orbital angular momentum operators in a GRIN fibre}
\subsection{Principles}
We must develop a quantum field theory for a coherent monochromatic ray for photons, propagating in a waveguide. 
Therefore, we need to take topological charge into account for allowing the vortexed beam with a specially non-trivial profile.
In order to make the argument based on a specific example, we consider a GRIN fibere, but the application to the other waveguide will be straightforward.
Here, we consider the fundamental principle to develop the theory.

First, we consider a monochromatic coherent state for photons \cite{Grynberg10,Fox06,Parker05}, 
\begin{eqnarray}
|\alpha_{\sigma} \rangle
&=&{\rm e}^{-\frac{|\alpha_{\sigma}|^2}{2}}
{\rm e}^{\alpha_{\sigma} \hat{a}_{\sigma}^{\dagger}}
|0\rangle , 
\end{eqnarray}
where ${\sigma}$ describes the polarisation state such as horizontal (H) and vertical (V) states.
$\alpha_{\sigma}$ is a complex number, which we will obtain, soon.
We can also choose other combinations of orthogonal states such as diagonal (D) and anti-diagonal (A) or left (L) and right (R) polarised states.
The quantum mechanical expectation value of the number operators by the coherent state becomes 
\begin{eqnarray}
\langle \alpha_{\sigma}|
\hat{a}_{\sigma}^{\dagger}
\hat{a}_{\sigma}
|\alpha_{\sigma} \rangle
&=&
|\alpha_{\sigma}|^2 \\
&=&
{\mathcal N}_{\sigma}
 , 
\end{eqnarray}
where ${\mathcal N}_{\sigma}$ is the average number of photons in the polarisation mode of $\sigma$ \cite{Grynberg10,Fox06,Parker05}.
From the total number of photons, we have a sum rule
\begin{eqnarray}
{\mathcal N}
&=&
{\mathcal N}_{\rm H}
+
{\mathcal N}_{\rm V}
, 
\end{eqnarray}
which is obtained by assigning 
\begin{eqnarray}
\alpha_{\rm H}
&=&
\sqrt{{\mathcal N}}
{\rm e}^{- i \delta/2}
\cos \alpha \\
\alpha_{\rm V}
&=&
\sqrt{{\mathcal N}}
{\rm e}^{+ i \delta/2}
\sin \alpha 
, 
\end{eqnarray}
where $\alpha$ is the auxiliary angle to split ${\mathcal N}$ into ${\mathcal N}_{\rm H}$ and ${\mathcal N}_{\rm V}$ by decomposing the electric field into 2 orthogonal components, and $\delta$ is the phase between two orthogonal modes.
The total state of the photonic state is described by a direct product as  
\begin{eqnarray}
|\alpha, \delta \rangle
&=&
|\alpha_{\rm H},\alpha_{\rm V}\rangle \\
&=&|\alpha_{\rm H}\rangle | \alpha_{\rm V}\rangle.
\end{eqnarray}

The electromagnetic field, expected from the coherent state, must be compatible with Maxwell equations and, thus, with the Helmholtz equation.
Both the electric field and the magnetic field are observalbes and obtained by taking the quantum-mechanical expectation values by the coherent state.
The dominant contribution for the complex electric field becomes
\begin{eqnarray}
\bm{\hat{\mathcal{E}}}({\bf r},t) \approx
\sqrt{
  \frac{2 \hbar \omega}{\epsilon V}
  }
{\Psi}({\bf r},t)
\left(
  \hat{a}_{\rm H}
  \hat{\bf x}
  +\hat{a}_{\rm V}
  \hat{\bf y}
\right),
 \end{eqnarray}
where ${\Psi}({\bf r},t)$ works as a wavefunction to describe the orbital part of photons.
If we take quantum-mechanical average of $\bm{\hat{\mathcal{E}}}({\bf r},t)$, we obtain the complex electric field 
\begin{eqnarray}
\bm{{E}}({\bf r},t)
&=&
\left (
  \begin{array}{c}
    {E}_{x} \\
    {E}_{y}
  \end{array}
\right) \\
&=&
\langle \alpha, \delta |
\bm{\hat{\mathcal{E}}}({\bf r},t)
| \alpha, \delta \rangle
\\
&=&
\sqrt{
  \frac{2 \hbar \omega}{\epsilon V}
  }
{\Psi}({\bf r},t)
\langle \alpha, \delta |
\left(
  \hat{a}_{\rm H}
  \hat{\bf x}
  +\hat{a}_{\rm V}
  \hat{\bf y}
\right)
| \alpha, \delta \rangle
\\
&=&
E_{0}
{\Psi}({\bf r},t)
\left (
  \begin{array}{c}
    {\rm e}^{-i\delta/2}\cos \alpha \  \\
    {\rm e}^{+i\delta/2} \sin \alpha \ 
  \end{array}
\right),
 \end{eqnarray}
where $E_0=\sqrt{2 \hbar \omega {\mathcal N}/(\epsilon V)}$ as before, and the vectorial part represents the Jones vector
\begin{eqnarray}
\langle \alpha, \delta | {\rm Jones} \rangle
&=&
\left (
  \begin{array}{c}
    {\rm e}^{-i\delta/2}\cos \alpha \  \\
    {\rm e}^{+i\delta/2} \sin \alpha \ 
  \end{array}
\right),
 \end{eqnarray}
which describes the spin state of photons \cite{Yariv97,Goldstein11,Gil16}.

As we have shown in the previous sections, the results should not depend on the choice of the gauge. 
We will chose the Coulomb gauge, such that $\bm{{E}}({\bf r},t)$ should satisfy the Helmholtz equation  (Table I), which is equivalent to imposing ${\Psi}({\bf r},t)$ to satisfy the Hemholtz equation,
\begin{eqnarray}
\nabla^2 {\Psi}({\bf r},t)
=
\mu_0 \epsilon (r) 
\frac{\partial^2}{\partial t^2}{\Psi}({\bf r},t).
\end{eqnarray}
This means that the orbital wavefunction of a photon is described by the Hemholtz equation rather than the Schr\"odinger equation. 
In a free space, this simply gives the plane-wave, but in a material with the spacial profile of the dielectric constant, the solution can be highly non-trivial, depending on the symmetry of the system and boundary conditions.
For a monochromatic ray, we can assume a simple Plank-Einstein relationship of $E=\hbar \omega$, such that the wavefunction is described by a single mode of the angular frequency of $\omega$ as ${\Psi}({\bf r},t)={\Psi}({\bf r}){\rm e}^{-i \omega t}$, and we obtain
\begin{eqnarray}
\nabla^2 {\Psi}({\bf r})
=
-\omega^2 
\mu_0 \epsilon (r) 
{\Psi}({\bf r}).
\end{eqnarray}

\subsection{Hermite-Gauss and Laguerre-Gauss modes}
In a GRIN waveguide, we can assume $\mu_0 \epsilon (r)=(1-g^2 r^2)/v_0^2$ and 
$
\Psi ({\bf r})
=
\psi (x,y)
{\rm e}^{i(kz-\omega t)}
$, 
which allows to de-couple the plane-wave propagation along $z$ with the mode confinement in $(x,y)$, which is governed by
\begin{eqnarray}
\left(
  \left(
\partial_x^2 
+
\partial_y^2 
-\frac{\omega^2 g^2}{v_0^2}r^2
  \right)
+
  \left(
-k^2
+
\frac{\omega^2}{v_0^2}
  \right)
\right)
\psi
=0 \nonumber \\
\end{eqnarray}

In the cartesian coordinate, we can assume
\begin{eqnarray}
\left(
\partial_x^2 
+
\partial_y^2 
-\frac{\omega^2 g^2}{v_0^2}r^2
\right)
\psi
=
-2 \frac{g}{v_0} (l+m+1) \omega \psi,
\end{eqnarray}
which gives the Hermite-Gauss mode \cite{Yariv97}
\begin{eqnarray}
\psi(x,y)
&=&
H_l
\left(
  \sqrt{2}\frac{x}{w_0}
\right)
H_m
\left(
  \sqrt{2}\frac{y}{w_0}
\right)
{\rm  e}^{-\frac{r^2}{w_0^2}},
\end{eqnarray}
where $w_0=\sqrt{2/(gk)}$ and $H_l$ is the Hrmite polynomial.

In a cylindrical coordinate, 
\begin{eqnarray}
&&\left(
\partial_r^2 
+
\frac{1}{r}
\partial_r 
+
\frac{1}{r^2}
\partial_{\phi}^2 
-\frac{\omega^2 g^2}{v_0^2}r^2
\right)
\psi \nonumber \\
&&=
-2 \frac{g}{v_0} (2n+|m|+1) \omega \psi, 
\end{eqnarray}
which gives the Laguerre-Gauss mode
\begin{eqnarray}
\psi(r,\phi)
&=&
\left(
\frac{\sqrt{2}r}{w_0}
\right)^{|m|}
L_n^{|m|} 
\left(
2
\left(
  \frac{r}{w_0}
\right)^2 
\right) 
{\rm  e}^{-\frac{r^2}{w_0^2}}
{\rm  e}^{im\phi}. \nonumber \\
\end{eqnarray}

The dispersion relationship for the Hermite-Gauss mode is given by a frequency shift, $\delta w_0=v_0 g$, as
\begin{eqnarray}
\omega^2  -2 \delta w_0 (l+m+1) \omega - v_0^2 k^2 =0,
\end{eqnarray}
and the corresponding equation for the Laguerre-Gauss mode is obtained by replacing $l=2n$ and $m \rightarrow |m|$.
This can re-written by using the Plank-Einstein relationship for the energy ($E=\hbar \omega$) and momentum ($p=\hbar k$) of a photon, 
\begin{eqnarray}
E^2  -2 (\hbar \delta  w_0) (l+m+1) E - (v_0 p)^2 =0,
\end{eqnarray}
which yields
\begin{eqnarray}
E
=
\Delta
\pm
\sqrt{\Delta^2+ (v_0 p)^2},
\end{eqnarray}
where the energy gap $\Delta$ is 
\begin{eqnarray}
\Delta
&=&\hbar \delta  w_0 (l+m+1)  \\
&=&m^{*} v_0^2,
\end{eqnarray}
which implies that the photon confined in a waveguide is massive due to the broken symmetry  \cite{Nambu59,Anderson58,Goldstone62,Higgs64}.
The mass increases with the increase of the orbital angular momentum $m$ and the radial quantum number of $n$.
In a weak coupling limit ($g\rightarrow 0$), the effective mass of $m^{*}$ vanishes. 
We should choose the solution of the positive energy for the confined mode, propagating the waveguide, and thus we obtain
\begin{eqnarray}
E
=
\Delta
+
\sqrt{\Delta^2+ (v_0 p)^2}.
\end{eqnarray}
Below, we will focus on the Laguerre-Gauss mode with a cylindrical symmetry.
We normalise the wavefunction as 
\begin{eqnarray}
\int 
d^3{\bf r}
|\Psi|^2
=
V, 
\end{eqnarray}
and the normalised solution becomes
\begin{eqnarray}
\psi(r,\phi)
&=&
\sqrt{
\frac{2}{\pi}
\frac{n!}{(n+|m|)!}
}
\left(
\frac{\sqrt{2}r}{w_0}
\right)^{|m|}
\nonumber \\
&&
L_n^{|m|} 
\left(
2
\left(
  \frac{r}{w_0}
\right)^2 
\right) 
{\rm  e}^{-\frac{r^2}{w_0^2}}
{\rm  e}^{i m \phi}, \nonumber \\
\end{eqnarray}
where the volume is given by $V=w_0^2 L_{z}$.
The amplitude of the electric field for the ray is given by $E_0=\sqrt{2 \hbar \omega {\cal N}/(\epsilon w_0^2 L_z)}$ and the amplitude per photon is $e_0=\sqrt{2 \hbar \omega/(\epsilon w_0^2 L_z)}$.

Now, we will examine the complex electric field operator in more detail.
According to our classical considerations for a Laguerre-Gauss beam, it was essential to take the small longitudinal component for ensuring the vortexed beam sustained by topological charge.
This corresponds to add the longitudinal component, $\hat{\mathcal{E}}_z $, as 
\begin{eqnarray}
\bm{\hat{\mathcal{E}}}(x,y,z,t)
&=&
\hat{\mathcal{E}}_x 
\hat{\bf x}
+
\hat{\mathcal{E}}_y 
\hat{\bf y}
+
\hat{\mathcal{E}}_z 
\hat{\bf z} \\
&=&
e_0
\psi
{\rm e}^{i \beta}
\left(
  \hat{a}_{\rm H}
  \hat{\bf x}
  +\hat{a}_{\rm V}
  \hat{\bf y}
\right)
+
\hat{\mathcal{E}}_z 
\hat{\bf z}
\end{eqnarray}
for obtaining a self-consistent result in the Coulomb gauge (Table I), for which
\begin{eqnarray}
\hat{\bf E}= - \partial_{t} \hat{\bf A}
\end{eqnarray}
must be satisfied. 
The latter corresponds to the identity for the complex vector potential operator, 
\begin{eqnarray}
\bm{\mathcal{\hat{A}}}= 
\frac{1}{i \omega}
\bm{\mathcal{\hat{E}}}.
\end{eqnarray}
By inserting this into ${\bf \nabla} \cdot \bm{\mathcal{\hat{A}}}=0$, we obtain
\begin{eqnarray}
\nabla
\cdot
\bm{\mathcal{\hat{A}}}
&=& 
\frac{e_0}{i \omega}
\left(
  \partial_x \Psi 
    \hat{a}_{\rm H}
+
  \partial_y \Psi 
    \hat{a}_{\rm V}
\right)
+
\frac{1}{i \omega}
  \partial_z 
\hat{\mathcal{E}}_z \\
&=&0,
\end{eqnarray}
which gives the longitudinal component of the operator as
\begin{eqnarray}
\hat{\mathcal{E}}_z 
= 
-
e_0
\frac{v_0}{i \omega}
\left(
  \partial_x \Psi 
    \hat{a}_{\rm H}
+
  \partial_y \Psi 
    \hat{a}_{\rm V}
\right),
\end{eqnarray}
where we have used 
\begin{eqnarray}
\frac{1}{i \omega}
  \partial_z 
&=&
\frac{k}{ \omega} \\
&\approx&
\frac{1}{v_0},
\end{eqnarray}
which is valid in the weak confinement limit, $g\rightarrow 0$.

Consequently, we obtain 
\begin{eqnarray}
\bm{\mathcal{\hat{E}}}
= 
e_0
\left(
\Psi 
\hat{a}_{\rm H}
,
\Psi 
\hat{a}_{\rm V}
,
-
\frac{v_0}{i \omega}
\left(
  \partial_x \Psi 
    \hat{a}_{\rm H}
+
  \partial_y \Psi 
    \hat{a}_{\rm V}
\right)
\right),
\end{eqnarray}
whose conjugate becomes
\begin{eqnarray}
\bm{\mathcal{\hat{E}}} ^{\dagger}
= 
e_0
\left(
\Psi^{*} 
\hat{a}_{\rm H}^{\dagger}
,
\Psi^{*} 
\hat{a}_{\rm V}^{\dagger}
,
+
\frac{v_0}{i \omega}
\left(
  \partial_x \Psi^{*} 
    \hat{a}_{\rm H}^{\dagger}
+
  \partial_y \Psi^{*} 
    \hat{a}_{\rm V}^{\dagger}
\right)
\right). \nonumber \\
\end{eqnarray}
The electric field operator is also obtained as 
\begin{eqnarray}
\hat{\bf E}
&=&
\frac{1}{2}
\left(
\bm{\mathcal{\hat{E}}}
+
\bm{\mathcal{\hat{E}}} ^{\dagger}
\right)\\
&=&
\frac{e_0}{2}
\left(
\Psi 
\hat{a}_{\rm H}
+
\Psi^{*} 
\hat{a}_{\rm H}^{\dagger}
,
\Psi 
\hat{a}_{\rm V}
+
\Psi^{*} 
\hat{a}_{\rm V}^{\dagger}
,
\right. \nonumber \\
&&
\left.
-
\frac{v_0}{i \omega}
\left(
  \partial_x \Psi 
    \hat{a}_{\rm H}
-  \partial_x \Psi^{*} 
    \hat{a}_{\rm H}^{\dagger}
+
  \partial_y \Psi 
    \hat{a}_{\rm V}
-
  \partial_y \Psi^{*} 
    \hat{a}_{\rm V}^{\dagger}
\right)
\right), \nonumber \\
\end{eqnarray}
whose quantum-mechanical expectation value must always be real, which is guaranteed by $\hat{\bf E}^{\dagger}
=
\hat{\bf E}$ and the electric field of a photon is observable.

On the other hand, the conjugate of the complex vector potential operator satisfies
\begin{eqnarray}
\bm{\mathcal{\hat{A}}}^{\dagger}
= 
-\frac{1}{i \omega}
\bm{\mathcal{\hat{E}}}^{\dagger},
\end{eqnarray}
which yields
\begin{eqnarray}
\hat{\bf A}
&=&
\frac{1}{2}
\left(
\bm{\mathcal{\hat{A}}}
+
\bm{\mathcal{\hat{A}}} ^{\dagger}
\right)\\
&=&
\frac{1}{i \omega}
\frac{1}{2}
\left(
\bm{\mathcal{\hat{E}}}
-
\bm{\mathcal{\hat{E}}} ^{\dagger}
\right)\\
&=&
\frac{1}{i \omega}
\frac{e_0}{2}
\left(
\Psi 
\hat{a}_{\rm H}
-
\Psi^{*} 
\hat{a}_{\rm H}^{\dagger}
,
\Psi 
\hat{a}_{\rm V}
-
\Psi^{*} 
\hat{a}_{\rm V}^{\dagger}
,
\right. \nonumber \\
&&
\left.
-
\frac{v_0}{i \omega}
\left(
  \partial_x \Psi 
    \hat{a}_{\rm H}
+  \partial_x \Psi^{*} 
    \hat{a}_{\rm H}^{\dagger}
+
  \partial_y \Psi 
    \hat{a}_{\rm V}
+
  \partial_y \Psi^{*} 
    \hat{a}_{\rm V}^{\dagger}
\right)
\right). \nonumber \\
\end{eqnarray}
This satisfies the transversality condition of the Coulomb gauge
\begin{eqnarray}
{\bf \nabla} \cdot \hat{\bf A}=0.
\end{eqnarray}

It is also straightforward to calculate
\begin{eqnarray}
\hat{\bf B}=\nabla \times \hat{\bf A}
\end{eqnarray}
by assuming 
$|\partial_x^2 \Psi|, 
|\partial_y^2 \Psi|, 
|\partial_{x}\partial_{y} \Psi|
\ll
|\partial_z \partial_x \Psi|, 
|\partial_z \partial_y \Psi|$, which is justified for a ray predominantly propagating along $z$ in the waveguide.
We obtain
\begin{eqnarray}
\hat{\bf B}
&=&
\frac{1}{v_0}
\frac{e_0}{2}
\left(
-(
\Psi 
\hat{a}_{\rm V}
+
\Psi^{*} 
\hat{a}_{\rm V}^{\dagger}
)
,
\Psi 
\hat{a}_{\rm H}
+
\Psi^{*} 
\hat{a}_{\rm H}^{\dagger}
,
\right. \nonumber \\
&&
\left.
\frac{1}{i k}
\left(
  \partial_x \Psi 
    \hat{a}_{\rm V}
-  \partial_x \Psi^{*} 
    \hat{a}_{\rm V}^{\dagger}
-
  \partial_y \Psi 
    \hat{a}_{\rm H}
+
  \partial_y \Psi^{*} 
    \hat{a}_{\rm H}^{\dagger}
\right)
\right),
\nonumber \\
\end{eqnarray}
which guarantees that the magnetic induction is also observable, $\hat{\bf B}^{\dagger}=\hat{\bf B}$.
We also confirm that the transversality condition, 
\begin{eqnarray}
\hat{\bf E} \cdot \hat{\bf B}=0,
\end{eqnarray}
 (Fig. 2) is also satisfied for a vortexed beam, because
\begin{eqnarray}
\hat{B}_x
&=& -
\frac{1}{v_0}
\hat{E}_y \\
\hat{B}_y
&=&
\frac{1}{v_0}
\hat{E}_x.
\end{eqnarray}

By using the obtained $\hat{\bf E}$ and $\hat{\bf B}$, we obtain the Hamiltonina for a vortexed ray, as 
\begin{eqnarray}
\hat{H}
&=&
\hbar \omega\left [
\left(
\hat{a}_{\rm H}^{\dagger}\hat{a}_{\rm H}
+
\frac{1}{2}
\right)
+
\left(
\hat{a}_{\rm V}^{\dagger}\hat{a}_{\rm V}
+
\frac{1}{2}
\right)
\right ],
\end{eqnarray}
where we have used 
$|\partial_x^2 \Psi|, 
|\partial_y^2 \Psi|, 
|\partial_{x}\partial_{y} \Psi|
\ll
|\partial_z \partial_x \Psi|, 
|\partial_z \partial_y \Psi|$, again.

By taking the quantum-mechanical average using the coherent state, we obtain the total energy of photons, 
\begin{eqnarray}
\bar{U}_{\rm tot}
=
\langle \hat{H} \rangle 
=
\langle \alpha, \delta | \hat{H} |  \alpha, \delta \rangle 
=\hbar \omega ({\cal N}+1),
\end{eqnarray}
and the energy density of the electromagnetic waves becomes
\begin{eqnarray}
\bar{U}_{\rm Field}
=
\frac{\bar{U}_{\rm tot}}{V}
=
\hbar \omega 
\left(
\frac{{\cal N}+1}{V}
\right)
=
\hbar \omega 
\left(
n
+
\frac{1}{V}
\right),
\end{eqnarray}
where the photon density is given by $n={\cal N}/V$.

\subsection{Momentum and angular momentum operators for photons}
We define the complex momentum operator, 
\begin{eqnarray}
\bm{\mathcal{\hat{P}}}_{\rm Field}
=
\int
d^3 {\bf r} \ 
\epsilon
\hat{\bf E}
\times
\hat{\bf B},
\end{eqnarray}
whose conjugate is
\begin{eqnarray}
\bm{\mathcal{\hat{P}}}_{\rm Field}^{\dagger}
&=&
\int
d^3 {\bf r} \ 
\epsilon
\left(
\hat{\bf E}
\times
\hat{\bf B}
\right)^{\dagger} \\
&=&
\int
d^3 {\bf r} \ 
\epsilon
\left(
\hat{\bf B}
\times
\hat{\bf E}
\right) \\
&\neq&
\bm{\mathcal{\hat{P}}}_{\rm Field}.
\end{eqnarray}
Therefore, $\bm{\mathcal{\hat{P}}}_{\rm Field}$ is not observable. 
Nevertheless, the momentum operator,
\begin{eqnarray}
\hat{\bf P}_{\rm Field}
=
\frac{1}{2}
\left(
\bm{\mathcal{\hat{P}}}_{\rm Field}
+
\bm{\mathcal{\hat{P}}}_{\rm Field}^{\dagger}
\right),
\end{eqnarray}
is observable, because $\hat{\bf P}_{\rm Field}=\hat{\bf P}_{\rm Field}^{\dagger}$.
We can also define the momentum-density operator, $\hat{\bf p}_{\rm Field}$, before the integration as
\begin{eqnarray}
\hat{\bf p}_{\rm Field}
=
\frac{1}{2}
\epsilon
\left(
\left(
\hat{\bf E}
\times
\hat{\bf B}
\right)
+
\left(
\hat{\bf E}
\times
\hat{\bf B}
\right)^{\dagger} 
\right),
\end{eqnarray}
whose average over space becomes
\begin{eqnarray}
\hat{\bf P}_{\rm Field}
=
\frac{1}{V}
\int
d^3 {\bf r} \ 
\hat{\bf p}_{\rm Field}.
\end{eqnarray}

The major component along $z$ is obtained as 
\begin{eqnarray}
\hat{P_z}
&=&
\int
d^3 {\bf r}
\frac{\epsilon}{v_0}
(
\hat{E}_x
\hat{E}_x
+
\hat{E}_y
\hat{E}_y
)
\\
&=&
\hbar k
\left [
\left(
\hat{n}_{\rm H}
+
\frac{1}{2}
\right)
+
\left(
\hat{n}_{\rm V}
+
\frac{1}{2}
\right)
\right ],
\end{eqnarray}
as we expected.
In a similar way, we calculate 
\begin{eqnarray}
\hat{\mathcal P}_x
&&=
\frac{\hbar}{2 i V}
\int
d^3 {\bf r}
\left [
(
\Psi^{*} \partial_x \Psi - \Psi \partial_x \Psi^{*}
)
\left(
\hat{n}_{\rm H}
+
\frac{1}{2}
+
\hat{n}_{\rm V}
+
\frac{1}{2}
\right)
\right.
\nonumber \\
&&
\left.
+(
\Psi^{*} \partial_y \Psi + \Psi \partial_y \Psi^{*}
)
\left(
\hat{a}_{\rm H}^{\dagger} \hat{a}_{\rm V}
-
\hat{a}_{\rm V}^{\dagger} \hat{a}_{\rm H}
\right) 
+
(
\Psi^{*} \partial_y \Psi + \Psi \partial_y \Psi^{*}
)
\right ], \nonumber \\
\end{eqnarray}
where the last term of $\Psi^{*} \partial_y \Psi + \Psi \partial_y \Psi^{*}$ will be cancelled when we calculate 
\begin{eqnarray}
\hat{P}_x
&=&
\frac{1}{2}
(
\hat{\mathcal P}_x
+
\hat{\mathcal P}_x^{\dagger}
)
\\
&=&
\frac{\hbar}{2 i V}
\int
d^3 {\bf r}
\left [
(
\Psi^{*} \partial_x \Psi - \Psi \partial_x \Psi^{*}
)
\left(
\hat{n}_{\rm H}
+
\frac{1}{2}
+
\hat{n}_{\rm V}
+
\frac{1}{2}
\right)
\right.
\nonumber \\
&&
\left.
+(
\Psi^{*} \partial_y \Psi + \Psi \partial_y \Psi^{*}
)
\left(
\hat{a}_{\rm H}^{\dagger} \hat{a}_{\rm V}
-
\hat{a}_{\rm V}^{\dagger} \hat{a}_{\rm H}
\right)
\right ].
\end{eqnarray}
We can simplify the integrands as
\begin{eqnarray}
\Psi^{*} \partial_x \Psi - \Psi \partial_x \Psi^{*}
&=&
2i
\Im
[
\Psi^{*} \partial_x \Psi 
]
\\
&=&
2i
\Im
\left [
\Psi^{*} 
  \left(
\cos \phi \partial_r
-
\frac{1}{r}
\sin \phi
\partial_\phi
  \right)
 \Psi 
\right] \nonumber \\
&=&
-
\frac{2im}{r}
\sin \phi,
\end{eqnarray}
and 
\begin{eqnarray}
\Psi^{*} \partial_y \Psi + \Psi \partial_y \Psi^{*}
&=&
2
\Re
[
\Psi^{*} \partial_y \Psi 
] \\
&=&
2
\Re
\left[
\Psi^{*} 
  \left(
\sin \phi \partial_r
+
\frac{1}{r}
\cos \phi
\partial_\phi
  \right)
\Psi 
\right]
\nonumber \\
&=&
2
\Re
\left[
\Psi^{*} 
  \left(
\sin \phi \partial_r
  \right)
\Psi 
\right]\\
&=&
\sin \phi 
\left(
u^{*} 
\partial_r
u
+
u 
\partial_r
u^{*}
\right)
\\
&=&
\sin \phi
\partial_r |u|^2.
\end{eqnarray}
Then, we obtain
\begin{eqnarray}
\hat{P}_x
&=&
\frac{1}{V}
\int
d^3 {\bf r}
\left [
(
-\frac{\hbar m}{r}
\sin \phi
)
\left(
\hat{n}_{\rm H}
+
\frac{1}{2}
+
\hat{n}_{\rm V}
+
\frac{1}{2}
\right)
\right.
\nonumber \\
&&
\left.
+
\frac{\hbar}{2 i}
\sin \phi
\partial_r |u|^2
\left(
\hat{a}_{\rm H}^{\dagger} \hat{a}_{\rm V}
-
\hat{a}_{\rm V}^{\dagger} \hat{a}_{\rm H}
\right)
\right ].
\end{eqnarray}
We obtain
\begin{eqnarray}
\hat{p}_x
&=&
-\frac{\hbar m}{r}
\sin \phi
\left(
\hat{n}_{\rm H}
+
\hat{n}_{\rm V}
+1
\right) \nonumber \\
&&+
\frac{\hbar}{2 i}
\sin \phi
\partial_r |u|^2
\left(
\hat{a}_{\rm H}^{\dagger} \hat{a}_{\rm V}
-
\hat{a}_{\rm V}^{\dagger} \hat{a}_{\rm H}
\right).
\end{eqnarray}

Similarly, we calculate
\begin{eqnarray}
\hat{\mathcal P}_y
&&=
\frac{\hbar}{2 i V}
\int
d^3 {\bf r}
\left [
(
\Psi^{*} \partial_y \Psi - \Psi \partial_y \Psi^{*}
)
\left(
\hat{n}_{\rm H}
+
\frac{1}{2}
+
\hat{n}_{\rm V}
+
\frac{1}{2}
\right)
\right.
\nonumber \\
&&
\left.
-(
\Psi^{*} \partial_x \Psi + \Psi \partial_x \Psi^{*}
)
\left(
\hat{a}_{\rm H}^{\dagger} \hat{a}_{\rm V}
-
\hat{a}_{\rm V}^{\dagger} \hat{a}_{\rm H}
\right)
+
(
\Psi^{*} \partial_x \Psi + \Psi \partial_x \Psi^{*}
)
\right ], \nonumber \\
\end{eqnarray}
and then, we obtain
\begin{eqnarray}
\hat{P}_y
&=&
\frac{1}{2}
(
\hat{\mathcal P}_y
+
\hat{\mathcal P}_y^{\dagger}
)
\\
&=&
\frac{\hbar}{2 i V}
\int
d^3 {\bf r}
\left [
(
\Psi^{*} \partial_y \Psi - \Psi \partial_y \Psi^{*}
)
\left(
\hat{n}_{\rm H}
+
\frac{1}{2}
+
\hat{n}_{\rm V}
+
\frac{1}{2}
\right)
\right.
\nonumber \\
&&
\left.
-
(
\Psi^{*} \partial_x \Psi + \Psi \partial_x \Psi^{*}
)
\left(
\hat{a}_{\rm H}^{\dagger} \hat{a}_{\rm V}
-
\hat{a}_{\rm V}^{\dagger} \hat{a}_{\rm H}
\right)
\right ],
\end{eqnarray}
for which, we calculate the integrands,
\begin{eqnarray}
\Psi^{*} \partial_y \Psi - \Psi \partial_y \Psi^{*}
&=&
2i
\Im
[
\Psi^{*} \partial_y \Psi 
]
\\
&=&
2i
\Im
\left [
\Psi^{*} 
  \left(
\sin \phi \partial_r
+
\frac{1}{r}
\cos \phi
\partial_\phi
  \right)
 \Psi 
\right]
\nonumber \\
&=&
\frac{2im}{r}
\cos \phi,
\end{eqnarray}
and
\begin{eqnarray}
\Psi^{*} \partial_x \Psi + \Psi \partial_x \Psi^{*}
&=&
2
\Re
[
\Psi^{*} \partial_x \Psi 
] \\
&=&
2
\Re
\left[
\Psi^{*} 
  \left(
\cos \phi \partial_r
-
\frac{1}{r}
\cos \phi
\partial_\phi
  \right)
\Psi 
\right]
\\
&=&
2
\Re
\left[
\Psi^{*} 
  \left(
\cos \phi \partial_r
  \right)
\Psi 
\right]
=
\cos \phi 
\left(
u^{*} 
\partial_r
u
+
u 
\partial_r
u^{*}
\right) \nonumber \\
&=&
\cos \phi
\partial_r |u|^2.
\end{eqnarray}
Then, we obtain
\begin{eqnarray}
\hat{P}_y
&=&
\frac{1}{V}
\int
d^3 {\bf r}
\left [
\frac{\hbar m}{r}
\sin \phi
\left(
\hat{n}_{\rm H}
+
\frac{1}{2}
+
\hat{n}_{\rm V}
+
\frac{1}{2}
\right)
\right.
\nonumber \\
&&
\left.
-
\frac{\hbar}{2 i}
\cos \phi
\partial_r |u|^2
\left(
\hat{a}_{\rm H}^{\dagger} \hat{a}_{\rm V}
-
\hat{a}_{\rm V}^{\dagger} \hat{a}_{\rm H}
\right)
\right ],
\end{eqnarray}
whose integrand becomes
\begin{eqnarray}
\hat{p}_y
&=&
\frac{\hbar m}{r}
\cos \phi
\left(
\hat{n}_{\rm H}
+
\hat{n}_{\rm V}
+1
\right) \nonumber \\
&&-
\frac{\hbar}{2 i}
\cos \phi
\partial_r |u|^2
\left(
\hat{a}_{\rm H}^{\dagger} \hat{a}_{\rm V}
-
\hat{a}_{\rm V}^{\dagger} \hat{a}_{\rm H}
\right).
\end{eqnarray}

If we move to the cylindrical coordinate $(r,\phi,z)$, we obtain the momentum-density operators
\begin{eqnarray}
\hat{p}_r
&=&
0\\
\hat{p}_\phi
&=&
\frac{\hbar m}{r}
\left(
\hat{n}_{\rm H}
+
\hat{n}_{\rm V}
+1
\right)
-
\frac{\hbar}{2 i}
\partial_r |u|^2
\left(
\hat{a}_{\rm H}^{\dagger} \hat{a}_{\rm V}
-
\hat{a}_{\rm V}^{\dagger} \hat{a}_{\rm H}
\right) \nonumber \\ \\
\hat{p}_z
&=&
\hbar k
\left(
\hat{n}_{\rm H}
+
\hat{n}_{\rm V}
+1
\right).
\end{eqnarray}

Finally, we can calculate the angular momentum-density operators by assuming 
\begin{eqnarray}
\hat{\bf m} 
&=&
{\bf r}
\times
\hat{\bf p}.
\end{eqnarray}
In the Cartesian coordinate, $(x,y,z)=(r \cos \phi, r \sin \phi,z)$, we obtain
\begin{eqnarray}
\hat{m}_x
&=&
\left(
\hbar k
r
\sin \phi
-\frac{\hbar m}{r}
z
\cos \phi
\right)
\left(
\hat{n}_{\rm H}
+
\hat{n}_{\rm V}
+1
\right)
\nonumber \\
&&
-
\frac{\hbar}{2 i}
z
\cos \phi
\partial_r |u|^2
\left(
\hat{a}_{\rm H}^{\dagger} \hat{a}_{\rm V}
-
\hat{a}_{\rm V}^{\dagger} \hat{a}_{\rm H}
\right) \\
\hat{m}_y
&=&
\left(
-\hbar k
r
\cos \phi
-\frac{\hbar m}{r}
z
\sin \phi
\right)
\left(
\hat{n}_{\rm H}
+
\hat{n}_{\rm V}
+1
\right)
\nonumber \\
&&
+
\frac{\hbar}{2 i}
z
\sin \phi
\partial_r |u|^2
\left(
\hat{a}_{\rm H}^{\dagger} \hat{a}_{\rm V}
-
\hat{a}_{\rm V}^{\dagger} \hat{a}_{\rm H}
\right)\\
\hat{m}_z
&=&
\hbar m
\left(
\hat{n}_{\rm H}
+
\hat{n}_{\rm V}
+1
\right)
-
\frac{\hbar}{2 i}
r
\partial_r |u|^2
\left(
\hat{a}_{\rm H}^{\dagger} \hat{a}_{\rm V}
-
\hat{a}_{\rm V}^{\dagger} \hat{a}_{\rm H}
\right)
\nonumber \\
&=&
\hbar m
\left(
\hat{n}_{\rm H}
+
\hat{n}_{\rm V}
+1
\right)
+
\frac{\hbar}{i}
\left(
\hat{a}_{\rm H}^{\dagger} \hat{a}_{\rm V}
-
\hat{a}_{\rm V}^{\dagger} \hat{a}_{\rm H}
\right),
\end{eqnarray}
where we have used
\begin{eqnarray}
-\int_0^{\infty} r dr r \frac{1}{2} \partial_r |u|^2
&=&
\left[
-\frac{r^2}{2} |u|^2
\right]_0^{\infty}
+
\int_0^{\infty} dr r  |u|^2 \nonumber \\
&=&1
\end{eqnarray}
at the last line.

In the cylindrical coordinate, $\hat{\bf m}$ becomes
\begin{eqnarray}
\hat{m}_r
&=&
-
\hbar m
\frac{z}{r}
\left(
\hat{n}_{\rm H}
+
\hat{n}_{\rm V}
+1
\right)
\nonumber \\
&&
-
\frac{\hbar}{2 i}
z
\cos (2 \phi)
\partial_r |u|^2
\left(
\hat{a}_{\rm H}^{\dagger} \hat{a}_{\rm V}
-
\hat{a}_{\rm V}^{\dagger} \hat{a}_{\rm H}
\right)\\
\hat{m}_\phi
&=&
-\hbar k
r
\left(
\hat{n}_{\rm H}
+
\hat{n}_{\rm V}
+1
\right)
\nonumber \\
&&
+
\frac{\hbar}{2 i}
z
\sin (2 \phi)
\partial_r |u|^2
\left(
\hat{a}_{\rm H}^{\dagger} \hat{a}_{\rm V}
-
\hat{a}_{\rm V}^{\dagger} \hat{a}_{\rm H}
\right) \\
\hat{m}_z
&=&
\hbar m
\left(
\hat{n}_{\rm H}
+
\hat{n}_{\rm V}
+1
\right)
+
\frac{\hbar}{i}
\left(
\hat{a}_{\rm H}^{\dagger} \hat{a}_{\rm V}
-
\hat{a}_{\rm V}^{\dagger} \hat{a}_{\rm H}
\right) \\
&=&
\hat{l}_z
+
\hat{s}_z,
\end{eqnarray}
where the last line is especially important, since we finally obtained orbital and spin angular momentum operators
\begin{eqnarray}
\hat{l}_z
&=&
\hbar m
\left(
\hat{n}_{\rm H}
+
\hat{n}_{\rm V}
+1
\right)\\
\hat{s}_z
&=&
\frac{\hbar}{i}
\left(
\hat{a}_{\rm H}^{\dagger} \hat{a}_{\rm V}
-
\hat{a}_{\rm V}^{\dagger} \hat{a}_{\rm H}
\right) ,
\end{eqnarray}
respectively.

When we integrate over space, we realise
\begin{eqnarray}
\int
d \phi
\ 
\cos \phi
=
\int
d \phi
\ 
\sin \phi
=0.
\end{eqnarray}
Thus, we obtain 
\begin{eqnarray}
\hat{P}_x
&=&
\hat{P}_y
=
0 \\
\hat{P}_z
&=&
\hbar k
\left(
\hat{n}_{\rm H}
+
\hat{n}_{\rm V}
+1
\right).
\end{eqnarray}

For the angular momentum operator, defined by
\begin{eqnarray}
\hat{\bf M}
&=&
\frac{1}{V}
\int
d^3 {\bf r}
\ 
\hat{\bf m} \\
&=&
\frac{1}{V}
\int
d^3 {\bf r}
\ 
{\bf r}
\times
\hat{\bf P},
\end{eqnarray}
we obtain
\begin{eqnarray}
\hat{M}_x
&=&
\hat{M}_y
=
0 \\
\hat{M}_z
&=&
\hat{L}_z
+
\hat{S}_z,
\end{eqnarray}
where
\begin{eqnarray}
\hat{L}_z
&=&
\hbar m
\left(
\hat{n}_{\rm H}
+
\hat{n}_{\rm V}
+1
\right) \\
\hat{S}_z
&=&
\hbar
\left(
\hat{n}_{\rm L}
-
\hat{n}_{\rm R}
\right).
\end{eqnarray}
For the spin operator, the number operators of left and right circular states are used, which are defined as $\hat{n}_{\rm L}=\hat{a}_{\rm L}^{\dagger} \hat{a}_{\rm L}$ and $\hat{n}_{\rm R}=\hat{a}_{\rm R}^{\dagger} \hat{a}_{\rm R}$, respectively, where the field operators are obtained by unitary transformations, 
\begin{eqnarray}
\left (
  \begin{array}{c}
\hat{a}_{\rm L}^{\dagger} 
\\
\hat{a}_{\rm R}^{\dagger} 
  \end{array}
\right)
=
\frac{1}{\sqrt{2}}
\left (
  \begin{array}{cc}
1 & i \\
1 & -i
  \end{array}
\right)
\left (
  \begin{array}{c}
\hat{a}_{\rm H}^{\dagger} 
\\
\hat{a}_{\rm V}^{\dagger} 
  \end{array}
\right)
\end{eqnarray}
and
\begin{eqnarray}
\left (
  \begin{array}{c}
\hat{a}_{\rm L}
\\
\hat{a}_{\rm R} 
  \end{array}
\right)
=
\frac{1}{\sqrt{2}}
\left (
  \begin{array}{cc}
1 & -i \\
1 & i
  \end{array}
\right)
\left (
  \begin{array}{c}
\hat{a}_{\rm H}
\\
\hat{a}_{\rm V} 
  \end{array}
\right).
\end{eqnarray}

Here, we could split the total angular momentum operator into orbital and spin angular momentum operators without the apparent gauge dependence.
We could perform a gauge transformation for photons, but due to the absence of charge for photons, the gauge field will not couple to the change of the angular momentum operators.
The gauge independence is obvious in our expressions, because the number of photons should not depend on the choice of the gauge, otherwise the total energy of the system can change depending on the arbitrary choice of the gauge.

It is interesting to be aware that there exists contributions from zero-point oscillations in the orbital angular momentum for a ray propagating towards one direction.
Such a zero-point fluctuation is absent for spin.

Another interesting point is that we could obtain only the angular momentum operators along the direction of the propagation from simple analogy from the classical counter part defined by $\hat{\bf m} =
{\bf r}
\times
\hat{\bf p}$.
This does not prove that there is no perpendicular components for spin and orbital angular momentum.
In fact, the perpendicular components of spin states can be described by the superposition state of left and right circular polarised states. 
We emphasise this point and discuss the full components of spin and orbital angular momentum operators, latter.

\section{Origin of photonic spin angular momentum}
Before proceeding to consider the full orbital angular momentum operators, further, in this section, we discuss the origin of the photonic spin angular momentum for a coherent monochromatic ray without an orbital angular momentum in a general waveguide (Fig. 3).
Spin of a photon is an inherent quantum degree of freedom, which should be described quantum-mechanically rather than classically.
In the absence of the orbital angular momentum ($m=0$), we should not have any issue to regard the total angular momentum is exclusively coming from spin.
Therefore, the situation would be simpler than the splitting of spin and orbital angular momentum.
We check the derivation of the last section for the case of $m=0$ in detail to understand spin of photons.

\begin{figure}[h]
\begin{center}
\includegraphics[width=5cm]{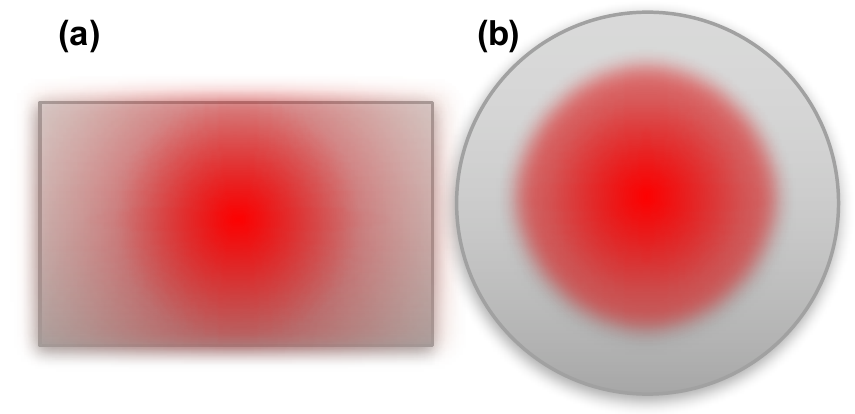}
\caption{
Examples of waveguides of (a) a rectangle shape and (b) a cylindrical shape. 
The mode profile is essential to confine lights inside waveguides, so that a plane-wave cannot be a good approximate mode.
The Gaussian profile of $\psi(x,y)=\psi(r) \propto{\rm e}^{-\frac{r^2}{w_0^2}}$ is used to describe the Hermite-Gauss mode for (a) and the Laguerre-Gauss mode for (b) in a GRIN waveguide.
}
\end{center}
\end{figure}

For photons propagating in a waveguide, it is essential to take the mode profile \cite{Yariv97} into account, which means $|\partial_r \Psi|\neq 0$.
On the other hand, we will employ the paraxial approximation, 
$
|\partial_x^2 \Psi|, 
|\partial_y^2 \Psi|, 
|\partial_x\partial_y \Psi|
\ll
|\partial_z \partial_x \Psi|, 
|\partial_z \partial_y \Psi|
$, 
which is justified for a ray propagating in a waveguide, because the propagation is predominantly along one direction of $z$.
We assume a generic form of the complex electric field operator,
\begin{eqnarray}
\bm{\hat{\mathcal{E}}}(x,y,z,t)
&=&
e_0
\psi
{\rm e}^{i \beta}
\left(
  \hat{a}_{\rm H}
  \hat{\bf x}
  +\hat{a}_{\rm V}
  \hat{\bf y}
\right)
+
\hat{\mathcal{E}}_z 
\hat{\bf z}.
\end{eqnarray}

From the Coulomb gauge condition, 
\begin{eqnarray}
{\bf \nabla} \cdot \bm{\mathcal{\hat{A}}}=0,
\end{eqnarray}
we obtain the longitudinal component,
\begin{eqnarray}
\hat{\mathcal{E}}_z 
= 
-
e_0
\frac{v_0}{i \omega}
\left(
  \partial_x \Psi 
    \hat{a}_{\rm H}
+
  \partial_y \Psi 
    \hat{a}_{\rm V}
\right),
\end{eqnarray}
which was not considered in the plane-wave expansions.
The existence of this small longitudinal component is responsible to obtain the spin angular momentum operator, properly.
Then, we obtain the same expression for $\hat{\bf E}$, $\hat{\bf B}$, $\hat{\bf A}$, $\hat{H}$, and $\hat{P}_z$.

On the other hand, in the absence of the angular orbital momentum, the mode profile is described by a real function, $\psi(x,y) \in \mathbb{R}$, except for the global phase of ${\rm e}^{i \beta}$.
Consequently, we obtain
\begin{eqnarray}
\hat{P}_x
&=&
\frac{1}{V}
\int
d^3 {\bf r}
\left [
\frac{\hbar}{2 i}
\sin \phi
\partial_r |u|^2
\left(
\hat{a}_{\rm H}^{\dagger} \hat{a}_{\rm V}
-
\hat{a}_{\rm V}^{\dagger} \hat{a}_{\rm H}
\right)
\right ] \\
\hat{P}_y
&=&
\frac{1}{V}
\int
d^3 {\bf r}
\left [
-
\frac{\hbar}{2 i}
\cos \phi
\partial_r |u|^2
\left(
\hat{a}_{\rm H}^{\dagger} \hat{a}_{\rm V}
-
\hat{a}_{\rm V}^{\dagger} \hat{a}_{\rm H}
\right)
\right ], \nonumber \\
\end{eqnarray}
which correspond to 
\begin{eqnarray}
\hat{p}_x
&=&
\frac{\hbar}{2 i}
\sin \phi
\partial_r |u|^2
\left(
\hat{a}_{\rm H}^{\dagger} \hat{a}_{\rm V}
-
\hat{a}_{\rm V}^{\dagger} \hat{a}_{\rm H}
\right)\\
\hat{p}_y
&=&
-
\frac{\hbar}{2 i}
\sin \phi
\partial_r |u|^2
\left(
\hat{a}_{\rm H}^{\dagger} \hat{a}_{\rm V}
-
\hat{a}_{\rm V}^{\dagger} \hat{a}_{\rm H}
\right)\\
\hat{p}_z
&=&
\hbar k
\left(
\hat{n}_{\rm H}
+
\hat{n}_{\rm V}
+
1
\right)\\
\end{eqnarray}
in a Cartesian coordinate, and 
\begin{eqnarray}
\hat{p}_r
&=&
0
\\
\hat{p}_{\phi}
&=&
-
\frac{\hbar}{2 i}
\partial_r |u|^2
\left(
\hat{a}_{\rm H}^{\dagger} \hat{a}_{\rm V}
-
\hat{a}_{\rm V}^{\dagger} \hat{a}_{\rm H}
\right)\\
\hat{p}_z
&=&
\hbar k
\left(
\hat{n}_{\rm H}
+
\hat{n}_{\rm V}
+
1
\right)\\
\end{eqnarray}
in a cylindrical coordinate.

Then, we calculate 
\begin{eqnarray}
\hat{m}_x
&=&
\hbar k
r
\sin \phi
\left(
\hat{n}_{\rm H}
+
\hat{n}_{\rm V}
+1
\right)
\nonumber \\
&&
-
\frac{\hbar}{2 i}
z
\cos \phi
\partial_r |u|^2
\left(
\hat{a}_{\rm H}^{\dagger} \hat{a}_{\rm V}
-
\hat{a}_{\rm V}^{\dagger} \hat{a}_{\rm H}
\right) \\
\hat{m}_y
&=&
-\hbar k
r
\cos \phi
\left(
\hat{n}_{\rm H}
+
\hat{n}_{\rm V}
+1
\right)
\nonumber \\
&&
+
\frac{\hbar}{2 i}
z
\sin \phi
\partial_r |u|^2
\left(
\hat{a}_{\rm H}^{\dagger} \hat{a}_{\rm V}
-
\hat{a}_{\rm V}^{\dagger} \hat{a}_{\rm H}
\right)\\
\hat{m}_z
&=&
\frac{\hbar}{i}
\left(
\hat{a}_{\rm H}^{\dagger} \hat{a}_{\rm V}
-
\hat{a}_{\rm V}^{\dagger} \hat{a}_{\rm H}
\right),
\end{eqnarray}
in a Cartesian coordinate, and 
\begin{eqnarray}
\hat{m}_r
&=&
-
\frac{\hbar}{2 i}
z
\cos (2 \phi)
\partial_r |u|^2
\left(
\hat{a}_{\rm H}^{\dagger} \hat{a}_{\rm V}
-
\hat{a}_{\rm V}^{\dagger} \hat{a}_{\rm H}
\right)\\
\hat{m}_\phi
&=&
-\hbar k
r
\left(
\hat{n}_{\rm H}
+
\hat{n}_{\rm V}
+1
\right)
\nonumber \\
&&
+
\frac{\hbar}{2 i}
z
\sin (2 \phi)
\partial_r |u|^2
\left(
\hat{a}_{\rm H}^{\dagger} \hat{a}_{\rm V}
-
\hat{a}_{\rm V}^{\dagger} \hat{a}_{\rm H}
\right) \\
\hat{m}_z
&=&
\frac{\hbar}{i}
\left(
\hat{a}_{\rm H}^{\dagger} \hat{a}_{\rm V}
-
\hat{a}_{\rm V}^{\dagger} \hat{a}_{\rm H}
\right) \\
&=&
\hat{s}_z.
\end{eqnarray}

After the integration, finally, we obtain 
\begin{eqnarray}
\hat{P}_x
&=&
\hat{P}_y
=
0 \\
\hat{P}_z
&=&
\hbar k
\left(
\hat{n}_{\rm H}
+
\hat{n}_{\rm V}
+1
\right),
\end{eqnarray}
as before, while 
\begin{eqnarray}
\hat{M}_x
&=&
\hat{M}_y
=
0 \\
\hat{M}_z
&=&
\hat{S}_z,
\end{eqnarray}
where the total angular momentum along $z$ is solely described by the spin angular momentum
\begin{eqnarray}
\hat{S}_z
&=&
\hbar
\left(
\hat{n}_{\rm L}
-
\hat{n}_{\rm R}
\right),
\end{eqnarray}
as we expected, and the orbital angular momentum vanishes.
We also confirmed that the final result depends solely on the difference of number of photons between left and right circularly polarised photons, such that $\hat{S}_z$ is independent on the choice of the gauge.
Therefore, our results are independent on the mode profile, and the expression of $\hat{S}_z$ is validated for an arbitrary mode profile as far as the mode is propagating predominantly along one direction.

\section{Principle of rotational symmetry for photonic spin states}
In the previous sections, we have obtained the spin operator along the direction of the propagation as,  
\begin{eqnarray}
\hat{S}_{z}
&=&
\hbar 
\left(
  \begin{array}{cc}
     \hat{a}_{\rm L}^{\dagger}, & 
     \hat{a}_{\rm R}^{\dagger} 
  \end{array}
\right)
\left(
  \begin{array}{cc}
1 & 0 \\
0 & -1
  \end{array}
\right) 
\left(
  \begin{array}{c}
     \hat{a}_{\rm L} \\
     \hat{a}_{\rm R}
  \end{array}
\right) \\
&=&
\hbar 
     \bm{\hat{\psi}}_{\rm LR}^{\dagger}
\sigma_3
\bm{\hat{\psi}}_{\rm LR},
\end{eqnarray}
where $\bm{\hat{\psi}}_{\rm LR}^{\dagger}=( \hat{a}_{\rm L}^{\dagger}, \hat{a}_{\rm R}^{\dagger})$ and $\bm{\hat{\psi}}_{\rm LR}$ are creation and annihilation operators in a chiral spinor representation by using the analogy with the classical mechanics, $\hat{\bf m}={\bf r} \times \hat{\bf p}$.
While we do not know the exact reason why we could obtain a reasonable expression of $\hat{S}_z$, while the calculated angular momentum along the direction perpendicular to the propagation became zero, $\hat{M}_x=\hat{M}_y
=0$.
This does not necessarily mean that the quantum field operators of $\hat{S}_x$ and $\hat{S}_y$ vanish, because the spin states of photons, polarised perpendicular to the direction of the propagation, can be described by superposition states of left and right circularly polarised states.
Clearly, the correspondence from the classical mechanics, using $\hat{\bf m}={\bf r} \times \hat{\bf p}$, was not enough to derive $\hat{S}_x$ and $\hat{S}_y$, such that we need a guiding principle for spin operators.

Here, we impose {\it the principle of rotational invariance for photonic polarisation states to describe the propagation in a waveguide with a cylindrical symmetry or a free space}.
We know that there exists 2 orthogonal polarised states for describing the photonic state, and we choose left and right circularly polarised states as basis states, for example.
Then, we use SU$(2)$ Lie-Algebra \cite{Georgi99}, and spin should work as a generator of rotation \cite{Dirac30, Baym69,Sakurai14,Sakurai67}, 
\begin{eqnarray}
\hat{\mathcal{D}}({\bf \hat{n}},\delta \phi)
&=\exp 
\left (
-i 
{\bm \sigma}\cdot {\bf \hat{n}}
\left (
\frac{\delta \phi}{2}
\right)
\right),
\end{eqnarray}
where ${\bf \hat{n}}$ describes the unit vector ($|\hat{\bf \hat{n}}|=1$) pointing towards the rotational axis, $\delta \phi$ is the angle of rotation, for which the anti-clock-wise rotation (left rotaion), seen from the top of the rotational axis, is taken to be positive (Fig. 1), and ${\bm \sigma}=(\sigma_1,\sigma_2,\sigma_3)$ describes the Pauli matrices
\begin{eqnarray}
\sigma_1=
\left(
  \begin{array}{cc}
0 & 1 \\
1 & 0
  \end{array}
\right),
\sigma_2=
\left(
  \begin{array}{cc}
0 & -i \\
i & 0
  \end{array}
\right) , 
\sigma_3=
\left(
  \begin{array}{cc}
1 & 0 \\
0 & -1
  \end{array}
\right), \nonumber \\
\end{eqnarray}
which satisfy the commutation relationship $\left [ \sigma_{i},\sigma_{j} \right ]=2 i \epsilon_{ijk}\sigma_{k}$ and the anti-commutation relationship $\left \{\sigma_{i},\sigma_{j}\right \}=2\delta_{ij}{\bf 1}$, where $\epsilon_{ijk}$ is Levi-Civita symbol for a completely antisymmetric tensor and $\delta_{ij}$ is the Kronecker delta, for components $i,j,k=1,2,3$ or $x,y,z$, and ${\bf 1}$ is the $2\times 2$ identity matrix, 
\begin{eqnarray}
{\bf 1}=
\left(
  \begin{array}{cc}
1 & 0 \\
0 & 1
  \end{array}
\right). \nonumber \\
\end{eqnarray}
Then, we obtain $\hat{S}_x$ simply by rotating $\hat{S}_z$ with the amount of $\pi/2$ along $y$ as, 
\begin{eqnarray}
\hat{S}_x
&=&
\hat{\mathcal{D}}
\left(
{\bf \hat{y}},
  \frac{\pi}{4}
\right)
\hat{S}_z
\hat{\mathcal{D}}^{\dagger} 
\left(
{\bf \hat{y}},
  \frac{\pi}{4}
\right) \\
&=&
{\bm \psi}_{\rm LR}^{\dagger}
\sigma_1
{\bm \psi}_{\rm LR}.
\end{eqnarray}
Similarly, we obtain $\hat{S}_y$ by rotating $-\pi/2$ along $x$ as, 
\begin{eqnarray}
\hat{S}_y
&=&
\hat{\mathcal{D}}
\left(
{\bf \hat{x}},
  -\frac{\pi}{4}
\right)
\hat{S}_z
\hat{\mathcal{D}}^{\dagger} 
\left(
{\bf \hat{x}},
  -\frac{\pi}{4}
\right) \\
&=&
{\bm \psi}_{\rm LR}^{\dagger}
\sigma_2
{\bm \psi}_{\rm LR}.
\end{eqnarray}

We also define 
\begin{eqnarray}
\hat{S}_0
&=&
\hbar 
{\bm \psi}_{\rm LR}^{\dagger}
{\bf 1}
{\bm \psi}_{\rm LR},
\end{eqnarray}
to account for the total number of coherent photons for each polarised components.
This also accounts for the time averaging of incoherent lights, which we are not discussing, here.

The general polarisation coherent state in the chiral basis is described by Bloch state \cite{Dirac30, Baym69,Sakurai14,Sakurai67}
\begin{eqnarray}
\langle \theta, \phi  | {\rm Bloch}\rangle
&=&
\left (
  \begin{array}{c}
    {\rm e}^{-i\frac{\phi}{2}}  \cos \left( \frac{\theta}{2} \right)    \\
    {\rm e}^{+i\frac{\phi}{2}}\sin \left( \frac{\theta}{2} \right)  
  \end{array}
\right),
\end{eqnarray}
where $\theta$ is the polar angle and $\phi$ is the azimuthal angle.
By taking the quantum mechanical average of spin operators $\bm{\hat{\mathcal S}}=(\hat{S}_0,\hat{S}_x,\hat{S}_y,\hat{S}_z)$ over the coherent Bloch state, we obtain the expectation values
\begin{eqnarray}
\bm{\mathcal S}
&=&
\langle \bm{\hat{\mathcal S}} \rangle \\
&=&
\left (
  \begin{array}{c}
S_0 \\
S_1 \\
S_2 \\
S_3 
  \end{array}
\right )\\
&=&
\hbar {\mathcal N}
\left (
  \begin{array}{c}
1 \\
\sin \theta \cos \phi \\
\sin \theta \sin \phi \\
\cos \theta \\
  \end{array}
\right ),
\end{eqnarray}
which means that the {\it Stokes parameters \cite{Goldstein11,Gil16,Pedrotti07,Hecht17,Payne52,Fano54,Collett70,Delbourgo77,Luis02,Luis07,Bjork10} to describe the polarisation state of coherent photons were actually the quantum-mechanical expectation values of spin for photons}.

We can also go back to the original horizontal-vertical (HV) basis by the unitary transformation, which we obtained from the classical correspondence using $\hat{\bf m}={\bf r} \times \hat{\bf p}$ as,
\begin{eqnarray}
\hat{S}_{z}
&=&
\hbar 
\left(
  \begin{array}{cc}
     \hat{a}_{\rm H}^{\dagger}, & 
     \hat{a}_{\rm V}^{\dagger} 
  \end{array}
\right)
\left(
  \begin{array}{cc}
0 & -i \\
i & 0
  \end{array}
\right) 
\left(
  \begin{array}{c}
     \hat{a}_{\rm H} \\
     \hat{a}_{\rm V}
  \end{array}
\right) \\
&=&
\hbar 
\bm{\hat{\psi}}_{\rm HV}^{\dagger}
\sigma_2
\bm{\hat{\psi}}_{\rm HV},
\end{eqnarray}
where $\bm{\hat{\psi}}_{\rm HV}^{\dagger}=(\hat{a}_{\rm H}^{\dagger},\hat{a}_{\rm V}^{\dagger})$ and $\bm{\hat{\psi}}_{\rm HV}$ are the spinor representations of creation and annihilation operators in HV-basis.
For this basis, we should assign ${\bm \sigma}=(\sigma_3,\sigma_1,\sigma_2)$ by the cyclic exchange.
Then, we obtain 
\begin{eqnarray}
\hat{S}_{x}
&=&
\hbar 
\bm{\hat{\psi}}_{\rm HV}^{\dagger}
\sigma_3
\bm{\hat{\psi}}_{\rm HV},
\end{eqnarray}
and  
\begin{eqnarray}
\hat{S}_{y}
&=&
\hbar 
\bm{\hat{\psi}}_{\rm HV}^{\dagger}
\sigma_1
\bm{\hat{\psi}}_{\rm HV}.
\end{eqnarray}

We can also re-write 
\begin{eqnarray}
\hat{S}_{x}
&=&
\hbar
\left(
\hat{n}_{\rm H}
-
\hat{n}_{\rm V}
\right) \\
\hat{S}_{y}
&=&
\hbar
\left(
\hat{a}_{\rm H}^{\dagger} \hat{a}_{\rm V}
+
\hat{a}_{\rm V}^{\dagger} \hat{a}_{\rm H}
\right) \\
\hat{S}_{z}
&=&
\hbar
\left(
-i
\hat{a}_{\rm H}^{\dagger} \hat{a}_{\rm V}
+i
\hat{a}_{\rm V}^{\dagger} \hat{a}_{\rm H}
\right).
\end{eqnarray}
The quantum mechanical expectation values using coherent state of $|\alpha_{\rm H},\alpha_{\rm V} \rangle$ are immediately calculated as
\begin{eqnarray}
\langle \hat{S}_{x} \rangle
&=&
\hbar
\left(
{\mathcal N}_{\rm H}
-
{\mathcal N}_{\rm V}
\right) \\
&=&
\hbar {\mathcal N}
\left(
\cos^2 \alpha
-
\sin^2 \alpha
\right) \\
&=&
\hbar {\mathcal N}
\cos( 2 \alpha) \\
\langle \hat{S}_{y} \rangle
&=&
\hbar
\left(
{\alpha}_{\rm H}^{*} {\alpha}_{\rm V}
+
{\alpha}_{\rm V}^{*} {\alpha}_{\rm H}
\right) \\
&=&
\hbar {\mathcal N}
\left(
{\rm e}^{i \delta} \cos \alpha \sin \alpha
+
{\rm e}^{-i \delta} \cos \alpha \sin \alpha
\right) \\
&=&
\hbar {\mathcal N}
\cos{\delta} \sin (2\alpha)
\\
\langle \hat{S}_{z} \rangle
&=&
\hbar
\left(
-i
{\alpha}_{\rm H}^{*} {\alpha}_{\rm V}
+i
{\alpha}_{\rm V}^{*} {\alpha}_{\rm H}
\right) \\
&=&
\hbar {\mathcal N}
\left(
-i {\rm e}^{i \delta} \cos \alpha \sin \alpha
+
i {\rm e}^{-i \delta} \cos \alpha \sin \alpha
\right) \\
&=&
\hbar {\mathcal N}
\sin{\delta} \sin (2\alpha).
\end{eqnarray}

We can also use the Jones vector to calculate the spin expectation values by using the coherent state, and we obtain
\begin{eqnarray}
\langle \bm{\hat{\mathcal S}} \rangle 
&=
\hbar {\mathcal N}
\left (
  \begin{array}{c}
1 \\
    \cos (2 \alpha) \\
    \sin (2 \alpha) \cos \delta \\
    \sin (2 \alpha) \sin \delta 
  \end{array}
\right),
\end{eqnarray}
which is consistent with the above results obtained in the chiral representation.
The spacial components of Stokes parameters, ${\bf S}=(S_1,S_2,S_3)$, are usually shown in Poincar\'e sphere.
In the Jones vector description, the polar angle $\gamma=2\alpha$ is measured from $S_1$ axis and the azimuthal angle $\delta$ is measured from $S_2$ in the $S_2$-$S_3$ plane.

We can also confirm the sum rule 
\begin{eqnarray}
S_0 =\sqrt{S_1^2+S_2^2+S_3^2} 
\end{eqnarray}
for the expectation values in the coherent spin states.

We also obtained the commutation relationships \cite{Payne52,Fano54,Collett70,Delbourgo77,Luis02,Luis07,Bjork10} for spin operators as 
\begin{eqnarray}
\left[
\hat{S}_{x},
\hat{S}_{y}
\right]
&=&
2i
\hbar
\hat{S}_{z}, \\
\left[
\hat{S}_{y},
\hat{S}_{z}
\right]
&=&
2i
\hbar
\hat{S}_{x}, \\
\left[
\hat{S}_{z},
\hat{S}_{x}
\right]
&=&
2i
\hbar
\hat{S}_{y}, 
\end{eqnarray}
which are valid for both chiral and Jones bases.
Therefore, we obtained the spin operators for all components as generators of rotations for polarisation state of a coherent monochromatic ray of photons.

Now, we are ready to discuss what was $\hat{S}_z$ obtained from $\hat{\bf m} ={\bf r} \times\hat{\bf p}$.
If we focus on the spatial components of the Stokes operators,  ${\bf \hat{S}}=(\hat{S}_x,\hat{S}_y,\hat{S}_z)$,  it is equivalent to the helicity operator \cite{Sakurai67,Barnett12}, which is defined as the projection of the spin operators to the unit vector along the direction of the propagation, $\hat{\bf k}={\bf k}/k$, as 
\begin{eqnarray}
 \hat{h}_z 
&=&{\bf \hat{S}} \cdot \hat{\bf k} \\
&=&\hat{S}_z.
\end{eqnarray}
The helicity operator naturally sets the direction of the quantisation axis of spin aligned to the direction of the propagation. 
Nevertheless, this does not exclude the other polarisation states nor the spin components, perpendicular to the direction of the propagation.
The spin expectation values are observables, as clearly established as polarimetry \cite{Goldstein11,Gil16}.
Please also note that the expectation values of spin components are independent on the value of the quantum orbital angular momentum, $m$, because we have allowed the vortexed ray with non-zero topological charge.
In that sense, our results show that the spin angular momentum is independent on the orbital angular momentum.
Therefore, our framework is a natural extension of a standard QED theory to account for the spacial profile of the orbital wavefunction of photons, and we found that the spin angular momentum was not affected by the orbital angular momentum.

It is amazing to consider why Stokes and Poincar\'e \cite{Stokes51,Poincare92,Max99,Jackson99,Yariv97} could establish the descriptions of polarisation states using these parameters before the discoveries of quantum mechanics and the quantum field theories.
It is also astonishing to be aware that Stokes and Poincar\'e \cite{Stokes51,Poincare92} properly assigned the correct order parameters $ \bm{\mathcal S}=\langle \bm{\hat{\mathcal S}} \rangle=(S_0,S_x,S_y,S_z)$ in the 4-dimensional time-space coordinate, before the discovery of Einstein's theory of relativity, the Ginzburg-Landau theory of phase transitions, and the invention of a laser.

\section{Higher-order Poincar\'e sphere}
Now, we will extend our discussions for quantum-mechanical nature of orbital angular momentum for photons.
In order to make the argument specific, we consider a GRIN fibre under a cylindrical symmetry, again, but the extension to a more general waveguide is straightforward, as we discussed in sections for obtaining spin operators.
In the preceding sections, we obtained the orbital angular momentum along the direction of the propagation as, 
\begin{eqnarray}
\hat{L}_z
&=&
\hbar m
\left(
\hat{n}_{\rm H}
+
\hat{n}_{\rm V}
+1
\right).
\end{eqnarray}
There is no doubt that $\hat{L}_z$ describes the quantum orbital angular momentum along the direction of the propagation, because the expectation value becomes
\begin{eqnarray}
\langle \hat{L}_z \rangle
&=&
\hbar m
\left(
{\mathcal N}
+1
\right).
\end{eqnarray}
This means that the orbital angular momentum is not dependent on the polarisation state, as far as the average number of total photons, $\mathcal N$, is fixed.

Our next challenge is to identify the corresponding transverse operators, which should satisfy the commutation relationship.
In conjunction with the argument for spin operators, $\hat{L}_z$ must also be the helicity operator of orbital angular momentum, 
\begin{eqnarray}
\hat{L}_z = \hat{\bf L} \cdot \hat{\bf k},
\end{eqnarray}
if we could successfully define the orbital angular momentum operator, $\hat{\bf L}=(\hat{L}_x,\hat{L}_y,\hat{L}_z)$.

For further consideration of the orbital angular momentum, we should consider the orbital wavefunction, 
\begin{eqnarray}
\psi_n^m(r,\phi,z)
&=&
\left \langle 
r,\phi,z
|
n,m
\right \rangle \\
&=&
\frac{1}{w_0}
\sqrt{
\frac{2}{\pi}
\frac{n!}{(n+m)!}
}
\left(
\frac{\sqrt{2}r}{w_0}
\right)^{m}
\nonumber \\
&&
L_n^{m} 
\left(
2
\left(
  \frac{r}{w_0}
\right)^2 
\right) 
{\rm  e}^{-\frac{r^2}{w_0^2}}
{\rm  e}^{i m \phi}
{\rm  e}^{i kz},
\end{eqnarray}
and its energy dispersion
\begin{eqnarray}
E_n^m(k)
&=&
\hbar \omega_n^m(k) \\
&=&
\Delta_n^m
+
\sqrt{(\Delta_n^m)^2+ (\hbar v_0 k)^2},
\end{eqnarray}
where the energy gap, 
\begin{eqnarray}
\Delta_n^m
&=&\hbar \delta  w_0 (2n+|m|+1),  
\end{eqnarray}
is dependent on the quantum numbers $n$ and $m$.
From this dispersion, we recognise that the frequency depends on $n$ and $|m|$, such that the coupling between modes with different quantum numbers would not be coherently maintained for a long-distance propagation, because the phase and group velocities are different.
For a monochromatic ray, considered in this work, we will not discuss the coupling between modes with different energies.
We also neglect the coupling between modes with the different values of $n$, such that the coupling within the same $n$ is considered, which is not explicitly shown below for simplicity.
On the other hand, the modes with $m$ and $-m$ are degenerate, such that the coherent coupling among these modes are allowed. 
Moreover, these modes are orthogonal, 
\begin{eqnarray}
\langle -m | m \rangle
&=&
\int_0^{2\pi} \frac{d \phi}{2 \pi} {\rm e}^{+i m \phi}{\rm e}^{+i m \phi} \\
&=&0,
\end{eqnarray}
for $m \neq 0$.
Therefore, we can consider the coherent coupling between $| m \rangle$ and $| - m \rangle$, which is described by SU$(2)$, and phases and amplitudes of these orthogonal components will determine the quantum mechanical average of the orbital angular momentum, similar to the Stokes parameters in the Poincar\'e sphere.

First, we consider the consequence of the coupling between $| m \rangle$ and $| - m \rangle$ for the angular momentum along $z$, which should become
\begin{eqnarray}
\hat{L}_z^{m} 
&=& \hbar m 
\sum_{\sigma}
\left(
\hat{a}_{m \sigma}^{\dagger}\hat{a}_{m\sigma}-\hat{a}_{-m \sigma}^{\dagger}\hat{a }_{-m \sigma}
\right) \\
&=&
\hbar m
\sum_{\sigma}
\left(
  \begin{array}{cc}
     \hat{a}_{m \sigma}^{\dagger}, & 
     \hat{a}_{-m \sigma}^{\dagger} 
  \end{array}
\right)
\left(
  \begin{array}{cc}
1 & 0 \\
0 & -1
  \end{array}
\right) 
\left(
  \begin{array}{c}
     \hat{a}_{m \sigma} \\
     \hat{a}_{-m\sigma}
  \end{array}
\right) \nonumber \\
&=&
\hbar m
\sum_{\sigma}
\bm{\hat{\psi}}_{m\sigma}^{\dagger}
\sigma_3^{m}
\bm{\hat{\psi}}_{m\sigma},
\end{eqnarray}
where $\sigma_3^{m}=\sigma_3$ works for SU$(2)$ space of the $|m|$-th orbital angular momentum, $\bm{\hat{\psi}}_{m\sigma}^{\dagger}=(\hat{a}_{m\sigma}^{\dagger},\hat{a}_{-m\sigma}^{\dagger})$ and $\bm{\hat{\psi}}_{m\sigma}$ are spinor representations of the photonic field operators for $m$ and $-m$, and creation and annihilation operators with the angular momentum $m$ and the spin $\sigma$ are defined $\hat{a}_{m\sigma}^{\dagger}$ and $\hat{a}_{m\sigma}$, respectively.
The relationship between the single particle wavefunction and the operator, $\hat{a}_{m \sigma}^{\dagger}$, is given by
\begin{eqnarray}
\psi_n^m(r,\phi,z)
&=&
\langle 
r,\phi,z
|
\hat{a}_{m\sigma}^{\dagger}
|
0 \rangle \\
&=&
\langle 
r,\phi,z
|
n,m
\rangle, 
\end{eqnarray}
which is independent on the polarisation state, $\sigma$.
Therefore, the single particle wavefunction describes the orbital degree of freedom including the orbital angular momentum.
We realised the zero-point oscillations have not contributed to $\hat{L}_z^{m}$, because the contributions from the opposite angular momentum cancel out.

Then, we apply the same principle for spin to the orbital angular momentum, that {\it photonic votexed states are rotationally invariant for the light propagation in a waveguide with a cylindrical symmetry or a free space}.
This means that we can allow arbitrary superposition states between $| m \rangle$ and $| - m \rangle$ defined by their relative phases and amplitudes.
This allows us to use the SU$(2)$-Lie algebra for describing the orbital angular momentum operators, which are represented as
\begin{eqnarray}
\hat{L}_0^{m} 
&=&
\hbar m
\sum_{\sigma}
\bm{\hat{\psi}}_{m\sigma}^{\dagger}
{\bf 1}^{m}
\bm{\hat{\psi}}_{m\sigma} \\
\hat{L}_x^{m} 
&=&
\hbar m
\sum_{\sigma}
\bm{\hat{\psi}}_{m\sigma}^{\dagger}
\sigma_1^{m}
\bm{\hat{\psi}}_{m\sigma} \\
\hat{L}_y^{m} 
&=&
\hbar m
\sum_{\sigma}
\bm{\hat{\psi}}_{m\sigma}^{\dagger}
\sigma_2^{m}
\bm{\hat{\psi}}_{m\sigma}, 
\end{eqnarray}
where $({\bf 1}^{m},\sigma^{m}_{x},\sigma^{m}_{y},\sigma^{m}_{z})=({\bf 1},\sigma_1,\sigma_2,\sigma_3)$ is applied to the Hilbert space spanned by $| m \rangle$ and $| - m \rangle$.
This means that we are focussing on the direct product space of orbital and spin, described by ${\rm SU}(2) \otimes {\rm SU}(2)$.

Within this Hilbert space, we realise that the helicity operator is obtained as
\begin{eqnarray}
\hat{L}_z^{m} 
&=& \hat{\bf L}^{m} \cdot \hat{\bf k} \\
&=& \hbar m 
\sum_{\sigma}
\left(
\hat{n}_{m\sigma}-\hat{n}_{-m\sigma}
\right)
\end{eqnarray}
where the number operator is defined as $\hat{n}_{m\sigma}=\hat{a}_{m \sigma}^{\dagger}\hat{a}_{m\sigma}$ and we have defined spatial components of orbital angular momentum operators as an operational vector, $\hat{\bf L}^{m}=(\hat{L}_x^{m},\hat{L}_y^{m},\hat{L}_z^{m})$.

For example, if we take the quantum-mechanical average over the coherent spin state with the average number of photons ${\mathcal N}$, we obtain
\begin{eqnarray}
\hat{L}^{m}_x
&=&
\hbar m
{\mathcal N}
{\bf 1}^{m}
\\
\hat{L}^{m}_x
&=&
\hbar m
{\mathcal N}
\sigma^{m}_{x}
\\
\hat{L}^{m}_y
&=&
\hbar m
{\mathcal N}
\sigma^{m}_{y}
\\
\hat{L}^{m}_z
&=&
\hbar m
{\mathcal N}
\sigma^{m}_{z},
\end{eqnarray}
while we still expect non-trivial expectation values for the orbital angular momentum.
 
Moreover, if we assume the superposition state of the orbitals of $| m \rangle$ and $| - m \rangle$ with the polar angle of $\Theta$ and the azimuthal angle of $\Phi$ in the higher-order Poincar\'e sphere \cite{Padgett99,Milione11,Liu17,Erhard18}, the higher-order Bloch state becomes
\begin{eqnarray}
\langle \Theta, \Phi | {\rm Bloch} \rangle
&=&
{\rm e}^{-i \frac{\Phi}{2}} \cos (\Theta/2) \langle m | {\rm Bloch}\rangle \nonumber \\
&&+
{\rm e}^{+i \frac{\Phi}{2}} \sin  (\Theta/2)  \langle -m  | {\rm Bloch}\rangle  \nonumber \\
&=&
\left (
  \begin{array}{c}
    {\rm e}^{-i\Phi/2}\cos  (\Theta/2) \  \\
    {\rm e}^{+i\Phi/2} \sin  (\Theta/2) \ 
  \end{array}
\right),
\end{eqnarray}
which yields the expectation value of the orbital angular momentum as 
\begin{eqnarray}
\bm{\mathcal L}
&=&
\langle \bm{\hat{\mathcal L}} \rangle \\
&=&
\left (
  \begin{array}{c}
L_0 \\
L_1 \\
L_2 \\
L_3 
  \end{array}
\right )\\
&=&
\hbar m
{\mathcal N}
\left (
  \begin{array}{c}
1 \\
\sin \Theta \cos \Phi \\
\sin \Theta \sin \Phi \\
\cos \Theta \\
  \end{array}
\right ).
\end{eqnarray}
This shows that the vortexed photon with the topological charge of $m$ has an angular momentum of $\hbar m$ and the vectorial direction of the orbital angular momentum is proportional to the spatial vector, ${\bf L}=(L_1,L_2,L_3)$, shown in the higher-order Poincar\'e sphere.

We can also confirm the sum rule 
\begin{eqnarray}
L_0 =\sqrt{L_1^2+L_2^2+L_3^2} 
\end{eqnarray}
for the expectation values for the coherent vortexed states, similar to the spin state.

The commutation relationships for orbital angular momentum operators are obtained as
\begin{eqnarray}
\left[
\hat{L}_{x}^{m},
\hat{L}_{y}^{m}
\right]
&=&
2i
\hbar m
\hat{L}_{z}^{m}, \\
\left[
\hat{L}_{y}^{m},
\hat{L}_{z}^{m}
\right]
&=&
2i
\hbar m
\hat{L}_{x}^{m}, \\
\left[
\hat{L}_{z}^{m},
\hat{L}_{x}^{m}
\right]
&=&
2i
\hbar m
\hat{L}_{y}^{m}, 
\end{eqnarray}
where the unusual factor of $2$ is coming from the SU$(2)$ nature of the Hilbert space for coupling among $\hbar m$ and $-\hbar m$, which we are considering due to the energy coherence of the mode, similar to the case for spin operators.

More generally, the entire Hilbert space is described by the direct sum for states with different $m$, composed of $2m$ degrees of freedom from multiple SU$(2)$ spaces and $1$ degree of freedom from U$(1)$ for $m=0$, as $\left \{
{\rm SU}(2)
\oplus
\cdots
\oplus
{\rm SU}(2)
\oplus
{\rm U}(1)
\right \}
\otimes
{\rm SU}(2)$, where the last part of $\otimes{\rm SU}(2)$ describes the direct product to the spin space.

For the free space, in the limits of $g \rightarrow 1$ and $v_0 \rightarrow c$, the states of photons with different $m$ would degenerate due to the closing of the energy gap. 
In this case, the coherent superposition between states with different $m$ will be allowed. 
The total Hilbert space will become the direct product between the orbital Hilbert space and the spin Hilbert space, ${\rm SU}(2m+1) \otimes {\rm SU}(2)$ with $m\rightarrow \infty$, in principle.

\section{Conclusions}
We have reviewed the historical derivations of the angular momentum using classical electromagnetic waves of Laguerre-Gauss modes.
While extending the treatment towards the quantum field theory, we have found that the plane-wave expansions cannot sustain a vortex with topological charge, which also leads erroneous results of zero angular momentum and gauge dependent expressions.

The problem could be overcome by taking the small longitudinal component along the direction of the propagation due to the finite mode profile of the ray.
As a result, we obtained helicity operators for both spin and orbital angular momentum.
By accepting the principle of the rotational symmetries of photonic states in a waveguide with a cylindrical symmetry, we obtain the angular momentum operators as generators of rotations for both spin and orbital angular momentum.
We have also shown that the Stokes parameters in Poincar\'e sphere are actually quantum-mechanical averages of spin operators by coherent states.
We could extend this concept to the orbital angular momentum in higher-order Poincar\'e sphere.

In conclusion, spin and orbital angular momentum are intrinsic quantum degrees of freedom for photons.
We have shown that the splitting of spin and orbital angular momentum from the total orbital angular momentum is achievable for a coherent monochromatic ray of photons emitted from a laser source.
Therefore, spin and orbit can be treated independently.
We believe that our results will be valuable for various applications of spin and orbital angular momentum of photons, because fully quantum-mechanical degrees of freedom are available by using ubiquitous laser sources.

\section*{Acknowledgements}
This work is supported by JSPS KAKENHI Grant Number JP 18K19958.
The author would like to express sincere thanks to Prof I. Tomita for continuous discussions and encouragements.

\bibliography{Split_SAM_OAM}% Produces the bibliography via BibTeX.

\begin{thebibliography}{55}%
\makeatletter
\providecommand \@ifxundefined [1]{%
 \@ifx{#1\undefined}
}%
\providecommand \@ifnum [1]{%
 \ifnum #1\expandafter \@firstoftwo
 \else \expandafter \@secondoftwo
 \fi
}%
\providecommand \@ifx [1]{%
 \ifx #1\expandafter \@firstoftwo
 \else \expandafter \@secondoftwo
 \fi
}%
\providecommand \natexlab [1]{#1}%
\providecommand \enquote  [1]{``#1''}%
\providecommand \bibnamefont  [1]{#1}%
\providecommand \bibfnamefont [1]{#1}%
\providecommand \citenamefont [1]{#1}%
\providecommand \href@noop [0]{\@secondoftwo}%
\providecommand \href [0]{\begingroup \@sanitize@url \@href}%
\providecommand \@href[1]{\@@startlink{#1}\@@href}%
\providecommand \@@href[1]{\endgroup#1\@@endlink}%
\providecommand \@sanitize@url [0]{\catcode `\\12\catcode `\$12\catcode
  `\&12\catcode `\#12\catcode `\^12\catcode `\_12\catcode `\%12\relax}%
\providecommand \@@startlink[1]{}%
\providecommand \@@endlink[0]{}%
\providecommand \url  [0]{\begingroup\@sanitize@url \@url }%
\providecommand \@url [1]{\endgroup\@href {#1}{\urlprefix }}%
\providecommand \urlprefix  [0]{URL }%
\providecommand \Eprint [0]{\href }%
\providecommand \doibase [0]{https://doi.org/}%
\providecommand \selectlanguage [0]{\@gobble}%
\providecommand \bibinfo  [0]{\@secondoftwo}%
\providecommand \bibfield  [0]{\@secondoftwo}%
\providecommand \translation [1]{[#1]}%
\providecommand \BibitemOpen [0]{}%
\providecommand \bibitemStop [0]{}%
\providecommand \bibitemNoStop [0]{.\EOS\space}%
\providecommand \EOS [0]{\spacefactor3000\relax}%
\providecommand \BibitemShut  [1]{\csname bibitem#1\endcsname}%
\let\auto@bib@innerbib\@empty
%</preamble>
\bibitem [{\citenamefont {Newton}(2010)}]{Newton1730}%
  \BibitemOpen
  \bibfield  {author} {\bibinfo {author} {\bibfnamefont {I.}~\bibnamefont
  {Newton}},\ }\href@noop {} {\emph {\bibinfo {title} {Opticks}}}\ (\bibinfo
  {publisher} {London: William Innys},\ \bibinfo {year} {1730 (Project
  Gutenberg, 2010)})\BibitemShut {NoStop}%
\bibitem [{\citenamefont {Stokes}(1851)}]{Stokes51}%
  \BibitemOpen
  \bibfield  {author} {\bibinfo {author} {\bibfnamefont {G.~G.}\ \bibnamefont
  {Stokes}},\ }\bibfield  {title} {\bibinfo {title} {On the composition and
  resolution of streams of polarized light from different sources},\
  }\href@noop {} {\bibfield  {journal} {\bibinfo  {journal} {Trans. Cambridge
  Phil. Soc.}\ }\textbf {\bibinfo {volume} {9}},\ \bibinfo {pages} {399}
  (\bibinfo {year} {1851})}\BibitemShut {NoStop}%
\bibitem [{\citenamefont {Poincar$\rm\acute{e}$}(1892)}]{Poincare92}%
  \BibitemOpen
  \bibfield  {author} {\bibinfo {author} {\bibfnamefont {J.~H.}\ \bibnamefont
  {Poincar$\rm\acute{e}$}},\ }\href@noop {} {\emph {\bibinfo {title}
  {Th$\rm\acute{e}$orie math$\rm\acute{e}$matique de la
  lumi$\rm\grave{e}$re}}}\ (\bibinfo  {publisher} {G. Carr$\rm\acute{e}$},\
  \bibinfo {year} {1892})\BibitemShut {NoStop}%
\bibitem [{\citenamefont {Born}\ and\ \citenamefont {Wolf}(1999)}]{Max99}%
  \BibitemOpen
  \bibfield  {author} {\bibinfo {author} {\bibfnamefont {M.}~\bibnamefont
  {Born}}\ and\ \bibinfo {author} {\bibfnamefont {E.}~\bibnamefont {Wolf}},\
  }\href@noop {} {\emph {\bibinfo {title} {Principles of Optics}}}\ (\bibinfo
  {publisher} {Cambridge University Press},\ \bibinfo {year}
  {1999})\BibitemShut {NoStop}%
\bibitem [{\citenamefont {Jackson}(1999)}]{Jackson99}%
  \BibitemOpen
  \bibfield  {author} {\bibinfo {author} {\bibfnamefont {J.~D.}\ \bibnamefont
  {Jackson}},\ }\href@noop {} {\emph {\bibinfo {title} {Classical
  Electrodynamics}}}\ (\bibinfo  {publisher} {John Wiley \& Sons},\ \bibinfo
  {year} {1999})\BibitemShut {NoStop}%
\bibitem [{\citenamefont {Yariv}\ and\ \citenamefont {Yeh}(1997)}]{Yariv97}%
  \BibitemOpen
  \bibfield  {author} {\bibinfo {author} {\bibfnamefont {Y.}~\bibnamefont
  {Yariv}}\ and\ \bibinfo {author} {\bibfnamefont {P.}~\bibnamefont {Yeh}},\
  }\href@noop {} {\emph {\bibinfo {title} {Photonics: optical electronics in
  modern communications}}}\ (\bibinfo  {publisher} {Oxford University Press},\
  \bibinfo {year} {1997})\BibitemShut {NoStop}%
\bibitem [{\citenamefont {Dirac}(1930)}]{Dirac30}%
  \BibitemOpen
  \bibfield  {author} {\bibinfo {author} {\bibfnamefont {P.~A.~M.}\
  \bibnamefont {Dirac}},\ }\href@noop {} {\emph {\bibinfo {title} {The
  Principle of Quantum Mechanics}}}\ (\bibinfo  {publisher} {Oxford University
  Press},\ \bibinfo {year} {1930})\BibitemShut {NoStop}%
\bibitem [{\citenamefont {Baym}(1969)}]{Baym69}%
  \BibitemOpen
  \bibfield  {author} {\bibinfo {author} {\bibfnamefont {G.}~\bibnamefont
  {Baym}},\ }\href@noop {} {\emph {\bibinfo {title} {Lectures on Quantum
  Mechanics}}}\ (\bibinfo  {publisher} {Westview Press},\ \bibinfo {year}
  {1969})\BibitemShut {NoStop}%
\bibitem [{\citenamefont {Sakurai}\ and\ \citenamefont
  {Napolitano}(2014)}]{Sakurai14}%
  \BibitemOpen
  \bibfield  {author} {\bibinfo {author} {\bibfnamefont {J.~J.}\ \bibnamefont
  {Sakurai}}\ and\ \bibinfo {author} {\bibfnamefont {J.~J.}\ \bibnamefont
  {Napolitano}},\ }\href@noop {} {\emph {\bibinfo {title} {Modern Quantum
  Mechanics}}}\ (\bibinfo  {publisher} {Pearson},\ \bibinfo {year}
  {2014})\BibitemShut {NoStop}%
\bibitem [{\citenamefont {Sakurai}(1967)}]{Sakurai67}%
  \BibitemOpen
  \bibfield  {author} {\bibinfo {author} {\bibfnamefont {J.~J.}\ \bibnamefont
  {Sakurai}},\ }\href@noop {} {\emph {\bibinfo {title} {Advanced Quantum
  Mechanics}}}\ (\bibinfo  {publisher} {Addison-Wesley Publishing Company},\
  \bibinfo {year} {1967})\BibitemShut {NoStop}%
\bibitem [{\citenamefont {Goldstein}(2011)}]{Goldstein11}%
  \BibitemOpen
  \bibfield  {author} {\bibinfo {author} {\bibfnamefont {D.~H.}\ \bibnamefont
  {Goldstein}},\ }\href@noop {} {\emph {\bibinfo {title} {Polarized Light}}}\
  (\bibinfo  {publisher} {CRC Press},\ \bibinfo {year} {2011})\BibitemShut
  {NoStop}%
\bibitem [{\citenamefont {Gil}\ and\ \citenamefont {Ossikovski}(2016)}]{Gil16}%
  \BibitemOpen
  \bibfield  {author} {\bibinfo {author} {\bibfnamefont {J.~J.}\ \bibnamefont
  {Gil}}\ and\ \bibinfo {author} {\bibfnamefont {R.}~\bibnamefont
  {Ossikovski}},\ }\href@noop {} {\emph {\bibinfo {title} {Polarized Light and
  the Mueller Matrix Approach}}}\ (\bibinfo  {publisher} {CRC Press},\ \bibinfo
  {year} {2016})\BibitemShut {NoStop}%
\bibitem [{\citenamefont {Pedrotti}\ \emph {et~al.}(2007)\citenamefont
  {Pedrotti}, \citenamefont {Pedrotti},\ and\ \citenamefont
  {Pedrotti}}]{Pedrotti07}%
  \BibitemOpen
  \bibfield  {author} {\bibinfo {author} {\bibfnamefont {F.~L.}\ \bibnamefont
  {Pedrotti}}, \bibinfo {author} {\bibfnamefont {L.~M.}\ \bibnamefont
  {Pedrotti}},\ and\ \bibinfo {author} {\bibfnamefont {L.~S.}\ \bibnamefont
  {Pedrotti}},\ }\href@noop {} {\emph {\bibinfo {title} {Introduction to
  Optics}}}\ (\bibinfo  {publisher} {Pearson Education},\ \bibinfo {year}
  {2007})\BibitemShut {NoStop}%
\bibitem [{\citenamefont {Hecht}(2017)}]{Hecht17}%
  \BibitemOpen
  \bibfield  {author} {\bibinfo {author} {\bibfnamefont {E.}~\bibnamefont
  {Hecht}},\ }\href@noop {} {\emph {\bibinfo {title} {Optics}}}\ (\bibinfo
  {publisher} {Pearson Education},\ \bibinfo {year} {2017})\BibitemShut
  {NoStop}%
\bibitem [{\citenamefont {Jones}(1941)}]{Jones41}%
  \BibitemOpen
  \bibfield  {author} {\bibinfo {author} {\bibfnamefont {R.~C.}\ \bibnamefont
  {Jones}},\ }\bibfield  {title} {\bibinfo {title} {A new calculus for the
  treatment of optical systems i. description and discussion of the calculus},\
  }\href {https://doi.org/10.1364/JOSA.31.000488} {\bibfield  {journal}
  {\bibinfo  {journal} {J. Opt. Soc. Am.}\ }\textbf {\bibinfo {volume} {31}},\
  \bibinfo {pages} {488} (\bibinfo {year} {1941})}\BibitemShut {NoStop}%
\bibitem [{\citenamefont {Payne}(1952)}]{Payne52}%
  \BibitemOpen
  \bibfield  {author} {\bibinfo {author} {\bibfnamefont {W.~T.}\ \bibnamefont
  {Payne}},\ }\bibfield  {title} {\bibinfo {title} {Elementary spinor theory},\
  }\href {https://doi.org/10.1119/1.1933190} {\bibfield  {journal} {\bibinfo
  {journal} {Am. J. Phys.}\ }\textbf {\bibinfo {volume} {20}},\ \bibinfo
  {pages} {253} (\bibinfo {year} {1952})}\BibitemShut {NoStop}%
\bibitem [{\citenamefont {Collett}(1970)}]{Collett70}%
  \BibitemOpen
  \bibfield  {author} {\bibinfo {author} {\bibfnamefont {E.}~\bibnamefont
  {Collett}},\ }\bibfield  {title} {\bibinfo {title} {Stokes parameters for
  quantum systems},\ }\href@noop {} {\bibfield  {journal} {\bibinfo  {journal}
  {Am. J. Phys.}\ }\textbf {\bibinfo {volume} {38}} (\bibinfo {year}
  {1970})}\BibitemShut {NoStop}%
\bibitem [{\citenamefont {Luis}(2002)}]{Luis02}%
  \BibitemOpen
  \bibfield  {author} {\bibinfo {author} {\bibfnamefont {A.}~\bibnamefont
  {Luis}},\ }\bibfield  {title} {\bibinfo {title} {Degree of polarization in
  quantum optics},\ }\href {https://doi.org/10.1103/PhysRevA.66.013806}
  {\bibfield  {journal} {\bibinfo  {journal} {Phys. Rev. A}\ }\textbf {\bibinfo
  {volume} {66}},\ \bibinfo {pages} {013806} (\bibinfo {year}
  {2002})}\BibitemShut {NoStop}%
\bibitem [{\citenamefont {Luis}(2007)}]{Luis07}%
  \BibitemOpen
  \bibfield  {author} {\bibinfo {author} {\bibfnamefont {A.}~\bibnamefont
  {Luis}},\ }\bibfield  {title} {\bibinfo {title} {Polarization distributions
  and degree of polarization for quantum gaussian light fields},\ }\href@noop
  {} {\bibfield  {journal} {\bibinfo  {journal} {Opt. Comm.}\ }\textbf
  {\bibinfo {volume} {273}},\ \bibinfo {pages} {173} (\bibinfo {year}
  {2007})}\BibitemShut {NoStop}%
\bibitem [{\citenamefont {Bj$\rm\ddot{o}$rk}\ \emph {et~al.}(2010)\citenamefont
  {Bj$\rm\ddot{o}$rk}, \citenamefont {S$\rm\ddot{o}$derholm}, \citenamefont
  {S$\rm\acute{a}$nchez-Soto}, \citenamefont {Klimov}, \citenamefont {Ghiu},
  \citenamefont {Marian},\ and\ \citenamefont {Marian}}]{Bjork10}%
  \BibitemOpen
  \bibfield  {author} {\bibinfo {author} {\bibfnamefont {G.}~\bibnamefont
  {Bj$\rm\ddot{o}$rk}}, \bibinfo {author} {\bibfnamefont {J.}~\bibnamefont
  {S$\rm\ddot{o}$derholm}}, \bibinfo {author} {\bibfnamefont {L.~L.}\
  \bibnamefont {S$\rm\acute{a}$nchez-Soto}}, \bibinfo {author} {\bibfnamefont
  {A.~B.}\ \bibnamefont {Klimov}}, \bibinfo {author} {\bibfnamefont
  {I.}~\bibnamefont {Ghiu}}, \bibinfo {author} {\bibfnamefont {P.}~\bibnamefont
  {Marian}},\ and\ \bibinfo {author} {\bibfnamefont {T.~A.}\ \bibnamefont
  {Marian}},\ }\bibfield  {title} {\bibinfo {title} {Quantum degrees of
  polarization},\ }\href {https://doi.org/10.1016/j.optcom.2010.04.088}
  {\bibfield  {journal} {\bibinfo  {journal} {Opt. Comm.}\ }\textbf {\bibinfo
  {volume} {283}},\ \bibinfo {pages} {4440} (\bibinfo {year}
  {2010})}\BibitemShut {NoStop}%
\bibitem [{\citenamefont {d.~Castillo}\ and\ \citenamefont
  {Garc$\rm\acute{i}$a}(2011)}]{Castillo11}%
  \BibitemOpen
  \bibfield  {author} {\bibinfo {author} {\bibfnamefont {G.~F.~T.}\
  \bibnamefont {d.~Castillo}}\ and\ \bibinfo {author} {\bibfnamefont {I.~R.}\
  \bibnamefont {Garc$\rm\acute{i}$a}},\ }\bibfield  {title} {\bibinfo {title}
  {The {J}ones vector as a spinor and its representation on the
  {P}oincar$\rm\acute{e}$ sphere},\ }\href@noop {} {\bibfield  {journal}
  {\bibinfo  {journal} {Rev. Mex. Fis.}\ }\textbf {\bibinfo {volume} {57}},\
  \bibinfo {pages} {406} (\bibinfo {year} {2011})}\BibitemShut {NoStop}%
\bibitem [{\citenamefont {Sotto}\ \emph
  {et~al.}(2018{\natexlab{a}})\citenamefont {Sotto}, \citenamefont {Tomita},
  \citenamefont {Debnath},\ and\ \citenamefont {Saito}}]{Sotto18}%
  \BibitemOpen
  \bibfield  {author} {\bibinfo {author} {\bibfnamefont {M.}~\bibnamefont
  {Sotto}}, \bibinfo {author} {\bibfnamefont {I.}~\bibnamefont {Tomita}},
  \bibinfo {author} {\bibfnamefont {K.}~\bibnamefont {Debnath}},\ and\ \bibinfo
  {author} {\bibfnamefont {S.}~\bibnamefont {Saito}},\ }\bibfield  {title}
  {\bibinfo {title} {Polarization rotation and mode splitting in photonic
  crystal line-defect waveguides},\ }\href
  {https://doi.org/10.3389/fphy.2018.00085} {\bibfield  {journal} {\bibinfo
  {journal} {Front. Phys.}\ }\textbf {\bibinfo {volume} {6}},\ \bibinfo {pages}
  {85} (\bibinfo {year} {2018}{\natexlab{a}})}\BibitemShut {NoStop}%
\bibitem [{\citenamefont {Sotto}\ \emph
  {et~al.}(2018{\natexlab{b}})\citenamefont {Sotto}, \citenamefont {Debnath},
  \citenamefont {Khokhar}, \citenamefont {Tomita}, \citenamefont {Thomson},\
  and\ \citenamefont {Saito}}]{Sotto18b}%
  \BibitemOpen
  \bibfield  {author} {\bibinfo {author} {\bibfnamefont {M.}~\bibnamefont
  {Sotto}}, \bibinfo {author} {\bibfnamefont {K.}~\bibnamefont {Debnath}},
  \bibinfo {author} {\bibfnamefont {A.~Z.}\ \bibnamefont {Khokhar}}, \bibinfo
  {author} {\bibfnamefont {I.}~\bibnamefont {Tomita}}, \bibinfo {author}
  {\bibfnamefont {D.}~\bibnamefont {Thomson}},\ and\ \bibinfo {author}
  {\bibfnamefont {S.}~\bibnamefont {Saito}},\ }\bibfield  {title} {\bibinfo
  {title} {Anomalous zero-group-velocity photonic bonding states with local
  chirality},\ }\href {https://doi.org/10.1364/JOSAB.35.002356} {\bibfield
  {journal} {\bibinfo  {journal} {J. Opt. Soc. Am. B}\ }\textbf {\bibinfo
  {volume} {35}},\ \bibinfo {pages} {2356} (\bibinfo {year}
  {2018}{\natexlab{b}})}\BibitemShut {NoStop}%
\bibitem [{\citenamefont {Sotto}\ \emph {et~al.}(2019)\citenamefont {Sotto},
  \citenamefont {Debnath}, \citenamefont {Tomita},\ and\ \citenamefont
  {Saito}}]{Sotto19}%
  \BibitemOpen
  \bibfield  {author} {\bibinfo {author} {\bibfnamefont {M.}~\bibnamefont
  {Sotto}}, \bibinfo {author} {\bibfnamefont {K.}~\bibnamefont {Debnath}},
  \bibinfo {author} {\bibfnamefont {I.}~\bibnamefont {Tomita}},\ and\ \bibinfo
  {author} {\bibfnamefont {S.}~\bibnamefont {Saito}},\ }\bibfield  {title}
  {\bibinfo {title} {Spin-orbit coupling of light in photonic crystal
  waveguides},\ }\href {https://doi.org/10.1103/PhysRevA.99.053845} {\bibfield
  {journal} {\bibinfo  {journal} {Phys. Rev. A}\ }\textbf {\bibinfo {volume}
  {99}},\ \bibinfo {pages} {053845} (\bibinfo {year} {2019})}\BibitemShut
  {NoStop}%
\bibitem [{\citenamefont {Allen}\ \emph {et~al.}(1992)\citenamefont {Allen},
  \citenamefont {Beijersbergen}, \citenamefont {Spreeuw},\ and\ \citenamefont
  {Woerdman}}]{Allen92}%
  \BibitemOpen
  \bibfield  {author} {\bibinfo {author} {\bibfnamefont {L.}~\bibnamefont
  {Allen}}, \bibinfo {author} {\bibfnamefont {M.~W.}\ \bibnamefont
  {Beijersbergen}}, \bibinfo {author} {\bibfnamefont {R.~J.~C.}\ \bibnamefont
  {Spreeuw}},\ and\ \bibinfo {author} {\bibfnamefont {J.~P.}\ \bibnamefont
  {Woerdman}},\ }\bibfield  {title} {\bibinfo {title} {Orbital angular momentum
  of light and the transformation of {L}aguerre-{G}aussian laser modes},\
  }\href {https://doi.org/10.1103/PhysRevA.45.8185} {\bibfield  {journal}
  {\bibinfo  {journal} {Phys. Rev. A}\ }\textbf {\bibinfo {volume} {45}},\
  \bibinfo {pages} {8185} (\bibinfo {year} {1992})}\BibitemShut {NoStop}%
\bibitem [{\citenamefont {v.~Enk}\ and\ \citenamefont
  {Nienhuis}(1994)}]{Enk94}%
  \BibitemOpen
  \bibfield  {author} {\bibinfo {author} {\bibfnamefont {S.~J.}\ \bibnamefont
  {v.~Enk}}\ and\ \bibinfo {author} {\bibfnamefont {G.}~\bibnamefont
  {Nienhuis}},\ }\bibfield  {title} {\bibinfo {title} {Commutation rules and
  eigenvalues of spin and orbital angular momentum of radiation fields},\
  }\href {https://doi.org/10.1080/09500349414550911} {\bibfield  {journal}
  {\bibinfo  {journal} {J. Mod. Opt.}\ }\textbf {\bibinfo {volume} {41}},\
  \bibinfo {pages} {963} (\bibinfo {year} {1994})}\BibitemShut {NoStop}%
\bibitem [{\citenamefont {Leader}\ and\ \citenamefont
  {Lorc$\rm\acute{e}$}(2014)}]{Leader14}%
  \BibitemOpen
  \bibfield  {author} {\bibinfo {author} {\bibfnamefont {E.}~\bibnamefont
  {Leader}}\ and\ \bibinfo {author} {\bibfnamefont {C.}~\bibnamefont
  {Lorc$\rm\acute{e}$}},\ }\bibfield  {title} {\bibinfo {title} {The angular
  momentum controversy: {W}hat's it all about and does it matter?},\ }\href
  {https://doi.org/10.1016/j.physrep.2014.02.010} {\bibfield  {journal}
  {\bibinfo  {journal} {Phys. Rep.}\ }\textbf {\bibinfo {volume} {541}},\
  \bibinfo {pages} {163} (\bibinfo {year} {2014})}\BibitemShut {NoStop}%
\bibitem [{\citenamefont {Barnett}\ \emph
  {et~al.}(2016{\natexlab{a}})\citenamefont {Barnett}, \citenamefont {Allen},
  \citenamefont {Cameron}, \citenamefont {Gilson}, \citenamefont {Padgett},
  \citenamefont {Speirits},\ and\ \citenamefont {Yao}}]{Barnett16}%
  \BibitemOpen
  \bibfield  {author} {\bibinfo {author} {\bibfnamefont {S.~M.}\ \bibnamefont
  {Barnett}}, \bibinfo {author} {\bibfnamefont {L.}~\bibnamefont {Allen}},
  \bibinfo {author} {\bibfnamefont {R.~P.}\ \bibnamefont {Cameron}}, \bibinfo
  {author} {\bibfnamefont {C.~R.}\ \bibnamefont {Gilson}}, \bibinfo {author}
  {\bibfnamefont {M.~J.}\ \bibnamefont {Padgett}}, \bibinfo {author}
  {\bibfnamefont {F.~C.}\ \bibnamefont {Speirits}},\ and\ \bibinfo {author}
  {\bibfnamefont {A.~M.}\ \bibnamefont {Yao}},\ }\bibfield  {title} {\bibinfo
  {title} {On the natures of the spin and orbital parts of optical angualr
  momentum},\ }\href {https://doi.org/10.1088/2040-8978/18/6/064004} {\bibfield
   {journal} {\bibinfo  {journal} {J. Opt.}\ }\textbf {\bibinfo {volume}
  {18}},\ \bibinfo {pages} {064004} (\bibinfo {year}
  {2016}{\natexlab{a}})}\BibitemShut {NoStop}%
\bibitem [{\citenamefont {Grynberg}\ \emph {et~al.}(2010)\citenamefont
  {Grynberg}, \citenamefont {Aspect},\ and\ \citenamefont
  {Fabre}}]{Grynberg10}%
  \BibitemOpen
  \bibfield  {author} {\bibinfo {author} {\bibfnamefont {G.}~\bibnamefont
  {Grynberg}}, \bibinfo {author} {\bibfnamefont {A.}~\bibnamefont {Aspect}},\
  and\ \bibinfo {author} {\bibfnamefont {C.}~\bibnamefont {Fabre}},\
  }\href@noop {} {\emph {\bibinfo {title} {Introduction to Quantum Optics: From
  the Semi-classical Approach to Quantized Light}}}\ (\bibinfo  {publisher}
  {Cambridge University Press},\ \bibinfo {year} {2010})\BibitemShut {NoStop}%
\bibitem [{\citenamefont {Bliokh}\ \emph {et~al.}(2015)\citenamefont {Bliokh},
  \citenamefont {Rodr$\rm\acute{i}$guez-Fortu$\rm\tilde{n}$o}, \citenamefont
  {Nori},\ and\ \citenamefont {Zayats}}]{Bliokh15}%
  \BibitemOpen
  \bibfield  {author} {\bibinfo {author} {\bibfnamefont {K.~Y.}\ \bibnamefont
  {Bliokh}}, \bibinfo {author} {\bibfnamefont {F.~J.}\ \bibnamefont
  {Rodr$\rm\acute{i}$guez-Fortu$\rm\tilde{n}$o}}, \bibinfo {author}
  {\bibfnamefont {F.}~\bibnamefont {Nori}},\ and\ \bibinfo {author}
  {\bibfnamefont {A.~V.}\ \bibnamefont {Zayats}},\ }\bibfield  {title}
  {\bibinfo {title} {Spin-orbit interactions of light},\ }\href
  {https://doi.org/10.1038/NPHOTON.2015.201} {\bibfield  {journal} {\bibinfo
  {journal} {Nat. Photon.}\ }\textbf {\bibinfo {volume} {9}},\ \bibinfo {pages}
  {796} (\bibinfo {year} {2015})}\BibitemShut {NoStop}%
\bibitem [{\citenamefont {Chen}\ \emph {et~al.}(2008)\citenamefont {Chen},
  \citenamefont {L$\rm\ddot{u}$}, \citenamefont {Sun}, \citenamefont {Wang},\
  and\ \citenamefont {Goldman}}]{Chen08}%
  \BibitemOpen
  \bibfield  {author} {\bibinfo {author} {\bibfnamefont {X.~S.}\ \bibnamefont
  {Chen}}, \bibinfo {author} {\bibfnamefont {X.~F.}\ \bibnamefont
  {L$\rm\ddot{u}$}}, \bibinfo {author} {\bibfnamefont {W.~M.}\ \bibnamefont
  {Sun}}, \bibinfo {author} {\bibfnamefont {F.}~\bibnamefont {Wang}},\ and\
  \bibinfo {author} {\bibfnamefont {T.}~\bibnamefont {Goldman}},\ }\bibfield
  {title} {\bibinfo {title} {Spin and orbital angular momentum in gauge
  theories: Nucleon spin structure and multipole radiation revisited},\ }\href
  {https://doi.org/10.1103/PhysRevLett.100.232002} {\bibfield  {journal}
  {\bibinfo  {journal} {Phys. Rev. Lett.}\ }\textbf {\bibinfo {volume} {100}},\
  \bibinfo {pages} {232002} (\bibinfo {year} {2008})}\BibitemShut {NoStop}%
\bibitem [{\citenamefont {Ji}(2010)}]{Ji10}%
  \BibitemOpen
  \bibfield  {author} {\bibinfo {author} {\bibfnamefont {X.}~\bibnamefont
  {Ji}},\ }\bibfield  {title} {\bibinfo {title} {Comment on "{S}pin and orbital
  angular momentum in gauge theories: Nucleon spin structure and multipole
  radiation revisited"},\ }\href
  {https://doi.org/10.1103/PhysRevLett.104.039101} {\bibfield  {journal}
  {\bibinfo  {journal} {Phys. Rev. Lett.}\ }\textbf {\bibinfo {volume} {104}},\
  \bibinfo {pages} {039101} (\bibinfo {year} {2010})}\BibitemShut {NoStop}%
\bibitem [{\citenamefont {Kawakami}\ and\ \citenamefont
  {Nishizawa}(1968)}]{Kawakami68}%
  \BibitemOpen
  \bibfield  {author} {\bibinfo {author} {\bibfnamefont {S.}~\bibnamefont
  {Kawakami}}\ and\ \bibinfo {author} {\bibfnamefont {J.}~\bibnamefont
  {Nishizawa}},\ }\bibfield  {title} {\bibinfo {title} {An optical waveguide
  with the optimum distribution of the refractive index with reference to
  waveform distortion},\ }\href {https://doi.org/10.1109/TMTT.1968.1126797}
  {\bibfield  {journal} {\bibinfo  {journal} {IEEE Trans. Microw. Theory
  Techn.}\ }\textbf {\bibinfo {volume} {16}},\ \bibinfo {pages} {814} (\bibinfo
  {year} {1968})}\BibitemShut {NoStop}%
\bibitem [{\citenamefont {Simon}\ and\ \citenamefont
  {Mukunda}(1993)}]{Simon93}%
  \BibitemOpen
  \bibfield  {author} {\bibinfo {author} {\bibfnamefont {R.}~\bibnamefont
  {Simon}}\ and\ \bibinfo {author} {\bibfnamefont {N.}~\bibnamefont
  {Mukunda}},\ }\bibfield  {title} {\bibinfo {title} {Bargmann invariant and
  the geometry of g$\rm\ddot{u}$oy effect},\ }\bibfield  {journal} {\bibinfo
  {journal} {Phys. Rev. Lett.}\ }\textbf {\bibinfo {volume} {70}},\ \href
  {https://doi.org/10.1103/PhysRevLett.70.880} {10.1103/PhysRevLett.70.880}
  (\bibinfo {year} {1993})\BibitemShut {NoStop}%
\bibitem [{\citenamefont {Barnett}\ \emph
  {et~al.}(2016{\natexlab{b}})\citenamefont {Barnett}, \citenamefont
  {Babiker},\ and\ \citenamefont {Padgett}}]{Barnett16b}%
  \BibitemOpen
  \bibfield  {author} {\bibinfo {author} {\bibfnamefont {S.~M.}\ \bibnamefont
  {Barnett}}, \bibinfo {author} {\bibfnamefont {M.}~\bibnamefont {Babiker}},\
  and\ \bibinfo {author} {\bibfnamefont {M.~J.}\ \bibnamefont {Padgett}},\
  }\bibfield  {title} {\bibinfo {title} {Optical orbital angular momentum},\
  }\href {https://doi.org/10.1098/rsta.2015.0444} {\bibfield  {journal}
  {\bibinfo  {journal} {Phil. Trans. R. Soc. A}\ }\textbf {\bibinfo {volume}
  {375}},\ \bibinfo {pages} {20150444} (\bibinfo {year}
  {2016}{\natexlab{b}})}\BibitemShut {NoStop}%
\bibitem [{\citenamefont {Chuang}(2009)}]{Chuang09}%
  \BibitemOpen
  \bibfield  {author} {\bibinfo {author} {\bibfnamefont {S.~L.}\ \bibnamefont
  {Chuang}},\ }\href@noop {} {\emph {\bibinfo {title} {Physics of Photonic
  Devices}}}\ (\bibinfo  {publisher} {Wiley},\ \bibinfo {year}
  {2009})\BibitemShut {NoStop}%
\bibitem [{\citenamefont {Pancharatnam}(1956)}]{Pancharatnam56}%
  \BibitemOpen
  \bibfield  {author} {\bibinfo {author} {\bibfnamefont {S.}~\bibnamefont
  {Pancharatnam}},\ }\bibfield  {title} {\bibinfo {title} {Generalized theory
  of interference, and its applications},\ }\href@noop {} {\bibfield  {journal}
  {\bibinfo  {journal} {Proc. Indian Acad. Sci., Sect. A}\ }\textbf {\bibinfo
  {volume} {XLIV}},\ \bibinfo {pages} {398} (\bibinfo {year}
  {1956})}\BibitemShut {NoStop}%
\bibitem [{\citenamefont {Berry}(1984)}]{Berry84}%
  \BibitemOpen
  \bibfield  {author} {\bibinfo {author} {\bibfnamefont {M.~V.}\ \bibnamefont
  {Berry}},\ }\bibfield  {title} {\bibinfo {title} {Quantual phase factors
  accompanying adiabatic changes},\ }\bibfield  {journal} {\bibinfo  {journal}
  {Proc. R. Sco. Lond. A}\ }\textbf {\bibinfo {volume} {392}},\ \href
  {https://doi.org/10.1098/rspa.1984.0023} {10.1098/rspa.1984.0023} (\bibinfo
  {year} {1984})\BibitemShut {NoStop}%
\bibitem [{\citenamefont {Tomita}\ and\ \citenamefont {Cao}(1986)}]{Tomita86}%
  \BibitemOpen
  \bibfield  {author} {\bibinfo {author} {\bibfnamefont {A.}~\bibnamefont
  {Tomita}}\ and\ \bibinfo {author} {\bibfnamefont {R.~Y.}\ \bibnamefont
  {Cao}},\ }\bibfield  {title} {\bibinfo {title} {Observation of {B}erry's
  topological phase by use of an optical fiber},\ }\href
  {https://doi.org/10.1103/PhysRevLett.57.937} {\bibfield  {journal} {\bibinfo
  {journal} {Phys. Rev. Lett.}\ }\textbf {\bibinfo {volume} {57}},\ \bibinfo
  {pages} {937} (\bibinfo {year} {1986})}\BibitemShut {NoStop}%
\bibitem [{\citenamefont {Hamazaki}\ \emph {et~al.}(2006)\citenamefont
  {Hamazaki}, \citenamefont {Y}, \citenamefont {Oka},\ and\ \citenamefont
  {Morita}}]{Hamazaki06}%
  \BibitemOpen
  \bibfield  {author} {\bibinfo {author} {\bibfnamefont {J.}~\bibnamefont
  {Hamazaki}}, \bibinfo {author} {\bibfnamefont {M.}~\bibnamefont {Y}},
  \bibinfo {author} {\bibfnamefont {K.}~\bibnamefont {Oka}},\ and\ \bibinfo
  {author} {\bibfnamefont {R.}~\bibnamefont {Morita}},\ }\bibfield  {title}
  {\bibinfo {title} {Direct observation of pouy phase shift in a propagating
  optical vortex},\ }\href {https://doi.org/10.1364/OE.14.008382} {\bibfield
  {journal} {\bibinfo  {journal} {Opt. Exp,}\ }\textbf {\bibinfo {volume}
  {14}},\ \bibinfo {pages} {8382} (\bibinfo {year} {2006})}\BibitemShut
  {NoStop}%
\bibitem [{\citenamefont {Bliokh09}(2009)}]{Bliokh09}%
  \BibitemOpen
  \bibfield  {author} {\bibinfo {author} {\bibnamefont {Bliokh09}},\ }\bibfield
   {title} {\bibinfo {title} {Geometrodynamics of polarized light: {B}erry
  phase and spin {H}all effecct in a gradient-index medium},\ }\href
  {https://doi.org/10.1088/1464-4258/11/9/094009} {\bibfield  {journal}
  {\bibinfo  {journal} {J. Opt. A: Pure Appl Opt.}\ ,\ \bibinfo {pages}
  {094009}} (\bibinfo {year} {2009})}\BibitemShut {NoStop}%
\bibitem [{\citenamefont {Fox}(2006)}]{Fox06}%
  \BibitemOpen
  \bibfield  {author} {\bibinfo {author} {\bibfnamefont {M.}~\bibnamefont
  {Fox}},\ }\href@noop {} {\emph {\bibinfo {title} {Quantum Optics: An
  Introduction}}}\ (\bibinfo  {publisher} {Oxford University Press},\ \bibinfo
  {year} {2006})\BibitemShut {NoStop}%
\bibitem [{\citenamefont {Parker}(2005)}]{Parker05}%
  \BibitemOpen
  \bibfield  {author} {\bibinfo {author} {\bibfnamefont {M.~A.}\ \bibnamefont
  {Parker}},\ }\href@noop {} {\emph {\bibinfo {title} {Physics of
  Optoelectronics}}}\ (\bibinfo  {publisher} {Tylor \& Francis},\ \bibinfo
  {year} {2005})\BibitemShut {NoStop}%
\bibitem [{\citenamefont {Nambu}(1960)}]{Nambu59}%
  \BibitemOpen
  \bibfield  {author} {\bibinfo {author} {\bibfnamefont {Y.}~\bibnamefont
  {Nambu}},\ }\bibfield  {title} {\bibinfo {title} {Quasi-particles and gauge
  invariance in the theory of superconductivity},\ }\bibfield  {journal}
  {\bibinfo  {journal} {Phys. Rev.}\ }\textbf {\bibinfo {volume} {117}},\ \href
  {https://doi.org/10.1103/PhysRev.117.648} {10.1103/PhysRev.117.648} (\bibinfo
  {year} {1960})\BibitemShut {NoStop}%
\bibitem [{\citenamefont {Anderson}(1958)}]{Anderson58}%
  \BibitemOpen
  \bibfield  {author} {\bibinfo {author} {\bibfnamefont {P.~W.}\ \bibnamefont
  {Anderson}},\ }\bibfield  {title} {\bibinfo {title} {Random-phase
  approximation in the theory of superconductivity},\ }\href@noop {} {\bibfield
   {journal} {\bibinfo  {journal} {Phys. Rev.}\ }\textbf {\bibinfo {volume}
  {112}},\ \bibinfo {pages} {1900} (\bibinfo {year} {1958})}\BibitemShut
  {NoStop}%
\bibitem [{\citenamefont {Goldstone}\ \emph {et~al.}(1962)\citenamefont
  {Goldstone}, \citenamefont {Salam},\ and\ \citenamefont
  {Weinberg}}]{Goldstone62}%
  \BibitemOpen
  \bibfield  {author} {\bibinfo {author} {\bibfnamefont {J.}~\bibnamefont
  {Goldstone}}, \bibinfo {author} {\bibfnamefont {A.}~\bibnamefont {Salam}},\
  and\ \bibinfo {author} {\bibfnamefont {S.}~\bibnamefont {Weinberg}},\
  }\bibfield  {title} {\bibinfo {title} {Broken symmetries},\ }\href@noop {}
  {\bibfield  {journal} {\bibinfo  {journal} {Phy. Rev.}\ }\textbf {\bibinfo
  {volume} {127}},\ \bibinfo {pages} {965} (\bibinfo {year}
  {1962})}\BibitemShut {NoStop}%
\bibitem [{\citenamefont {Higgs}(1962)}]{Higgs64}%
  \BibitemOpen
  \bibfield  {author} {\bibinfo {author} {\bibfnamefont {P.~W.}\ \bibnamefont
  {Higgs}},\ }\bibfield  {title} {\bibinfo {title} {Broken symmetries, massless
  particles and gauge fields},\ }\href@noop {} {\bibfield  {journal} {\bibinfo
  {journal} {Phys. Lett}\ }\textbf {\bibinfo {volume} {12}},\ \bibinfo {pages}
  {132} (\bibinfo {year} {1962})}\BibitemShut {NoStop}%
\bibitem [{\citenamefont {Georgi}(1999)}]{Georgi99}%
  \BibitemOpen
  \bibfield  {author} {\bibinfo {author} {\bibfnamefont {H.}~\bibnamefont
  {Georgi}},\ }\href@noop {} {\emph {\bibinfo {title} {Lie Algebras in Particle
  Physics: from Isospin to Unified Theories (Frontiers in Physics)}}}\
  (\bibinfo  {publisher} {Westview Press},\ \bibinfo {year} {1999})\BibitemShut
  {NoStop}%
\bibitem [{\citenamefont {Fano}(1954)}]{Fano54}%
  \BibitemOpen
  \bibfield  {author} {\bibinfo {author} {\bibfnamefont {U.}~\bibnamefont
  {Fano}},\ }\bibfield  {title} {\bibinfo {title} {A stokes-parameter technique
  for the treatment of polarization in quantum mechnics},\ }\href@noop {}
  {\bibfield  {journal} {\bibinfo  {journal} {Phy. Rev.}\ }\textbf {\bibinfo
  {volume} {93}},\ \bibinfo {pages} {121} (\bibinfo {year} {1954})}\BibitemShut
  {NoStop}%
\bibitem [{\citenamefont {Delbourgo}(1977)}]{Delbourgo77}%
  \BibitemOpen
  \bibfield  {author} {\bibinfo {author} {\bibfnamefont {R.}~\bibnamefont
  {Delbourgo}},\ }\bibfield  {title} {\bibinfo {title} {Minimal uncertainty
  states for the rotaion and allied groups},\ }\href
  {https://doi.org/10.1088/0305-4470/10/11/012} {\bibfield  {journal} {\bibinfo
   {journal} {J. Phys. A: Math. Gen}\ }\textbf {\bibinfo {volume} {10}},\
  \bibinfo {pages} {1837} (\bibinfo {year} {1977})}\BibitemShut {NoStop}%
\bibitem [{\citenamefont {Barnett}\ \emph {et~al.}(2012)\citenamefont
  {Barnett}, \citenamefont {Cameron},\ and\ \citenamefont {Yao}}]{Barnett12}%
  \BibitemOpen
  \bibfield  {author} {\bibinfo {author} {\bibfnamefont {S.~M.}\ \bibnamefont
  {Barnett}}, \bibinfo {author} {\bibfnamefont {R.~P.}\ \bibnamefont
  {Cameron}},\ and\ \bibinfo {author} {\bibfnamefont {A.~M.}\ \bibnamefont
  {Yao}},\ }\bibfield  {title} {\bibinfo {title} {Duplex symmetry and its
  relation to the conservation of optical helicity},\ }\href
  {https://doi.org/10.1103/PhysRevA.86.013845} {\bibfield  {journal} {\bibinfo
  {journal} {Phys. Rev. A}\ }\textbf {\bibinfo {volume} {86}},\ \bibinfo
  {pages} {013845} (\bibinfo {year} {2012})}\BibitemShut {NoStop}%
\bibitem [{\citenamefont {Padgett}\ and\ \citenamefont
  {Courtial}(1999)}]{Padgett99}%
  \BibitemOpen
  \bibfield  {author} {\bibinfo {author} {\bibfnamefont {M.~J.}\ \bibnamefont
  {Padgett}}\ and\ \bibinfo {author} {\bibfnamefont {J.}~\bibnamefont
  {Courtial}},\ }\bibfield  {title} {\bibinfo {title}
  {Poincar$\rm\acute{e}$-sphere equivalent for light beams containing orbital
  angular momentum},\ }\href {https://doi.org/10.1364/OL.24.000430} {\bibfield
  {journal} {\bibinfo  {journal} {Opt. Lett.}\ }\textbf {\bibinfo {volume}
  {24}},\ \bibinfo {pages} {430} (\bibinfo {year} {1999})}\BibitemShut
  {NoStop}%
\bibitem [{\citenamefont {Milione}\ \emph {et~al.}(2011)\citenamefont
  {Milione}, \citenamefont {Sztul}, \citenamefont {Nolan},\ and\ \citenamefont
  {Alfano}}]{Milione11}%
  \BibitemOpen
  \bibfield  {author} {\bibinfo {author} {\bibfnamefont {G.}~\bibnamefont
  {Milione}}, \bibinfo {author} {\bibfnamefont {H.~I.}\ \bibnamefont {Sztul}},
  \bibinfo {author} {\bibfnamefont {D.~A.}\ \bibnamefont {Nolan}},\ and\
  \bibinfo {author} {\bibfnamefont {R.~R.}\ \bibnamefont {Alfano}},\ }\bibfield
   {title} {\bibinfo {title} {Higher-order poincar$\rm\acute{e}$ sphere, stokes
  parameters, and the angular momentum of light},\ }\href
  {https://doi.org/10.1103/PhysRevLett.107.053601} {\bibfield  {journal}
  {\bibinfo  {journal} {Phys. Rev. Lett.}\ }\textbf {\bibinfo {volume} {107}},\
  \bibinfo {pages} {053601} (\bibinfo {year} {2011})}\BibitemShut {NoStop}%
\bibitem [{\citenamefont {Liu}\ \emph {et~al.}(2017)\citenamefont {Liu},
  \citenamefont {Liu}, \citenamefont {Ke}, \citenamefont {Liu}, \citenamefont
  {Shu}, \citenamefont {Luo},\ and\ \citenamefont {Wen}}]{Liu17}%
  \BibitemOpen
  \bibfield  {author} {\bibinfo {author} {\bibfnamefont {Z.}~\bibnamefont
  {Liu}}, \bibinfo {author} {\bibfnamefont {Y.}~\bibnamefont {Liu}}, \bibinfo
  {author} {\bibfnamefont {Y.}~\bibnamefont {Ke}}, \bibinfo {author}
  {\bibfnamefont {Y.}~\bibnamefont {Liu}}, \bibinfo {author} {\bibfnamefont
  {W.}~\bibnamefont {Shu}}, \bibinfo {author} {\bibfnamefont {H.}~\bibnamefont
  {Luo}},\ and\ \bibinfo {author} {\bibfnamefont {S.}~\bibnamefont {Wen}},\
  }\bibfield  {title} {\bibinfo {title} {Generation of arbitrary vector vortex
  beams on hybrid-order poincar$\rm\acute{e}$ sphere},\ }\href
  {https://doi.org/10.1364/PRJ.5.000015} {\bibfield  {journal} {\bibinfo
  {journal} {Photon. Res.}\ }\textbf {\bibinfo {volume} {5}},\ \bibinfo {pages}
  {15} (\bibinfo {year} {2017})}\BibitemShut {NoStop}%
\bibitem [{\citenamefont {Erhard}\ \emph {et~al.}(2018)\citenamefont {Erhard},
  \citenamefont {Fickler}, \citenamefont {Krenn},\ and\ \citenamefont
  {Zeilinger}}]{Erhard18}%
  \BibitemOpen
  \bibfield  {author} {\bibinfo {author} {\bibfnamefont {M.}~\bibnamefont
  {Erhard}}, \bibinfo {author} {\bibfnamefont {R.}~\bibnamefont {Fickler}},
  \bibinfo {author} {\bibfnamefont {M.}~\bibnamefont {Krenn}},\ and\ \bibinfo
  {author} {\bibfnamefont {A.}~\bibnamefont {Zeilinger}},\ }\bibfield  {title}
  {\bibinfo {title} {Twisted photons: new quantum perspectives in high
  dimensions},\ }\bibfield  {journal} {\bibinfo  {journal} {Light: Science \&
  Applications}\ }\textbf {\bibinfo {volume} {7}},\ \href
  {https://doi.org/10.1038/lsa.2017.146} {10.1038/lsa.2017.146} (\bibinfo
  {year} {2018})\BibitemShut {NoStop}%
\end{thebibliography}%

\end{document}